 \documentclass[final,5p,times,twocolumn,authoryear]{elsarticle}

\usepackage{amsmath}
\usepackage{amssymb}
\usepackage{lipsum}
\usepackage{bm, color, comment}
\usepackage{hyperref}
\usepackage{multicol}
\usepackage{here}
\usepackage{xcolor}
\hypersetup{
    colorlinks=true,
    citecolor=blue,
    linkcolor=red,
    urlcolor=orange,
}

\journal{New Astronomy}

\begin{document}

\begin{frontmatter}



\title{Novel Hydrodynamic Schemes Capturing Shocks and Contact Discontinuities and Comparison Study with Existing Methods}


\author[first]{Takuhiro Yuasa}
\author[last]{Masao Mori}
\affiliation[first]{organization={Graduate School of Pure and Applied Sciences, University of Tsukuba},
            addressline={1-1-1, Tennoudai}, 
            city={Tsukuba},
            postcode={305-8577}, 
            state={Ibaraki},
            country={Japan}}

\affiliation[last]{organization={Center for Computational Sciences, University of Tsukuba},
            addressline={1-1-1, Tennoudai}, 
            city={Tsukuba},
            postcode={305-8577}, 
            state={Ibaraki},
            country={Japan}}

\begin{abstract}
We present a new hydrodynamic scheme named Godunov Density-Independent Smoothed Particle Hydrodynamics (GDISPH), that can accurately handle shock waves and contact discontinuities without any manually tuned parameters. 
    This is in contrast to the standard formulation of smoothed particle hydrodynamics (SSPH), which requires the parameters for an artificial viscosity term to handle the shocks and struggles to accurately handle the contact discontinuities due to unphysical repulsive forces, resulting in surface tension that disrupts pressure equilibrium and suppresses fluid instabilities. While Godunov SPH (GSPH) can handle the shocks without the parameters by using solutions from a Riemann solver, it still cannot fully handle the contact discontinuities. Density-Independent Smoothed Particle Hydrodynamics (DISPH), one of several schemes proposed to handle contact discontinuities more effectively than SSPH, demonstrates superior performance in our tests involving strong shocks and contact discontinuities. However, DISPH still requires the artificial viscosity term. We integrate the Riemann solver into DISPH in several ways, yielding some patterns of GDISPH. The results of standard tests such as the one-dimensional Riemann problem, pressure equilibrium, Sedov-Taylor, and Kelvin-Helmholtz tests are favourable to GDISPH Case 1 and GDISPH Case 2, as well as DISPH. We conclude that GDISPH Case 1 \textcolor{black}{has an advantage over GDISPH Case 2, }effectively handling shocks and contact discontinuities \textcolor{black}{without the need for specific parameters or introducing any additional numerical diffusion.}
\end{abstract}



\begin{keyword}
methods: numerical \sep hydrodynamics \sep shock waves \sep waves



\end{keyword}

\end{frontmatter}




\section{Introduction}
\label{sec:int}
Structure formation in the Universe is one of the fundamental challenges in astronomy and astrophysics. After the Big Bang, baryons condense due to global and continuous motions excited by gravitational interactions, eventually giving rise to various structures such as stars and galaxies. It is known that the compressible fluid approximation is well established in such astrophysical flows, and shock waves and contact discontinuities frequently emerge. In this context, smoothed particle hydrodynamics (SPH) has played an essential role as a tool for investigating astrophysical flows. However, there is still much room for improvement.

SPH is a mesh-free, Lagrangian scheme to solve the evolution of fluid using particles.
It has advantages over grid-based numerical methods in terms of its ease of programming, \textcolor{black}{Lagrangian character}, incorporating gravity interactions and chemical reactions, and free boundaries.
SPH was originally introduced by \citet{Lucy1977} and \citet{Gingold_Monaghan1977} and has been widely used in the field of computational astrophysics because of its simplicity and usefulness.
SPH has also found widespread use in other areas of science and engineering.
A broad discussion of the standard formulation of SPH (SSPH) can be found in the review by \citet{Springel2010}.
While these striking developments, several problems have been identified with SSPH. Especially we focus on the following two problems in this paper.
First, SSPH needs an artificial viscosity term with manually tuned parameters in order to add adequate viscosity in shock regions to handle shock waves correctly.
Second, SPH is known to have difficulty accurately capturing contact discontinuities due to the small non-physical jumps arising pre-/post- contact discontinuities.

So far, several formulae for artificial viscosity have been proposed, and in particular, the forms introduced by \citet{Monaghan1983} and \citet{Monaghan1997} are widely used in practical calculations using SPH.
When the artificial viscosity is too weak, post-shock oscillations occur; conversely, the shock fronts become too blunt when that is too strong. 
Furthermore, artificial viscosities commonly produce unnecessary viscosity outside of the shock regions, especially in areas with shear flows. This is because these viscosities incorrectly identify regions where particles are approaching each other as shock regions.
The Balsara switch, introduced by \citet{Balsara1995}, is a method to cure the problem of artificial viscosities in the shear flow regions. Because the Balsara switch only applies a coefficient that goes to zero in the shear flow regions to the artificial viscosity term, it can be easily implemented to various types of artificial viscosities and is used widely. \textcolor{black}{The time- and space-dependent artificial viscosity coefficient coupled with high-order divergence and vorticity for the Balsara switch, introduced by \citet{Beck2016}, shows an incredible performance in terms of suppressing the unnecessary viscosity in the shear flow regions.}

Godunov SPH (GSPH), developed by \citet{Inutsuka2002}, uses solutions of the Riemann problem, which has well-known algorithms (Riemann solver) to obtain analytical solutions, to evaluate the pressure gradient forces acting on each particle. 
This procedure is an excellent method for accurately treating the shock waves without the manual parameter adjustments required in SPH. In other words, this scheme has the effective viscosity, which is equivalent effect of automatically adding the appropriate viscosity to the shock regions.
However, GSPH introduced by \citet{Inutsuka2002} has to use the Gaussian kernel, which does not have compact support.
In addition, even if the truncated Gaussian kernel is used, the computational cost of GSPH can be significantly higher than that of SSPH.
On the other hand, several simpler versions of GSPH have hitherto been proposed.
\citet{Cha_Whitworth2003} propose a GSPH that is applicable to use arbitrary kernels and is computationally less expensive than the original GSPH \citep{Iwasaki2011}. GSPH still could have the problem of adding the unnecessary effective viscosity outside of the shock regions because when calculating the interaction between $i$-th and $j$-th particles, physical quantities of the two particles are used as initial values of the Riemann problem, and the effective viscosity is added through the solutions if the particles are in a relationship that forms the shocks.
In the shear flow regions, there are always pairs of particles that are approaching each other, which the Riemann solver might recognise as a pair of particles that cause the shocks.
Therefore, prescriptions that suppress the unnecessary viscosity, such as the Balsala switch, will also be effective in GSPH, but its practical implementation, however, is likely to be problematic.

At the contact discontinuities, SPH causes unphysical repulsive force, resulting in effective surface tension. This effective surface tension suppresses the developments of fluid instabilities and pressure equilibrium (\citet{Saitoh_Makino2013}, \citet{Read2010}, \citet{Price2008}).
There are three types of approaches to curing this problem. 
The first is to smooth the internal energy at the contact discontinuities so that the internal energy is as smooth as the density. 
The artificial thermal conductivity, first introduced by \citet{Price2008} to take this approach, is adopted in $Phantom$ \citep{Price2018}\textcolor{black}{, $Gadget3$  \textcolor{black}{updated by \cite{Beck2016}}, $Gasoline2$ \citep{Wadsley2017}, $SWIFT$ \citep{Schaller2016}, and $gizmo$ \citep{Hopkins2015}}.
The artificial thermal conductivity adds a physically non-existent dissipation to cure the problem at contact discontinuities and involves arbitrary parameters that require manual adjustment to avoid excessive smoothing. 
The second is to use another formula of SPH that is resistant to surface tension.
SPH with Geometric Density Average Force expression (SPH GDF), originally introduced by \citet{Monaghan1992} and used in $Gasoline 2$ \citep{Wadsley2017} is found to be good at handling the contact discontinuities better than SSPH.
Density-Independent SPH (DISPH), introduced by \citet{Saitoh_Makino2013}, is the third approach and can successfully get rid of the effective surface tension without any additional dissipation.
SSPH operates under the assumption that the local density distribution is differentiable, which is not physically accurate at contact discontinuities. Consequently, SSPH exhibits poor performance at these discontinuities. On the other hand, DISPH, introduced by \citep{Saitoh_Makino2013}, does not require this assumption, leading to improved performance over traditional SPH methods at 
the contact discontinuities. However, DISPH still depends on including the artificial viscosity term to adequately capture the shock waves.

In this paper, we present a new hydrodynamic scheme named Godunov DISPH (GDISPH), that can accurately handle the shocks and the contact discontinuities without the manually tuned parameters.
There are various degrees of freedom when integrating the Riemann solver into SPH. Several methods for integrating Riemann solvers into DISPH are proposed and the performance of each method is evaluated.
\textcolor{black}{For shear dominant-flow, we} devise a way to implement the Balsara switch into GDISPH.
The structure of this paper is as follows.
In Section \ref{sec:major}, we show several SPH-based schemes 
and their problems with the contact discontinuities and the artificial viscosities.
Section \ref{sec:incorporation} describes how to incorporate the Riemann solver into DISPH.
In Section \ref{sec:num}, 
we compare the results of test calculations with several existing methods and the three realised GDISPH methods such as GDISPH Case 1, GDISPH Case 2, and GDISPH Case 3.
In Section \ref{sec:conc}, a summary and discussion are presented.


\section{Review of Several SPH-based Methods and their \textcolor{black}{Problems}} 
\label{sec:major}
In this paper, we consider the following set of equations for non-radiating inviscid fluid:
\begin{equation}\label{eq:Major1}
    \frac{d\rho}{dt} = -\rho \nabla \cdot \bm{v},
\end{equation}

\begin{equation}\label{eq:Major2}
    \frac{d\bm{v}}{dt} = -\frac{1}{\rho} \nabla P,
\end{equation}
and
\begin{equation}\label{eq:Major3}
    \frac{du}{dt} = -\frac{P}{\rho} \nabla \cdot \bm{v},
\end{equation}
where $\rho$, $\bm{v}$, $P$, and $u$ are the density, velocity, pressure, and internal energy per unit mass of the fluid, respectively.
We consider an ideal gas, so the equation of state is defined as follows:
\begin{equation}\label{eq:Major4}
    P = (\gamma - 1)\rho u,
\end{equation}
where $\gamma$ is the specific heat ratio.
The following subsections show the SPH-based methods and their problems with contact discontinuities and artificial viscosity.

\subsection{Standard SPH}
\textcolor{black}{The standard SPH (SSPH), which was first introduced by \citet{Springel2002}, is one of the most widely used formulations in various codes \citep[e.g.][]{Springel2005,Price2018}.}
In this formulation, the momentum equation and the energy equation of the $i$-th particle are as follows:
\begin{equation}\label{eq:over1}
    m_i\frac{d\bm{v}_i}{dt} = - \displaystyle \sum^N_{j=1} m_i m_j \left[ f^{\textcolor{black}{\text{grad}}}_i \frac{P_i}{\rho^2_i} \nabla_i W_{ij}(h_i) + f^{\textcolor{black}{\text{grad}}}_j \frac{P_j}{\rho^2_j} \nabla_i W_{ij}(h_j)\right],
\end{equation}
and
\begin{equation}\label{eq:over2}
    \frac{du_i}{dt} = f^{\textcolor{black}{\text{grad}}}_i \frac{P_i}{\rho^2_i} \displaystyle \sum^N_{j=1} m_j \bm{v}_{ij} \cdot \nabla_i W_{ij}(h_i),
\end{equation}
where $N$ is the number of particles, \textcolor{black}{kernel} $W_{ij}(h) = W(|\bm{r}_i - \bm{r}_j|,h)$, \textcolor{black}{velocity difference between $i$-th and $j$-th particles} $\bm{v}_{ij} = \bm{v}_{i} - \bm{v}_{j}$, and the coefficients:
\begin{equation}
    f^{\textcolor{black}{\text{grad}}}_i = \left(1 + \frac{h_i}{D\rho_i}\frac{\partial \rho_i}{\partial h_i} \right)^{-1},
\end{equation}
where $D$ is the spacial dimension, \textcolor{black}{appear} by considering the spatial derivative of the smoothing length $h$.
Any physical quantity at any location is defined as follows:
\begin{equation}\label{eq:over21}
    f(\bm{r}) = \displaystyle \textcolor{black}{\sum^N_{j=1}} m_j \frac{f_j}{\rho_j} W(|\bm{r}-\bm{r}_j|,h(\bm{r})).
\end{equation}
Therefore, the density of the $i$-th particle is given by
\begin{equation}\label{eq:over3}
    \rho_i = \displaystyle \sum^N_{j=1} m_j W_{ij}(h_i).
\end{equation}
The smoothing length $h_i$ is updated so that 
\textcolor{black}{
\begin{equation}\label{eq:condic}
    \rho_i A (\xi h_i)^D = m_i N_{\text{ngb}},
\end{equation}
where $\xi$ is defined such that $W(x h_i, h_i) = 0$ if $x>\xi$ and $W(x h_i, h_i) \neq 0$ if $x<\xi$ under the use of compact supported kernel, and constant $A$ is defined such that $A (\xi h_i)^D$ is the volume of the $D$-dimensional sphere with the radius of $\xi h_i$ (e.g. $A=4/3 \pi$ when $D=3$ and $A=2$ when $D=1$), 
}
is satisfied for any particles at any time.
\textcolor{black}{There are several ways to impose the condition of equation (\ref{eq:condic}). The first, for example, is to iterate by updating the density using equation (\ref{eq:over3}) and then updating the smoothing length using equation (\ref{eq:condic}) until both the density and the smoothing length converge. \textcolor{black}{In this case, $N_{\text{ngb}}$ can be seen as effective neighbour number.}
\textcolor{black}{The second is to update the smoothing length such that the number of particles inside the radius of smoothing radius $\xi h_i$ centred on the $i$-th particle is nearly equal to $N_\text{ngb}$ and then update the density using equation (\ref{eq:over3}), even though this method does not strictly satisfy the condition (\ref{eq:condic}). In this case, $N_{\text{ngb}}$ can be seen as true neighbour number.}}

Usually, an artificial viscosity term 
is added to the momentum equation and the energy equation as additional terms.
In this paper, we mainly use the artificial viscosity introduced by \citet{Monaghan1997}.
The artificial viscosity terms for the momentum and energy equations are given by
\begin{equation}\label{eq:over4}
    \left. m_i\frac{d\bm{v}_i}{dt} \right |_{\textcolor{black}{\text{visc}}} = -\displaystyle \sum^N_{j=1} m_im_j \Pi_{ij} \nabla_i  \overline{W_{ij}},
\end{equation}
and
\begin{equation}\label{eq:over5}
    \left. \frac{du_i}{dt} \right |_{\textcolor{black}{\text{visc}}} = \frac{1}{2} \displaystyle \sum^N_{j=1} m_j \Pi_{ij} \bm{v}_{ij} \cdot \nabla_i \overline{W_{ij}}.
\end{equation}
Here $\nabla_i \overline{W_{ij}}$ and $\Pi_{ij}$ are defined as
\begin{equation}\label{eq:over4-1}
    \nabla_i \overline{W_{ij}} = \frac{\nabla_i W_{ij}(h_i) + \nabla_i W_{ij}(h_j)}{2},
\end{equation}
and
\begin{equation}\label{eq:over6}
    \begin{split}
        \Pi_{ij} &= \begin{cases}
            -\alpha_{AV} \frac{v^{\textcolor{black}{\text{sig}}}_{ij}w_{ij}}{\rho_i + \rho_j} & \text{if $\bm{v}_{ij}\cdot \bm{r}_{ij} < 0$,} \\
            0          & \text{if $\bm{v}_{ij}\cdot \bm{r}_{ij} \geq 0$,}
        \end{cases}\\
    \end{split}
\end{equation}
where \textcolor{black}{the estimate of the signal velocity between the two particles} $v^{\textcolor{black}{\text{sig}}}_{ij} = c_i + c_j - 3w_{ij}$ \textcolor{black}{with the sound speed as $c$}, \textcolor{black}{ $w_{ij} = \bm{v}_{ij} \cdot \bm{r}_{ij} / |\bm{r}_{ij}|$ with $\bm{r}_{ij} = \bm{r}_i - \bm{r}_j$}, and $\alpha_{\textcolor{black}{\text{AV}}}$ is the arbitrary parameter that adjusts the strength of the artificial viscosity.

SSPH has well-known several problems, and we focus on two problems.
First, SSPH needs the artificial viscosity term with manually tuned parameters
in order to add adequate viscosity in shock regions to handle shock waves correctly.
It is evident that the simulation outcomes can significantly differ based on these parameter values.
Hence, it becomes crucial to determine the optimal parameters that are substantial enough to capture the shocks effectively, yet subtle enough to preserve the overall solution's sharpness. Notably, these optimal parameters are problem-specific, varying with physical conditions.
In addition, it is a well-known fact that artificial viscosities act more than necessary in shear flows despite the outside of the shock regions.
This is very problematic, especially for the simulation of the Kelvin-Helmholtz instability and \textcolor{black}{Kepler} disk. \citet{Hosono2016} found that in the \textcolor{black}{Kepler} disk simulation, the angular momentum transfer due to the artificial viscosities at the inner edge is the primary reason why the breaking up of the disk happens. \textcolor{black}{\citet{Beck2016}'s method, which combines the use of the time- and space-dependent artificial viscosity coefficient with the higher accuracy of the divergence and vorticity for the Balsara switch, successfully suppresses the induced transport of angular momentum.}
\\
\textcolor{black}{The} Balsara switch, introduced by \citet{Balsara1995}, is a method to reduce
the strength of the artificial viscosity in the shear flow regions.
The Balsara switch replaces $\Pi_{ij}$ with $\Pi^{\textcolor{black}{\text{Balsara}}}_{ij}$ as follows:
\begin{equation}\label{eq:over7-1}
    \Pi^{\textcolor{black}{\text{Balsara}}}_{ij} = \frac{F^{\textcolor{black}{\text{Balsara}}}_i+F^{\textcolor{black}{\text{Balsara}}}_j}{2}\Pi_{ij},
\end{equation}
with
\begin{equation}\label{eq:over7}
    F^{\textcolor{black}{\text{Balsara}}}_i = \frac{|\nabla_i \cdot \bm{v}_i |}{|\nabla_i \cdot \bm{v}_i | + |\nabla_i \times \bm{v}_i| + 0.0001 c_i/h_i}.
\end{equation}
In the shock regions, $|\nabla_i \times \bm{v}_i| \ll |\nabla_i \cdot \bm{v}_i |$ is satisfied,
therefore $F^{\textcolor{black}{\text{Balsara}}}_i \to 1$. While in the shear regions, $|\nabla_i \cdot \bm{v}_i | \ll |\nabla_i \times \bm{v}_i|$ is satisfied, therefore $F^{\textcolor{black}{\text{Balsara}}}_i \to 0$.
\citet{Cullen_Dehnen2010} also introduced \textcolor{black}{a} method similar to the Balsara switch.

Second, SSPH has difficulty handling contact discontinuities.
SSPH assumes that the local density distribution is differentiable.
This assumption breaks down \textcolor{black}{both at the contact discontinuities and the shock regions}, where the density is discontinuous physically.
As a result, some concerns are raised there for SSPH.
\textcolor{black}{Here, we focus on the concerns at the contact discontinuities.}
For example, firstly, the density of each particle is evaluated through the kernel interpolation using equation (\ref{eq:over3}), 
while the internal energy is updated by the time integral using equation (\ref{eq:over2}).
This approach makes the density of the SPH particle smoother than the internal energy of the SPH particle at the contact discontinuities, where the density and the internal energy are discontinuous physically.
Therefore, the smoothness of the quantities at the contact discontinuities becomes inconsistent with each other,
causing pressure blips there since the pressure of each particle is calculated by the equation (\ref{eq:Major4}) \citep{Price2008}.
Secondly, in the derivation of SSPH, $\nabla \rho$ is used, which is mathematically undefined at density discontinuities. Therefore SSPH performance should decrease where density gradients are large \citep{Ritchie2001}.
Finally, SSPH approximates the volume element of each particle by $m/\rho$, so the
accuracy is reduced at the density discontinuities.
Consequently, these approaches result in suboptimal performance when handling contact discontinuities.

\citet{Price2008} introduced artificial thermal conductivity to eliminate the effective surface tension by smoothing the internal energy at the contact discontinuities so that the smoothness of the density and the internal energy becomes consistent.
The artificial thermal conductivity term is as follows:
\begin{equation}\label{eq:over7-2}
    \left. \frac{du_i}{dt} \right|_{\textcolor{black}{\text{cond}}} = \alpha_u \displaystyle \sum^N_{\textcolor{black}{j=1}} \frac{m_j}{\overline{\rho_{ij}}}v^{\textcolor{black}{\text{sig,u}}}_{ij}(u_i - u_j) \frac{\bm{r}_{ij}}{|\bm{r}_{ij}|}\cdot \nabla_i \overline{W_{ij}},
\end{equation}
where signal velocity $v^{\textcolor{black}{\text{sig,u}}}_{ij}$ and $\overline{\rho_{ij}}$ are defined by
\begin{equation}\label{eq:over7-3}
    v^{\textcolor{black}{\text{sig,u}}}_{ij} = \sqrt{\frac{|P_i - P_j|}{\overline{\rho_{ij}}}},
\end{equation}
and
\begin{equation}\label{eq:over7-4}
    \overline{\rho_{ij}} = \frac{\rho_i + \rho_j}{2},
\end{equation}
where $\alpha_u$ is the arbitrary parameter to adjust the strength of the artificial thermal conductivity.
The recommended value for $\alpha_u$ is $1$ \citep{Price2018}.
The artificial thermal conductivity is also used in $Phantom$ \citep{Price2018}.
It is very clear that the results of the simulation can vary depending on the value of the parameter \textcolor{black}{as demonstrated by \citet{Hosono20161b}}.
Therefore there is a need to fine-tune the parameter.

\subsection{SPH with Geometric Density Average Force Expression}
In \citet{Monaghan1992}, the following expression:
\begin{equation}
    \frac{\nabla P}{\rho} = \frac{1}{\rho^{2-\sigma}}\nabla\left( \frac{P}{\rho^{\sigma-1}} \right) + \frac{P}{\rho^{\sigma}}\nabla\left( \frac{1}{\rho^{1-\sigma}} \right). \label{eq:avg}
\end{equation}
was proposed as a somewhat technical way of dealing with the pressure gradient term. 
SPH with Geometric Density Average Force Expression (SPH GDF), corresponding to the blending parameter $\sigma=1$ in this equation, is the method used in $Gasoline 2$ \textcolor{black}{\citep{Wadsley2017}} since it is found to be good at handling the contact discontinuities better than SSPH.
\textcolor{black}{
The equation (\ref{eq:avg}) with $\sigma=1$  becomes trivial mathematically, but in an ad-hoc manner it can yield an interesting momentum equation.
The equation (\ref{eq:avg}) with $\sigma=1$ yields
\begin{equation}
    \frac{\nabla P}{\rho} = \frac{1}{\rho}\nabla P + \frac{P}{\rho}\nabla 1.
\end{equation}
In SPH, any physical quantity at any location is defined by equation (\ref{eq:over21}).
In addition, the spatial derivative of an arbitrary physical quantity is written as follows:
\begin{equation}\label{eq:3}
    \nabla f(\bm{r}) = \displaystyle \sum^N_{j=1} m_j \frac{f_j}{\rho_j} \nabla W(|\bm{r}-\bm{r}_j|,h(\bm{r})).
\end{equation}
Even though this is a strange technique, using equation (\ref{eq:3}), $\nabla 1$ can be written as follows:
\begin{equation}\label{eq:4}
    \nabla 1 = \displaystyle \sum^N_{j=1}  \frac{m_j}{\rho_j} \nabla W(|\bm{r}-\bm{r}_j|,h(\bm{r})).
\end{equation}
Using equation (\ref{eq:3}), $\nabla P$ becomes
\begin{equation}\label{eq:5}
    \nabla P = \displaystyle \sum^N_{j=1} m_j \frac{P_j}{\rho_j} \nabla W(|\bm{r}-\bm{r}_j|,h(\bm{r})).
\end{equation}
Using equation (\ref{eq:4}) and (\ref{eq:5}), the momentum equation for SPH becomes as follows:
\begin{equation}\label{eq:over1--2}
    m_i\frac{d\bm{v}_i}{dt} = - \displaystyle \sum^N_{j=1} m_i m_j \left[\frac{P_i+P_j} {\rho_i\rho_j}\nabla_i W_{ij}(h_i)\right],
\end{equation}
which has a similar form with \citet{Ritchie2001}'s and \citet{Wadsley2017}'s momentum equation.
}
In SPH GDF used in \citet{Wadsley2017}, the momentum equation and the energy equation are as follows:
\begin{equation}\label{eq:GDF1}
    m_i \frac{d\bm{v}_i}{dt} = -\displaystyle \sum^N_{\textcolor{black}{j=1}} m_i m_j \left( \frac{P_i+P_j}{\rho_i\rho_j}\right)\nabla_i \widetilde{W_{ij}},
\end{equation}
and
\begin{equation}\label{eq:GDF2}
    \frac{du_i}{dt} = \displaystyle \sum^N_{\textcolor{black}{j=1}} m_j \left( \frac{P_i}{\rho_i\rho_j}\right)\bm{v}_{ij} \cdot \nabla_i \widetilde{W_{ij}}.
\end{equation}
In $Gasoline 2$, the following form of the kernel is used:
\begin{equation}\label{eq:GDF0}
    W(|\bm{r}|,h) = \frac{1}{h^D} \textcolor{black}{B\left(\frac{|\bm{r}|}{h}\right)},
\end{equation}
where $D$ is the dimension \textcolor{black}{and $B$ is an arbitrary differentiable 
 function}. Note that commonly used kernels satisfy this form. Then, $\nabla_i \widetilde{W_{ij}}$ is defined as follows:
\begin{equation}\label{eq:GDF3}
    \nabla_i \widetilde{W_{ij}} = \frac{1}{2}f_i \nabla_i W(|\bm{r}_{ij}|, h_i) + \frac{1}{2}f_j \nabla_i W(|\bm{r}_{ij}|, h_j),
\end{equation}
and
\begin{equation}\label{eq:GDF4}
    f_i = \displaystyle \sum^N_{\textcolor{black}{j=1}} \frac{m_j}{\rho_i} \bm{r}^2_{ij}\textcolor{black}{B}^{'}\left (\frac{|\bm{r}_{ij}|}{h_i} \right ) / \displaystyle \sum^N_{\textcolor{black}{j=1}} \frac{m_j}{\rho_j} \bm{r}^2_{ij}\textcolor{black}{B}^{'}\left ( \frac{|\bm{r}_{ij}|}{h_i} \right ).
\end{equation}
Any physical quantity at any location is defined by equation (\ref{eq:over21}).
The density of each particle is given by equation (\ref{eq:over3}).
The artificial viscosity terms for the momentum and energy equations are as follows:
\begin{equation}\label{eq:GDF5}
    \left. m_i\frac{d\bm{v}_i}{dt} \right |_{\textcolor{black}{\text{visc}}} = -\displaystyle \sum^N_{j=1} m_i m_j \Pi_{ij} \nabla_i  \widetilde{W_{ij}},
\end{equation}
and 
\begin{equation}\label{eq:GDF6}
    \left. \frac{du_i}{dt} \right |_{\textcolor{black}{\text{visc}}} = \frac{1}{2} \displaystyle \sum^N_{j=1} m_j \Pi_{ij} \bm{v}_{ij} \cdot \nabla_i \widetilde{W_{ij}}.
\end{equation}

\textcolor{black}{\citet{Ritchie2001} stated that momentum equations derived from $\sigma=1$ can minimise the errors due to strong density gradients because $\nabla \rho$ term is not involved in the process of the derivation, showing the improved performance of SPH derived from $\sigma=1$ on the strong density gradient. $\nabla \rho$ mathematically becomes infinite at the contact discontinuity, while in SPH in general, $\rho$ is approximated by continuous functions, leading to having a finite value of $\nabla \rho$. Therefore, at the density discontinuity using $\nabla \rho$ can lead to huge error numerically. On the other hand, it can be understood that SSPH's momentum equation (\ref{eq:over1}) has similar form with the equation (\ref{eq:avg}) with $\sigma=2$, which has the $\nabla \rho$, degrading the accuracy of the contact discontinuities.}

\subsection{Godunov SPH}
GSPH, introduced by \citet{Inutsuka2002}, is a method that can handle the shocks without the manually tuned \textcolor{black}{artificial viscosity} parameters.
In GSPH, the force acting on each particle is determined by using the solutions of the Riemann problem, resulting in adding viscosity around the shock regions automatically without the parameters. 
\citet{Cha_Whitworth2003} introduced GSPH Case 3, which is simpler and has less computational cost than the original GSPH, which can be derived by applying the following approximation:
\begin{equation}
    \begin{split}\label{eq:over8}
        &\left[ \nabla_i - \nabla_j \right]W(|\bm{r}-\bm{r}_i|,h(\bm{r}))W(|\bm{r}-\bm{r}_j|,h(\bm{r})) \\
        &\quad\quad\quad\quad = \nabla_i W(|\bm{r}-\bm{r}_i|,h(\bm{r}))W(|\bm{r}-\bm{r}_j|,h(\bm{r}))\\ 
        & \quad\quad\quad\quad\quad\quad\quad\quad - W(|\bm{r}-\bm{r}_i|,h(\bm{r}))\nabla_j W(|\bm{r}-\bm{r}_j|,h(\bm{r}))\\
        &\quad\quad\quad\quad \approx \nabla_i W(|\bm{r}-\bm{r}_i|,h(\bm{r}))\delta(|\bm{r}-\bm{r}_j|) \\
        & \quad\quad\quad\quad\quad\quad\quad\quad - \delta(|\bm{r}-\bm{r}_i|)\nabla_j W(|\bm{r}-\bm{r}_j|,h(\bm{r})),
    \end{split}
\end{equation}
to the original GSPH \citep{Iwasaki2011}.
In GSPH Case 3, the momentum equation and the energy equation are given by
\begin{equation}\label{eq:over9}
    m_i\frac{d\bm{v}_i}{dt} = - \displaystyle \sum^N_{\textcolor{black}{j=1}} m_i m_j P^*_{ij} \left[ \frac{1}{\rho^2_i} \nabla_i W_{ij}(h_i) + \frac{1}{\rho^2_j}\nabla_i W_{ij}(h_j)\right],
\end{equation}
and
\begin{equation}\label{eq:over10}
    \frac{du_i}{dt} = - \displaystyle \sum^N_{\textcolor{black}{j=1}} m_j P^*_{ij} \left( \bm{v}^*_{ij} - \bm{v}_i \right) \cdot \left[ \frac{1}{\rho^2_i} \nabla_i W_{ij}(h_i) + \frac{1}{\rho^2_j}\nabla_i W_{ij}(h_j)\right],
\end{equation}
where $P^*_{ij}$ and $\bm{v}^*_{ij}$ are the pressure and the velocity from the star region, using physical quantities of the $i$-th particle and the $j$-th particle as \textcolor{black}{$\bm{\mathbb{W}}_R$} and \textcolor{black}{$\bm{\mathbb{W}}_L$} in \ref{ap:riemann}.
\textcolor{black}{
Physical quantities of the left side $\bm{\mathbb{W}}_L$ and that of the right side $\bm{\mathbb{W}}_R$ using $x=0$ as a partition is defined as follows:
\begin{equation}\label{eq:initial}
    \bm{\mathbb{W}} =\begin{cases}
                \bm{\mathbb{W}}_L & \text{if $x < 0$,} \\
                \bm{\mathbb{W}}_R & \text{if $x \geq 0$,}
            \end{cases}
\end{equation}
where
\begin{equation}
    \bm{\mathbb{W}}_L = \begin{pmatrix}
                    p_L \\ 
                    v_L \\ 
                    u_L 
                \end{pmatrix},
    \bm{\mathbb{W}}_R = \begin{pmatrix}
                    p_R \\ 
                    v_R \\ 
                    u_R 
                \end{pmatrix},
\end{equation}
and $p$, $v$, and $u$ are the pressure, velocity, and specific internal energy, respectively.
}
Note that we solve the 1D Riemann problem along the line joining the two particles with the $i$-th particle (respectively $j$-th particle) defining the right (left) state in the local coordinate system. 
For the input to the Riemann solver, \textcolor{black}{$v_R$ (respectively $v_L$)} is a component of \textcolor{black}{$\bm{v}_i$ ($\bm{v}_j$)} in the direction of $\bm{r}_i-\bm{r}_j$, 
so the result of the velocity $v^*$ is a component of $\bm{v}^*_{ij}$ in the direction of $\bm{r}_i-\bm{r}_j$.
Since $\nabla_i W_{ij}(h)$ is parallel to $\bm{r}_i-\bm{r}_j$,
we do not have to care about a component of $\bm{v}^*_{ij}$ in the perpendicular direction of $\bm{r}_i-\bm{r}_j$.
Therefore, we can set $P^*_{ij}$ to $p^*$ and $\bm{v}^*_{ij}$ to $v^{*}\left(\bm{r}_i-\bm{r}_j\right) /|\bm{r}_i-\bm{r}_j|$.

\citet{Cha_Whitworth2003} introduced several \textcolor{black}{types of GSPHs including GSPH Case 1, GSPH Case 2, and GSPH Case 3, and also performed the test calculations.}
While GSPH Case 1 and GSPH Case 2, where only the pressure solution of the Riemann problem is used, cause the pressure blips around the contact discontinuities, GSPH Case 3, where the pressure and velocity solution of the Riemann problem is used, have relatively less pressure blips.
Therefore, it is possible that the reason why GSPH by \citet{Cha_Whitworth2003} \textcolor{black}{suppresses} the pressure blips originates from the use of the velocity solution of the Riemann problem. Considering this from a different perspective, using the pressure from the Riemann solver, the GSPH is supposed to give the appropriate effective viscosity in the shock region while using the velocity from the Riemann solver, the effective thermal conductivity of the GSPH at the contact discontinuity obtained without the complicated adjustment parameters.

It is possible that GSPH recognises a pair of particles approaching each other as shock generators and adds effective viscosity through the solutions.
While the artificial viscosity that basically does the same thing has the problem of adding the unnecessary viscosity outside of the shock regions, especially for the shear flow regions, it is conceivable that GSPH has exactly the same problem.
The straightforward solution to this problem is to use the Balsara switch into GSPH, yet this is likely to be difficult because of the need to separate GSPH equations into effective viscous and inviscid terms.

\subsection{DISPH}
DISPH, developed by \citet{Saitoh_Makino2013}, is a method that can be done without setting manual parameters to deal adequately with the contact discontinuities. While SSPH assumes that the density is continuous and differentiable, which is not valid at the contact discontinuities, DISPH assumes that the pressure is continuous and differentiable, which is physically accurate at the contact discontinuities. \textcolor{black}{Note that this assumption still cannot be valid at the shock front.} In DISPH, the momentum equation and the energy equation are as follows:
\begin{equation}\label{eq:over11}
    m_i\frac{d\bm{v}_i}{dt} = - \left( \gamma - 1 \right)\displaystyle \sum^N_{\textcolor{black}{j=1}} U_i U_j \left[\frac{g^{\textcolor{black}{\text{grad}}}_i}{q_i} \nabla_i W_{ij}(h_i) + \frac{g^{\textcolor{black}{\text{grad}}}_j}{q_j} \nabla_i W_{ij}(h_j) \right],
\end{equation}
and
\begin{equation}\label{eq:over12}
    \frac{dU_i}{dt} = \left(\gamma - 1 \right) g^{\textcolor{black}{\text{grad}}}_i  \displaystyle \sum^N_{\textcolor{black}{j=1}} \frac{U_iU_j}{q_i} \bm{v}_{ij} \cdot \nabla_i W_{ij}(h_i),
\end{equation}
where $q$ is the internal energy density of each particle, $U = mu$ and the coefficients:
\begin{equation}\label{eq:over14}
    g^{\textcolor{black}{\text{grad}}}_i = \left( 1 + \frac{h_i}{Dq_i}\frac{\partial q_i}{\partial h_i} \right)^{-1},
\end{equation}
\textcolor{black}{appear} by considering the spatial derivative of the smoothing length.
Any physical quantity at any location is defined by
\begin{equation}\label{eq:over14_1}
    f(\bm{r}) = \displaystyle \sum_j U_j\frac{f_j}{q_j} W(|\bm{r}-\bm{r}_j|,h(\bm{r})).
\end{equation}
Therefore, the internal energy density of the $i$-th particle is defined by
\begin{equation}\label{eq:over13}
    q_i = \displaystyle \sum^N_{\textcolor{black}{j=1}} U_j W_{ij}(h_i).
\end{equation}
The smoothing length $h_i$ is updated so that 
\textcolor{black}{
\begin{equation}\label{eq:condi}
    \left(\frac{q_i}{u_i}\right) A(\xi h_i)^D = m_i N_{\text{ngb}},
\end{equation}
}
is satisfied for any particles at any time.
Pressure and the density of the $i$-th particle are given by
\begin{equation}\label{eq:over15}
    P_i = \left( \gamma - 1 \right) q_i,
\end{equation}
and
\begin{equation}\label{eq:over16}
    \rho_i = \frac{q_i}{u_i},
\end{equation}
respectively. One can use the same artificial viscosity with SSPH.
However, in \citet{Saitoh_Makino2013}, equation (\ref{eq:over3}) is used as the density in the artificial viscosity term because it is less noisy when there 
are huge pressure gradients
(Remember DISPH assumes that the pressure is continuous and differentiable).
In addition, when evaluating the density for plotting, using equation (\ref{eq:over3}) as the density is a better choice especially when there 
are the huge pressure gradients.
\citet{Hosono2016,Hosono20161b} and \cite{Saitoh_Makino2016} demonstrated advantages to and compatibility with SSPH by showing several tests calculations.

Note that there is a prescription for the huge pressure difference, called generalised DISPH \citep{Saitoh_Makino2013}; however, we do not deal with this method in this paper.
DISPH still needs the artificial viscosity term.
Therefore, one still needs to tune $\alpha_{AV}$ so that enough but not too excessive viscosity is added to handle \textcolor{black}{shocks}, depending on the simulation problem.


\section{Godunov DISPH}
\label{sec:incorporation}
The practical performance of the various schemes described in the previous section on standard test problems is discussed in detail in Section \ref{sec:num}, but up to this point, we have seen the advantages of two schemes; the GSPH can handle shock waves without special attention to artificial viscosity, and the DISPH can accurately represent contact discontinuities. 
\indent
Here, we propose the construction of a new scheme that utilises the advantages of both schemes by incorporating the Riemann solver into DISPH. 
There are several degrees of freedom to incorporate a Riemann solver in DISPH, two independent methods are considered here.
We name this new formulation Godunov DISPH (GDISPH).
A brief description of the Riemann solver is described in \ref{ap:riemann}.

The structure of this section is as follows.
In Section \ref{sec:FirstLaw}, \textcolor{black}{GDISPH Case 1} is derived through the first law of thermodynamics. Section \ref{sec:Inutsuka} describes derivations of \textcolor{black}{GDISPH Case 2 and Case 3} following a similar method of \citet{Inutsuka2002}.

\subsection{Derivation from First Law of Thermodynamics}
\label{sec:FirstLaw}
\textcolor{black}{This section shows the derivation of GDISPH Case 1 (The momentum equation (\ref{eq:law15}) and the energy equation (\ref{eq:law14})).}
Here, we first derive the energy equation of the $i$-th particle, then the momentum equation of that in a similar manner to \cite{Saitoh_Makino2013}.

The volume element of the $i$-th particle is given by
\begin{equation}\label{eq:law0}
    V_i = \frac{U_i}{q_i}.
\end{equation}
Any physical quantity at any location is defined as follows:
\begin{equation}\label{eq:law1}
    f(\bm{r}) = \displaystyle \sum_j U_j\frac{f_j}{q_j} W(|\bm{r}-\bm{r}_j|,h(\bm{r})).
\end{equation}
Therefore, the internal energy density of the $i$-th particle is
\begin{equation}\label{eq:law2}
    q_i = \displaystyle \sum^N_{\textcolor{black}{j=1}} U_j W_{ij}(h_i).
\end{equation}
We require that the mass in the kernel volume is constant, i.e.,
\textcolor{black}{
\begin{equation}\label{eq:law2_2}
    \frac{q_i}{U_i}h^D_i = \text{const}.
\end{equation}
}
Therefore, the total derivative of the smoothing length $h_i$ is given by:
\begin{equation}\label{eq:law2-3}
    \begin{split}
        \textcolor{black}{\text{d}}h_i &= \frac{\partial h_i}{\partial q_i}\textcolor{black}{\text{d}}q_i + \frac{\partial h_i}{\partial U_i}\textcolor{black}{\text{d}}U_i\\
        &= -\frac{h_i}{Dq_i}\textcolor{black}{\text{d}}q_i + \frac{h_i}{DU_i}\textcolor{black}{\text{d}}U_i.
    \end{split}
\end{equation}
The first law of thermodynamics in an adiabatic state is considered:
\begin{equation}\label{eq:law3}
    \textcolor{black}{\text{d}}U_i = \textcolor{black}{\mathcal{W}^{\text{Volume}}_{i}},
\end{equation}
where $\textcolor{black}{\mathcal{W}^{\text{Volume}}_{i}}$ is the infinitesimal work that the $i$-th particle receives through it's volume change $\textcolor{black}{\text{d}}V_i$ during the time interval $\textcolor{black}{\text{d}}t$.
Note that the infinitesimal work that the $i$-th particle receives through its movement of the centre of mass during the time $\textcolor{black}{\text{d}}t$ turns into its kinetic energy.
In the derivation of energy equations of SPH and DISPH,
$\textcolor{black}{\mathcal{W}^{\text{Volume}}_{i}}$ is effectively defined by
\begin{equation}\label{eq:law4}
    \textcolor{black}{\mathcal{W}^{\text{Volume}}_{i}} = - P_i \textcolor{black}{\text{d}} V_i.
\end{equation}
However, equation (\ref{eq:law4}) has the implicit assumption that the pressure on the $i$-th particle during the time interval
$\textcolor{black}{\text{d}}t$ is $P_i + \epsilon$, where $\epsilon$ is the first-order term.
As a result, $ (P_i + \epsilon)\textcolor{black}{\text{d}}V_i \approx P_i \textcolor{black}{\text{d}}V_i$ by neglecting the second-order term.
This assumption should become true accurately with arbitrary precision if there is an infinite number of particles, $h_i \to 0$, and $\textcolor{black}{\text{d}}t \to 0$ (the infinite spacial and time resolution).
Note that the same definition and assumption (but using the fluid elements instead of the particles) with an infinite number of infinitesimal fluid elements and $\textcolor{black}{\text{d}}t \to 0$, which turns the assumption true with arbitrary precision, is applied in deriving the fluid energy equation (\ref{eq:Major3}).
Therefore, the energy equations of SPH and DISPH using the infinite spacial and time resolution should converge to the energy equation (\ref{eq:Major3}).
However, in a realistic situation of numerical simulations where the number of particles, $V_i$, and $\textcolor{black}{\text{d}}t$ are finite values,
the pressure on the $i$-th particle should be different depending on the direction and the time.
Here, we let $\overline{P_{ix}}$ be a certain kind of the time-spatial averaged value of the pressure on the $i$-th particle from all directions centred on the $i$-th particle during the time $\textcolor{black}{\text{d}}t$.
Note that if the infinite spacial and time resolution are used, 
$\overline{P_{ix}}$ should become $P_i$ with arbitrary precision so those can lead to the conversion to the energy equation (\ref{eq:Major3}). The explicit form of $\overline{P_{ix}}$ is given later.
Then, we redefine $\textcolor{black}{\mathcal{W}^{\text{Volume}}_{i}}$ as follows:
\begin{equation}\label{eq:law5}
    \textcolor{black}{\mathcal{W}^{\text{Volume}}_{i}} = - \overline{P_{ix}} \textcolor{black}{\text{d}} V_i,
\end{equation}\label{eq:law6}
where the assumption that the pressure on the $i$-th particle during the time interval $\textcolor{black}{\text{d}}t$ is $\overline{P_{ix}} + \epsilon$, where $\epsilon$ is the first-order term, is applied.
Then, the energy equation of the $i$-th particle is given by
\begin{equation}
    \begin{split}
        \frac{dU_i}{dt} &= \frac{\textcolor{black}{\mathcal{W}^{\text{Volume}}_{i}}}{\textcolor{black}{\text{d}}t},\\
        &= - \overline{P_{ix}} \frac{d V_i}{dt}, \\
        &= - \overline{P_{ix}} \frac{d}{dt} \frac{U_i}{q_i}, \\
        &= - \overline{P_{ix}} \left( \frac{1}{q_i} \frac{dU_i}{dt} - \frac{U_i}{q^2_i}\frac{dq_i}{dt}\right).
    \end{split}
\end{equation}
Then,
\begin{equation}\label{eq:law7}
    \left( 1 + \frac{\overline{P_{ix}}}{q_i} \right)\frac{dU_i}{dt} = \overline{P_{ix}}\frac{U_i}{q^2_i}\frac{dq_i}{dt}.
\end{equation}
Taking the time derivative of $q_i$, we have
\begin{equation}\label{eq:law8}
    \begin{split}
        \frac{dq_i}{dt} &= \frac{d}{dt}\displaystyle \sum^N_{\textcolor{black}{j=1}} U_j W_{ij}(h_i),\\
        &= \displaystyle \sum^N_{\textcolor{black}{j=1}} \frac{dU_j}{dt} W_{ij}(h_i) + \displaystyle \sum^N_{\textcolor{black}{j=1}} U_j \frac{d W_{ij}(h_i)}{dt}.\\
    \end{split}
\end{equation}
The first term of the right-hand side of equation (\ref{eq:law8}) can be calculated as follows:
\begin{equation}\label{eq:law8-1}
    \begin{split}
        \displaystyle \sum^N_{\textcolor{black}{j=1}} \frac{dU_j}{dt} W_{ij}(h_i) &= \displaystyle \sum^N_{\textcolor{black}{j=1}} \frac{U_j}{q_j}\frac{q_j}{U_j}\frac{dU_j}{dt} W_{ij}(h_i),\\
        &= \frac{q_i}{U_i}\frac{dU_i}{dt},
    \end{split}
\end{equation}
where equation (\ref{eq:law1}) is used.
The second term of the right-hand side of equation (\ref{eq:law8}) can be calculated as follows:
\begin{align}\label{eq:law8-2}
        \notag&\displaystyle \sum^N_{\textcolor{black}{j=1}} U_j \frac{d W_{ij}(h_i)}{dt},\\
        \notag&= \displaystyle \sum^N_{\textcolor{black}{j=1}} U_j \left( \nabla_i W_{ij}(h_i) \cdot \bm{v}_{ij} + \frac{\partial W_{ij}(h_i)}{\partial h_i} \frac{d h_i}{dt}\right),\\
        &= \displaystyle \sum^N_{\textcolor{black}{j=1}} U_j \left( \nabla_i W_{ij}(h_i) \cdot \bm{v}_{ij} + \frac{\partial W_{ij}(h_i)}{\partial h_i} \left( -\frac{h_i}{Dq_i}\frac{dq_i}{dt} + \frac{h_i}{DU_i}\frac{dU_i}{dt}\right)\right),
\end{align}
where equation (\ref{eq:law2-3}) is used.
Substituting equation (\ref{eq:law8-1}) and equation (\ref{eq:law8-2}) into equation (\ref{eq:law8}), we get
\begin{equation}\label{eq:law9}
    \begin{split}
        \frac{dq_i}{dt} &= g^{\textcolor{black}{\text{grad}}}_i\left[\frac{q_i}{U_i} + \frac{h_i}{DU_i}\displaystyle \sum^N_{\textcolor{black}{j=1}} U_j\frac{\partial W_{ij}(h_i)}{\partial h_i}\right]\frac{dU_i}{dt}\\
        &\quad\quad\quad + g^{\textcolor{black}{\text{grad}}}_i \displaystyle \sum^N_{\textcolor{black}{j=1}} U_j \bm{v}_{ij}\cdot \nabla_i W_{ij}(h_i),\\
        &= g^{\textcolor{black}{\text{grad}}}_i\left[\frac{q_i}{U_i} + \frac{q_i}{U_i}\left( \frac{1}{g^{\textcolor{black}{\text{grad}}}_i} - 1 \right)\right]\frac{dU_i}{dt}\\
        &\quad\quad\quad + g^{\textcolor{black}{\text{grad}}}_i \displaystyle \sum^N_{\textcolor{black}{j=1}} U_j \bm{v}_{ij}\cdot \nabla_i W_{ij}(h_i),\\
        &= \frac{q_i}{U_i} \frac{dU_i}{dt} + g^{\textcolor{black}{\text{grad}}}_i \displaystyle \sum^N_{\textcolor{black}{j=1}} U_j \bm{v}_{ij}\cdot \nabla_i W_{ij}(h_i),
    \end{split}
\end{equation}
where $g^{\textcolor{black}{\text{grad}}}_i$ is the same coefficient \textcolor{black}{as in} equation (\ref{eq:over14}).
Substituting equation (\ref{eq:law9}) into equation (\ref{eq:law7}), we obtain
\begin{equation}\label{eq:law11}
    \begin{split}
        \left( 1 + \frac{\overline{P_{ix}}}{q_i} \right)\frac{dU_i}{dt} &= \frac{\overline{P_{ix}}}{q_i}\frac{dU_i}{dt} + g^{\textcolor{black}{\text{grad}}}_i\displaystyle \sum^N_{\textcolor{black}{j=1}} \frac{\overline{P_{ix}}U_iU_j}{q^2_i} \bm{v}_{ij}\cdot \nabla_i W_{ij}(h_i).
    \end{split}
\end{equation}
Then, we can obtain the time derivative of the internal energy of the $i$-th particle as follows:
\begin{equation}\label{eq:law12}
    \frac{dU_i}{dt} = g^{\textcolor{black}{\text{grad}}}_i\displaystyle \sum^N_{\textcolor{black}{j=1}} \frac{\overline{P_{ix}}U_iU_j}{q^2_i} \bm{v}_{ij}\cdot \nabla_i W_{ij}(h_i).
\end{equation}
Note that if $\overline{P_{ix}} = P_{i}$, this equation is the same as the original energy equation of DISPH.

\begin{figure*}[!t]
\centering
	\includegraphics[width=1\linewidth]{ 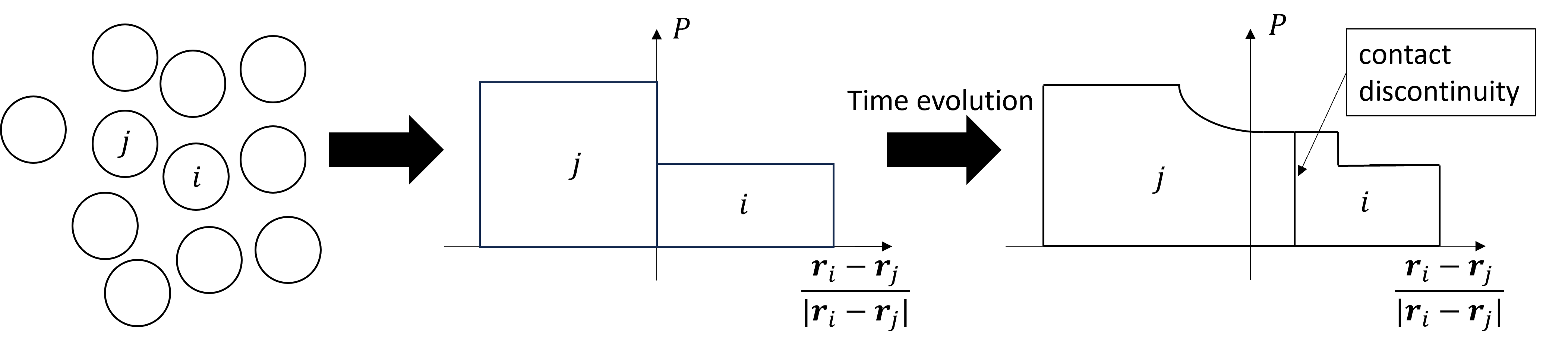}
    \caption{A schematic picture of the interaction between the $i$-th particle and the $j$-th particle.
    Consider the case of computing the interaction from the $j$th particle to the $i$th particle (left panel) and assume that the two particles are fluid elements in contact (centre panel). Solving the Riemann problem with these conditions as initial conditions, the solution is shown on the right panel. Since the contact discontinuity is the boundary between the two fluid elements, we use the pressure around the contact discontinuity as the time-averaged pressure that the $i$th particle receives from the $j$th particle.}
    \label{fig:interection_GDISPH}
\end{figure*}

In this paper, we take the average $P_{ix}$ by time-averaging the pressure that $i$-th particle receives from any surrounding particles and then spatial-averaging those values as follows: In the process of calculating the interaction between the $i$-th particle and the $j$-th particle, it is necessary to determine the time-averaged pressure that the $i$-th particle receives from the $j$-th particle.
As shown in Figure \ref{fig:interection_GDISPH}, to get the pressure, we assume that the two particles are the fluid elements that are in contact with each other.
As a result, we can use the time-averaged pressure that the $i$-th fluid element receives from the border between the two fluid elements as the time-averaged pressure that the $i$-th particle receives from the $j$-th particle.
To get the time-averaged value, we solve the Riemann \textcolor{black}{problem} and use the solution of the pressure around the contact discontinuity as the value since the discontinuity is literally the physical boundary between the two fluid elements.

Although arbitrariness exists in the method of a spatial average of pressures, the following method is adopted here,
\begin{equation}\label{eq:law13}
    \overline{P_{ix}}\displaystyle \sum^N_{\textcolor{black}{j=1}} \frac{U_iU_j}{q^2_i} \bm{v}_{ij}\cdot \nabla_i W_{ij}(h_i) = \displaystyle \sum^N_{\textcolor{black}{j=1}} \frac{P^*_{ij}U_iU_j}{q^2_i} \bm{v}_{ij}\cdot \nabla_i W_{ij}(h_i),
\end{equation}
where $P^*_{ij}$ denotes the pressure in the star region from the solutions of the 1D Riemann problem using the physical quantities of the $i$-th particle and the $j$-th particle as $\textcolor{black}{\bm{\mathbb{W}}_R}$ and $\textcolor{black}{\bm{\mathbb{W}}_L}$, respectively in \ref{ap:riemann}.
Note that we solve the 1D Riemann problem along the line joining two particles with the
$i$-th particle (respectively $j$-th particle) defining the right (left) state in the local coordinate system. 
For the input to the Riemann solver, \textcolor{black}{$v_R$ (respectively $v_L$)} is a component of \textcolor{black}{$\bm{v}_i$ ($\bm{v}_j$)} in the direction of $\bm{r}_i-\bm{r}_j$.
As a result, by substituting equation (\ref{eq:law13}) into equation (\ref{eq:law12}), 
The Riemann solver is incorporated into DISPH as follows:
\begin{equation}\label{eq:law14}
    \frac{dU_i}{dt} = g^{\textcolor{black}{\text{grad}}}_i\displaystyle \sum^N_{\textcolor{black}{j=1}} \frac{P^{*}_{ij}U_iU_j}{q^2_i} \bm{v}_{ij}\cdot \nabla_i W_{ij}(h_i),
\end{equation}
where $\overline{P_{ix}}$ in equation (\ref{eq:law12}) is just replaced by $P^{*}_{ij}$ so we do not have to pre-calculate the spatial average.
Next, according to the method of section 3.3. in \citet{Saitoh_Makino2013}, the momentum equation is derived by the conditions that \textcolor{black}{it} satisfies the law of action and reaction and the energy conservation with the energy equation (\ref{eq:law14}), yielding the following momentum equation:
\begin{equation}\label{eq:law15}
    m_i \frac{d\bm{v}_i}{dt} = -\displaystyle \sum^N_{\textcolor{black}{j=1}} \left [ g^{\textcolor{black}{\text{grad}}}_i \frac{P^{*}_{ij}  U_i U_j}{q^2_i} \nabla_i W_{ij}(h_i) + g^{\textcolor{black}{\text{grad}}}_j \frac{P^*_{ij}  U_i U_j}{q^2_j} \nabla_i W_{ij}(h_j) \right ]
\end{equation}
Note that $P^*_{ij} = P^*_{ji}$.
If $\overline{P_{ix}} = P_{i}$, this equation becomes the same as the original momentum equation in DISPH.
\textcolor{black}{We call the set of equation (\ref{eq:law15}) and equation (\ref{eq:law14}) GDISPH Case 1.}
\subsection{Derivation from Convolution}
\label{sec:Inutsuka}
\textcolor{black}{This section shows the derivation of GDISPH Case 2 (the momentum equation (\ref{eq:in39}) and the energy equation (\ref{eq:in40})) and GDISPH Case 3 (the momentum equation (\ref{eq:in41}) and the energy equation (\ref{eq:in42})).}
\textcolor{black}{
We tried to derive GDISPH following the similar method of \citet{Inutsuka2002}, yielding the two different types of momentum equation (\ref{eq:in7}) and (\ref{eq:in21}), and one energy equation (\ref{eq:in17}) but failed to get the pairs of equations that satisfy the law of action and reaction and energy conservation.
Note that \citet{Inutsuka2002} was able to derive those for GSPH (equation (\ref{eq:in7-1}) and equation (\ref{eq:in21-2})).
In this section, we dare to show the derivation for someone who is interested in that.
Then, the pairs of equations (the first pair is equation (\ref{eq:in10}) and equation (\ref{eq:in24}), and the second pair is equation (\ref{eq:in17}) and equation (\ref{eq:in25})) are created in an ad-hoc manner and, similar to GSPH Case 3, the approximation is applied to those two different formulations of GDISPH, yielding GDISPH Case 2 and GDISPH Case 3.}

The internal energy density field and its gradient are defined as follows:
\begin{equation}\label{eq:in1}
    q(\bm{r}) = \displaystyle \sum_j m_j u_j W(|\bm{r}-\bm{r}_j|,h(\bm{r})),
\end{equation}
\begin{equation}\label{eq:in1-1}
    \nabla q(\bm{r}) = \displaystyle \sum_j m_j u_j \nabla W(|\bm{r}-\bm{r}_j|,h(\bm{r})).
\end{equation}
The convolution of function $f(\bm{r})$ with the kernel function is defined by
\begin{equation}\label{eq:in3}
    \left<f\right>(\bm{r}) = \int f(\bm{r}^{'})W(|\bm{r}-\bm{r}^{'}|,h(\bm{r}^{'}))dV^{'}.
\end{equation}
In the following derivation, we ignore the differentiation of the smoothing length.

\subsubsection{Momentum Equation}
\label{sec:Inutsuka_EoM}
The following equation is adopted as the momentum equation for the $i$-th particle:
\begin{equation}\label{eq:in6}
    m_i \frac{d\bm{v}_i}{dt} = m_i \left<\frac{d\bm{v}}{dt}\right> (\bm{r}_i).
\end{equation}
Taking the convolution of equation (\ref{eq:Major2}) yields
\begin{equation}\label{eq:in4}
    \begin{split}
        \left<\frac{d\bm{v}}{dt}\right> (\bm{r}_i) &= \int \frac{d\bm{v}(\bm{r})}{dt}W(|\bm{r}-\bm{r}_i|,h(\bm{r}))dV, \\
        &= -\int \frac{1}{\rho(\bm{r})} \nabla P(\bm{r}) W(|\bm{r}-\bm{r}_i|,h(\bm{r}))dV.
    \end{split}
\end{equation}
These equations derive the following momentum equation:
\begin{equation}\label{eq:in7}
    \begin{split}
        &m_i \frac{d\bm{v}_i}{dt} \\
        &= m_i \displaystyle \sum_{j=1}^N m_ju_j \int \frac{P(\bm{r})u(\bm{r})}{q^2(\bm{r})} \nabla_j W(|\bm{r}-\bm{r}_j|,h(\bm{r})) W(|\bm{r} - \bm{r}_i|,h(\bm{r}))dV.
    \end{split}
\end{equation}
The mathematical manipulation of deriving equation (\ref{eq:in7}) from equation (\ref{eq:in6}) and equation (\ref{eq:in4}) is shown in \ref{ap:eom}.

Having failed to obtain the momentum equation that fulfils the law \textcolor{black}{of action and reaction}, we make that in an ad-hoc manner.
In \citet{Inutsuka2002}, from equation (\ref{eq:in6}) and equation (\ref{eq:in4}), the following momentum equation for SPH is derived:
\begin{equation}\label{eq:in7-1}
    \begin{split}
    m_i \frac{d\bm{v}_i}{dt} = -\displaystyle \sum_{j=1}^N &m_im_j \int \frac{P(\bm{r})}{\rho^2(\bm{r})} \cdot \\
    & \left( \nabla_i W(|\bm{r}-\bm{r}_i|,h(\bm{r}))W(|\bm{r}-\bm{r}_j|,h(\bm{r}))\right. \\
    &\left.- W(|\bm{r}-\bm{r}_i|,h(\bm{r}))\nabla_j W(|\bm{r}-\bm{r}_j|,h(\bm{r}))\right)dV.
    \end{split}
\end{equation}
With reference to the form of equation (\ref{eq:in7}) and equation (\ref{eq:in7-1}), the momentum equation for DISPH is created as follows:
\begin{equation}\label{eq:in10}
    \begin{split}
        m_i \frac{d\bm{v}_i}{dt} = -\displaystyle \sum_{j=1}^N &m_iu_i m_j u_j \int \frac{P(\bm{r})}{q^2(\bm{r})} \cdot \\
        & \left( \nabla_i W(|\bm{r}-\bm{r}_i|,h(\bm{r}))W(|\bm{r}-\bm{r}_j|,h(\bm{r}))\right. \\
        &\left.- W(|\bm{r}-\bm{r}_i|,h(\bm{r}))\nabla_j W(|\bm{r}-\bm{r}_j|,h(\bm{r}))\right)dV,
    \end{split}
\end{equation}
which agree with the law.

\citet{Inutsuka2002} also derived the momentum equation for Godunov SPH using the following Lagrangian but with $h(\bm{r}) = h$,
\begin{equation}\label{eq:in11-1}
    L \equiv \displaystyle \sum^N_i m_i \left[\frac{1}{2} \bm{\dot r^2}_i - \int u(\bm{r})W(|\bm{r}-\bm{r}_i|,h(\bm{r}))dV\right].
\end{equation}
We can then proceed to derive the equations of motion from the Euler-Lagrange equation,
\begin{equation}\label{eq:in12}
    \frac{d}{dt}\frac{\partial L}{\partial \bm{\dot r}_i} - \frac{\partial L}{\partial \bm{r}_i} = 0.
\end{equation}
This equation gives the following momentum equation:
\begin{equation}\label{eq:in17}
    \begin{split}
        &m_i\frac{d\bm{v}_i}{dt} = - \displaystyle \sum^N_{\textcolor{black}{j=1}} m_i u_i m_j \int P \frac{u}{q^2} \nabla_i W(|\bm{r}-\bm{r}_i|,h(\bm{r})) W(|\bm{r}-\bm{r}_j|,h(\bm{r}))dV\\
        &+ \displaystyle \sum^N_{\textcolor{black}{j=1}} m_i m_j u_j \int P\frac{u}{q^2} W(|\bm{r}-\bm{r}_i|,h(\bm{r})) \nabla_j W(|\bm{r}-\bm{r}_j|,h(\bm{r}))dV.
    \end{split}
\end{equation}
The manipulation of deriving
equation (\ref{eq:in17}) from equation (\ref{eq:in12}) is shown in \ref{ap:el}.
Note that this momentum equation fulfils the law of action and reaction.
\subsubsection{Energy Equation}
\label{sec:Inutsuka_EE}
We adopt the following equation as the energy equation for the $i$-th particle:
\begin{equation}\label{eq:in20}
    \frac{du_i}{dt} = \left<\frac{du}{dt}\right> (\bm{r}_i).
\end{equation}
Taking the convolution of equation (\ref{eq:Major3}) gives
\begin{equation}\label{eq:in18}
    \begin{split}
        \left< \frac{du}{dt} \right> (\bm{r}_i) &= \int \frac{du(\bm{r})}{dt} W(|\bm{r} - \bm{r}_i|,h(\bm{r}))dV\\
        &= - \int \frac{P(\bm{r})}{\rho(\bm{r})} \nabla \cdot \bm{v}(\bm{r}) W(|\bm{r} - \bm{r}_i|,h(\bm{r}))dV.
    \end{split}
\end{equation}
These equations derive the following energy equation:
\begin{equation}\label{eq:in21}
    \begin{split}
        m_i \frac{du_i}{dt} = \displaystyle \sum^N_{\textcolor{black}{j=1}}m_i m_j u_j \int &\frac{P(\bm{r})u(\bm{r})}{q^2(\bm{r})} \left[ \bm{v}(\bm{r}) - \bm{v}_i \right] \cdot \\
        &\nabla_j W(|\bm{r}-\bm{r}_j|,h(\bm{r})) W(|\bm{r}-\bm{r}_i|,h(\bm{r}))dV,\\    
    \end{split}
\end{equation}
which fails to satisfy total energy conservation with \textcolor{black}{neither equation (\ref{eq:in7}), equation (\ref{eq:in10}), or equation (\ref{eq:in17}).}
The mathematical manipulation of deriving
equation (\ref{eq:in21}) from equation (\ref{eq:in20}) and equation (\ref{eq:in18}) is shown in \ref{ap:ee}.

The failure leads us to make a new energy equation that satisfies the conservation in an ad-hoc manner.

In \citet{Inutsuka2002}, from equation (\ref{eq:in20}) and equation (\ref{eq:in18}), the following energy equation for SPH is derived:
\begin{equation}\label{eq:in21-2}
    \begin{split}
    m_i \frac{du_i}{dt} = -\displaystyle \sum_{j=1}^N &m_im_j \int \frac{P(\bm{r})}{\rho^2(\bm{r})}[\bm{v}(\bm{r}) - \bm{v}_i] \cdot \\
    & \left( \nabla_i W(|\bm{r}-\bm{r}_i|,h(\bm{r}))W(|\bm{r}-\bm{r}_j|,h(\bm{r}))\right. \\
    &\left.- W(|\bm{r}-\bm{r}_i|,h(\bm{r}))\nabla_j W(|\bm{r}-\bm{r}_j|,h(\bm{r}))\right)dV
    \end{split}
\end{equation}
Based on the form of equation (\ref{eq:in21}) and equation (\ref{eq:in21-2}), we create the energy equation for DISPH as follows:
\begin{equation}\label{eq:in24}
    \begin{split}
        m_i\frac{du_i}{dt} &= -\displaystyle \sum_{j=1}^N m_i u_i m_j u_j \int \frac{P(\bm{r})}{q^2(\bm{r})} \left[ \bm{v}(\bm{r}) - \bm{v}_i \right] \cdot \\
        & \left( \nabla_i W(|\bm{r}-\bm{r}_i|,h(\bm{r}))W(|\bm{r}-\bm{r}_j|,h(\bm{r}))\right. \\
        &\left.- W(|\bm{r}-\bm{r}_i|,h(\bm{r}))\nabla_j W(|\bm{r}-\bm{r}_j|,h(\bm{r}))\right)dV,
    \end{split}
\end{equation}
which satisfies the energy conservation with the momentum equation (\ref{eq:in10}).
Multiplying the contents of the integral in the momentum equation (\ref{eq:in17}) by $(\bm{v}(\bm{r})-\bm{v}_i)$ gives
the following energy equation:
\begin{equation}\label{eq:in25}
    \begin{split}
        &\frac{du_i}{dt} = \\
        &- \displaystyle \sum^N_{\textcolor{black}{j=1}} u_i m_j \int P \frac{u}{q^2} \left [ \bm{v}(\bm{r})-\bm{v}_i \right ] \cdot \nabla_i W(|\bm{r}-\bm{r}_i|,h(\bm{r})) W(|\bm{r}-\bm{r}_j|,h(\bm{r}))dV\\
        &+ \displaystyle \sum^N_{\textcolor{black}{j=1}} m_j u_j \int P\frac{u}{q^2} \left [ \bm{v}(\bm{r})-\bm{v}_i \right ] \cdot W(|\bm{r}-\bm{r}_i|,h(\bm{r})) \nabla_j W(|\bm{r}-\bm{r}_j|,h(\bm{r}))dV,
    \end{split}
\end{equation}
which satisfies the energy conservation with the momentum equation (\ref{eq:in17}).

\subsubsection{Incorporation of Riemann Solver}
\label{sec:InuRie}
So far, the two sets of DISPH, which agree with the law of action and reaction and energy conservation, have been obtained.
The first set is the momentum equation (\ref{eq:in10}) and the energy equation (\ref{eq:in24}).
The second set is the momentum equation (\ref{eq:in17}) and the energy equation (\ref{eq:in25}).
Here, we incorporate the Riemann solver into them.
To incorporate the Riemann solver into SPH, \citet{Inutsuka2002} implicitly assumed that the solutions of the Riemann solver $f^{RP}_{i,j}$ satisfy the following equation approximately:
\begin{equation}\label{eq:in26}
    \begin{split}
        \int \frac{f(\bm{r})}{\rho^2(\bm{r})}&W(|\bm{r}-\bm{r}_i|,h(\bm{r}))W(|\bm{r}-\bm{r}_j|,h(\bm{r}))dV \\
        &\approx f^{RP}_{i,j}\int \frac{1}{\rho^2(\bm{r})}W(|\bm{r}-\bm{r}_i|,h(\bm{r}))W(|\bm{r}-\bm{r}_j|,h(\bm{r}))dV,
    \end{split}
\end{equation}
where \citet{Inutsuka2002} defined $f^{RP}_{i,j}$ as the solution of the Riemann problem at the vicinity of the middle point of the $i$-th particle and the $j$-th particle.
A certain spatial-averaged values around the $i$-th particle and the $j$-th particle are used as $\textcolor{black}{\bm{\mathbb{W}}_R}$ and $\textcolor{black}{\bm{\mathbb{W}}_L}$ in \ref{ap:riemann},
and $f^{RP}_{i,j}$ is adopted as the solution at $x=0$ of the Riemann problem.
However, some implementations have been found to use the solutions in the star region of the Riemann problem.
There are some methods of determining $\textcolor{black}{\bm{\mathbb{W}}_R}$ and $\textcolor{black}{\bm{\mathbb{W}}_L}$ using the MUSCL method which was originally developed to improve the Godunov method, one of the finite volume methods, to second-order spatial accuracy \citep{Inutsuka2002,Iwasaki2011}.
\citet{Murante2011} used Godunov SPH using the MUSCL method with the flux limiter developed by \citet{VanLeer1979}.

Here, we extend \textcolor{black}{the \citet{Inutsuka2002}'s implicit assumption} so that there are weighted average values
$f^+_{i,j} , g^+_{i,j} \cdots z^+_{i,j}$ that satisfy the following equation:
\begin{equation}\label{eq:in32}
    \begin{split}
        &\int \frac{f(\bm{r})g(\bm{r})\cdots z(\bm{r})}{q^2(\bm{r})}W(|\bm{r}-\bm{r}_i|,h(\bm{r}))W(|\bm{r}-\bm{r}_j|,h(\bm{r}))dV \\
        &\approx f^+_{i,j}g^+_{i,j}\cdots z^+_{i,j} \int \frac{1}{q^2(\bm{r})}W(|\bm{r}-\bm{r}_i|,h(\bm{r}))W(|\bm{r}-\bm{r}_j|,h(\bm{r}))dV,
    \end{split}
\end{equation}
where $f^+_{i,j} = f^+_{j,i} , g^+_{i,j} = g^+_{j,i} \cdots z^+_{i,j}=z^+_{j,i}$ must be satisfied.
In this paper, we adopt the weighted average value $f^+_{i,j}$ as $f^{*}_{i,j}$ or $\frac{f(\bm{r}_i) + f(\bm{r}_j)}{2}$.
Note that $f^{*}_{i,j}$ denotes the solution in the star region from the Riemann solver using the physical quantities of the $i$-th particle and the $j$-th particle as $\textcolor{black}{\bm{\mathbb{W}}_R}$ and $\textcolor{black}{\bm{\mathbb{W}}_L}$ in  \ref{ap:riemann}.

Applying equation (\ref{eq:in32}) to the momentum equation (\ref{eq:in10}) and the energy equation (\ref{eq:in24}) yields the following set of equations:
\begin{equation}\label{eq:in33}
    \begin{split}
        m_i \frac{d\bm{v}_i}{dt} = -\displaystyle \sum_{j=1}^N &m_iu_i m_j u_j P^{+}_{i,j} \int \frac{1}{q^2(\bm{r})} \cdot \\
        & \left( \nabla_i W(|\bm{r}-\bm{r}_i|,h(\bm{r}))W(|\bm{r}-\bm{r}_j|,h(\bm{r}))\right. \\
        &\left.- W(|\bm{r}-\bm{r}_i|,h(\bm{r}))\nabla_j W(|\bm{r}-\bm{r}_j|,h(\bm{r}))\right)dV,
    \end{split}
\end{equation}

\begin{equation}\label{eq:in34}
    \begin{split}
        \frac{du_i}{dt} &= -\displaystyle \sum_{j=1}^N u_i m_j u_j P^{+}_{i,j} \left[ \bm{v}^{+}_{i,j} - \bm{v}_i \right] \cdot \int \frac{1}{q^2(\bm{r})} \\
        & \left( \nabla_i W(|\bm{r}-\bm{r}_i|,h(\bm{r}))W(|\bm{r}-\bm{r}_j|,h(\bm{r}))\right. \\
        &\left.- W(|\bm{r}-\bm{r}_i|,h(\bm{r}))\nabla_j W(|\bm{r}-\bm{r}_j|,h(\bm{r}))\right)dV,
    \end{split}
\end{equation}
while applying equation (\ref{eq:in32}) to the momentum equation (\ref{eq:in17}) and the energy equation (\ref{eq:in25}) yields the following set of equations:
\begin{equation}\label{eq:in35}
    \begin{split}
        &m_i\frac{d\bm{v}_i}{dt} = \\
        &- \displaystyle \sum^N_{\textcolor{black}{j=1}} m_i u_i m_j P^{+}_{i,j}u^{+}_{i,j} \int \frac{1}{q^2(\bm{r})} \nabla_i W(|\bm{r}-\bm{r}_i|,h(\bm{r})) W(|\bm{r}-\bm{r}_j|,h(\bm{r}))dV\\
        &+ \displaystyle \sum^N_{\textcolor{black}{j=1}} m_i m_j u_j P^{+}_{i,j}u^{+}_{i,j} \int \frac{1}{q^2(\bm{r})} W(|\bm{r}-\bm{r}_i|,h(\bm{r})) \nabla_j W(|\bm{r}-\bm{r}_j|,h(\bm{r}))dV,
    \end{split}
\end{equation}

\begin{equation}\label{eq:in36}
    \begin{split}
        &\frac{du_i}{dt} = - \displaystyle \sum^N_j u_i m_j P^{+}_{i,j}u^{+}_{i,j} \left[ \bm{v}^{+}_{i,j} - \bm{v}_i \right] \cdot \\
        & \quad \quad \quad \quad \quad \quad \quad \quad \quad\int \frac{1}{q^2(\bm{r})} \nabla_i W(\textcolor{black}{|}\bm{r}-\bm{r}_i\textcolor{black}{|},h(\bm{r})) W(\bm{r}-\bm{r}_j,h(\bm{r}))dV\\
        &+ \displaystyle \sum^N_j m_j u_j  P^{+}_{i,j}u^{+}_{i,j} \left[ \bm{v}^{+}_{i,j} - \bm{v}_i \right] \cdot \\ 
        &\quad \quad \quad \quad \quad \quad \quad \quad \quad\int \frac{1}{q^2(\bm{r})} W(\textcolor{black}{|}\bm{r}-\bm{r}_i\textcolor{black}{|},h(\bm{r})) \nabla_j W(\bm{r}-\bm{r}_j,h(\bm{r}))dV
    \end{split}
\end{equation}
These sets of equations fulfil the law of action and reaction and energy conservation.

\subsubsection{Cha's Case 3}
\label{sec:InuCase}
\textcolor{black}{
Here, we simplify the two sets of GDISPH (the first set is equation (\ref{eq:in33}) and equation (\ref{eq:in34}), and the second set is equation (\ref{eq:in35}) and equation (\ref{eq:in36})) by using an approximation.}
\citet{Cha_Whitworth2003} introduced the various type of GSPH equations.
GSPH Case 3 is derived by applying the following approximations into GSPH \citep{Iwasaki2011}:
\begin{equation}\label{eq:in37}
    \begin{split}
        \nabla_i W(|\bm{r}-\bm{r}_i|,h(\bm{r}))W(|\bm{r}-\bm{r}_j|,&h(\bm{r})) \\
        &\approx \nabla_i W(|\bm{r}-\bm{r}_i|,h(\bm{r})) \delta(\bm{r}-\bm{r}_j)
    \end{split}
\end{equation}

\begin{equation}\label{eq:in38}
    \begin{split}
        \nabla_j W(|\bm{r}-\bm{r}_i|,h(\bm{r}))W(|\bm{r}-\bm{r}_j|,&h(\bm{r})) \\
        &\approx \delta(\bm{r}-\bm{r}_i) \nabla_j W(|\bm{r}-\bm{r}_j|,h(\bm{r}))
    \end{split}
\end{equation}

By applying equation (\ref{eq:in37}) and equation (\ref{eq:in38}) into GDISPH, 
The following sets of equations are obtained.
The first set adopting $P^+_{i,j} = P^*_{i,j}$ and $\bm{v}^+_{i,j} = \frac{\bm{v}_i+\bm{v}_j}{2}$ is as follows (We call these GDISPH Case 2):
\begin{equation}\label{eq:in39}
    \begin{split}
        m_i\frac{d\bm{v}_i}{dt} &= - m_i \displaystyle \sum_{j=1}^N m_j u_j u_iP^*_{i,j}\\
        & \quad \left[ \frac{1}{q^2_i} \nabla_i W(\textcolor{black}{|}\bm{r}_i - \bm{r}_j\textcolor{black}{|},h_i) + \frac{1}{q^2_j} \nabla_i W(\textcolor{black}{|}\bm{r}_i - \bm{r}_j\textcolor{black}{|},h_j) \right],
    \end{split}
\end{equation}
\begin{equation}\label{eq:in40}
    \begin{split}
        \frac{du_i}{dt} &= -\displaystyle \sum_j^N m_j u_j u_iP^*_{i,j} \left( \frac{\bm{v}_i+\bm{v}_j}{2} - \bm{\dot r}_i\right)\\
        &\quad\quad \cdot \left[ \frac{1}{q^2_i} \nabla_i W(\textcolor{black}{|}\bm{r}_i - \bm{r}_j\textcolor{black}{|},h_i) + \frac{1}{q^2_j} \nabla_i W(\textcolor{black}{|}\bm{r}_i - \bm{r}_j\textcolor{black}{|},h_j) \right],
    \end{split}
\end{equation}
and the second set adopting $P^+_{i,j} = P^*_{i,j}$, $\bm{v}^+_{i,j} = \frac{\bm{v}_i+\bm{v}_j}{2}$, and $u^+_{i,j} = \frac{u_i+u_j}{2}$ is as follows (We call these GDISPH Case 3):
\begin{equation}\label{eq:in41}
    \begin{split}
        m_i\frac{d\bm{v}_i}{dt} &= - m_i \displaystyle \sum_{j=1}^N m_j P^*_{i,j} \frac{u_i+u_j}{2}\\
        & \quad \left[ \frac{u_j}{q^2_i} \nabla_i W(\textcolor{black}{|}\bm{r}_i - \bm{r}_j\textcolor{black}{|},h_i) + \frac{u_i}{q^2_j} \nabla_i W(\textcolor{black}{|}\bm{r}_i - \bm{r}_j\textcolor{black}{|},h_j) \right],
    \end{split}
\end{equation}
\begin{equation}\label{eq:in42}
    \begin{split}
        \frac{du_i}{dt} &= -\displaystyle \sum_j^N m_j P^*_{i,j} \frac{u_i+u_j}{2}\left( \frac{\bm{v}_i+\bm{v}_j}{2} - \bm{\dot r}_i\right)\\
        & \quad \left[ \frac{u_j}{q^2_i} \nabla_i W(\textcolor{black}{|}\bm{r}_i - \bm{r}_j\textcolor{black}{|},h_i) + \frac{u_i}{q^2_j} \nabla_i W(\textcolor{black}{|}\bm{r}_i - \bm{r}_j\textcolor{black}{|},h_j) \right],
    \end{split}
\end{equation}
where these sets of equations agree with the law of action and reaction and energy conservation.

\subsection{Incorporation of Balsara Switch}
Because the artificial viscosity adds unwanted viscosity outside of the shock regions, especially in the shear flow regions, the Balsara switch is necessary for SSPH and DISPH.
It is conceivable that GSPH and GDISPH effectively cause the same problem.
The straightforward solution to this problem is to use the Balsara switch in GSPH and GDISPH, yet this is likely to be difficult because of the need to separate the equations into effective viscous and inviscid terms. 
This is an issue that needs to be considered in future work.

In this section, we incorporate the Balsara switch into our GDISPH as follows.
For simplicity, let $D^{inv}_{ij}$ be defined as the inviscid force that $i$-th particle receives from $j$-th particle in DISPH.
Therefore, the momentum equation for DISPH (\ref{eq:over11}) can be rewritten by
\begin{equation}\label{eq:kh5}
    m_i \frac{d\bm{v}_i}{dt} = \displaystyle \sum_j D^{inv}_{ij},
\end{equation}
and let $G_{ij}$ as the force for any of our GDISPH.
Therefore, the momentum equation for GDISPH can be rewritten as follows:
\begin{equation}\label{eq:kh6}
    m_i \frac{d\bm{v}_i}{dt} = \displaystyle \sum_j G_{ij}.
\end{equation}
$G_{ij}$ can be decomposed by
\begin{equation}\label{eq:kh7}
    G_{ij} = G^{inv}_{ij} + G^{vis}_{ij},
\end{equation}
where $G^{inv}_{ij}$ and $G^{vis}_{ij}$ are the effective inviscid and the effective viscous term for GDISPH.
First, we assume that the effective inviscid term is almost the same as that for DISPH as follows:
\begin{equation}\label{eq:kh8}
    G^{inv}_{ij} = D^{inv}_{ij}.
\end{equation}
Therefore, $G^{vis}_{ij}$ is given by
\begin{equation}\label{eq:kh9}
    G^{vis}_{ij} = G_{ij} - D^{inv}_{ij}.
\end{equation}
Then, the Balsara switch is incorporated into the force of our GDISPH as follows:
\begin{equation}\label{eq:kh10}
    \begin{split}
        G^{'}_{ij} &= G^{inv}_{ij} + \frac{(F^{\textcolor{black}{\text{Balsara}}}_i + F^{\textcolor{black}{\text{Balsara}}}_j)}{2} G^{vis}_{ij},\\
        &= D^{inv}_{ij} + \frac{(F^{\textcolor{black}{\text{Balsara}}}_i + F^{\textcolor{black}{\text{Balsara}}}_j)}{2} \left( G_{ij} - D^{inv}_{ij} \right)
    \end{split}
\end{equation}
We also redefine the energy equation for our GDISPH in the same way.
As a result, for example, GDISPH Case 1 with Balsara switch is given by
\begin{equation}
    \begin{split}
        &m_i \frac{d\bm{v}_i}{dt} = -\displaystyle \sum^N_{\textcolor{black}{j=1}} \left [ g^{\textcolor{black}{\text{grad}}}_i \frac{P_i  U_i U_j}{q^2_i} \nabla_i W_{ij}(h_i) + g^{\textcolor{black}{\text{grad}}}_j \frac{P_j  U_i U_j}{q^2_j} \nabla_i W_{ij}(h_j) \right ] \\
        &\quad- \displaystyle \sum^N_{\textcolor{black}{j=1}} \frac{(F^{\textcolor{black}{\text{Balsara}}}_i + F^{\textcolor{black}{\text{Balsara}}}_j)}{2}\left [ g^{\textcolor{black}{\text{grad}}}_i \frac{(P^{*}_{ij}-P_i)  U_i U_j}{q^2_i} \nabla_i W_{ij}(h_i) \right.\\
        &\quad\quad\quad \left.+ g^{\textcolor{black}{\text{grad}}}_j \frac{(P^{*}_{ij}-P_j)  U_i U_j}{q^2_j} \nabla_i W_{ij}(h_j) \right ]
    \end{split}
\end{equation}
and
\begin{equation}
    \begin{split}
        \frac{dU_i}{dt} &= g^{\textcolor{black}{\text{grad}}}_i\displaystyle \sum^N_{\textcolor{black}{j=1}} \frac{P_iU_iU_j}{q^2_i} \bm{v}_{ij}\cdot \nabla_i W_{ij}(h_i) \\
        &+g^{\textcolor{black}{\text{grad}}}_i\displaystyle \sum^N_{\textcolor{black}{j=1}}  \frac{(F^{\textcolor{black}{\text{Balsara}}}_i + F^{\textcolor{black}{\text{Balsara}}}_j)}{2} 
 \frac{(P^*_{ij}-P_i)U_iU_j}{q^2_i} \bm{v}_{ij}\cdot \nabla_i W_{ij}(h_i).
    \end{split}
\end{equation}
\textcolor{black}{The Balsara switch can be incorporated into GSPH in a similar way.}

\section{Numerical Experiments}
\label{sec:num}
\begin{table}
	\centering
    \scalebox{0.85}[1]{
	\begin{tabular}{ccc} 
		\hline
		Scheme & RHS of Momentum eq. & RHS of Energy eq. \\
		\hline
		SSPH    & (\ref{eq:over1}) + (\ref{eq:over4}) & (\ref{eq:over2}) + (\ref{eq:over5})\\
        SSPH with ArtCond   & (\ref{eq:over1}) + (\ref{eq:over4}) & (\ref{eq:over2}) + (\ref{eq:over5}) + (\ref{eq:over7-2})\\
        SPH GDF   & (\ref{eq:GDF1}) + (\ref{eq:GDF5}) & (\ref{eq:GDF2}) + (\ref{eq:GDF6})\\
		GSPH Case 3   & (\ref{eq:over9}) & (\ref{eq:over10})\\
        GSPH Case 3-2  & (\ref{eq:over9}) & (\ref{eq:over10}) but $v^*_{ij} = \frac{(\bm{v}_i + \bm{v}_j)}{2}$\\
		DISPH  & (\ref{eq:over11}) + (\ref{eq:over4}) & (\ref{eq:over12}) + (\ref{eq:over5})\\
        GDISPH Case 1 & (\ref{eq:law15}) & (\ref{eq:law14})\\
        GDISPH Case 2 & (\ref{eq:in39})& (\ref{eq:in40})\\
        GDISPH Case 3 & (\ref{eq:in41}) & (\ref{eq:in42})\\
		\hline
	\end{tabular}
 }
    \caption{Formula of schemes used in the numerical experiments. Column 1: name of each scheme; Column 2: RHS of momentum equation; Column 3: RHS of energy equation;}
    \label{table:scheme}
\end{table}

This section shows the results of several tests using various schemes, including our GDISPH Case 1, GDISPH Case 2, and GDISPH Case 3.
Section \ref{sec:num_m} briefly describes the practical implementation of our code.
We show the results of the one-dimensional Riemann problem tests in Section \ref{sec:rie}, the pressure equilibrium tests in Section \ref{sec:hyd} which is the same test performed by \citet{Saitoh_Makino2013}, the Sedov-Taylor tests in Section \ref{sec:pol}, and the Kelvin-Helmholtz tests in Section \ref{sec:kh}.
\subsection{Numerical Method}
\label{sec:num_m}
\textcolor{black}{To examine a performance of the GDISPH, we developed a original code, adopting} the leap-frog method \textcolor{black}{(kick-drift-kick)} for time integration with the shared time step, which is given by
\begin{equation}\label{eq:num_m1}
    dt = \text{min}_i dt_i,
\end{equation}
where 
\begin{equation}\label{eq:num_m2}
    dt_i = C_{CFL} \frac{2h_i}{\text{max}_j v^{sig}_{ij}},
\end{equation}
and $C_{CFL} = 0.3$. 
\textcolor{black}{
The time integration process proceeds as follows:
\begin{align}
\bm{v}^{n+1/2}_i &= \bm{v}^{n}_i + \bm{a}^{n}_i \frac{dt}{2}, \\
u^{n+1/2}_i &= u^{n}_i + \dot{u}^{n}_i \frac{dt}{2}, \\
\bm{r}^{n+1}_i &= \bm{r}^{n}_i + \bm{v}^{n+1/2}_i dt, \label{eq:al:1} \\
^{*}\bm{v}^{n+1}_i &= \bm{v}^{n}_i + \bm{a}^{n}_i dt, \\
^{*}u^{n+1}_i &= u^{n}_i + \dot{u}^{n}_i dt, \label{eq:al:2} \\
\bm{a}^{n+1}_i &= \bm{a}(\bm{r}^{n+1}_i, ^{*}\bm{v}^{n+1}_i, ^{*}u^{n+1}_i), \\
\dot{u}^{n+1}_i &= \dot{u}(\bm{r}^{n+1}_i, ^{*}\bm{v}^{n+1}_i, ^{*}u^{n+1}_i), \\
\bm{v}^{n+1}_i &= \bm{v}^{n}_i + \bm{a}^{n+1}_i \frac{dt}{2}, \\
u^{n+1}_i &= u^{n}_i + \dot{u}^{n+1}_i \frac{dt}{2}.
\end{align}
The smoothing length $h$, the density $\rho$, and $f^{\text{grad}}$ is updated right after equation (\ref{eq:al:1}).
The energy density $q$, $F^{\text{Balsara}}$, and $g^{\text{grad}}$ is updated right after equation (\ref{eq:al:2}) using $^{*}\bm{v}^{n+1}_i$ and $^{*}u^{n+1}_i$.
}

\textcolor{black}{For any methods in two- and three-dimensional tests}, we set the smoothing length of the $i$-th particle $h_i$ so that there are $N_{ngb}$ particles within the region of radius $2h_i$ centred on the particle. 
$N_{ngb}$ is an arbitrary parameter. \textcolor{black}{Usually, $N_{ngb}$ is adjusted depending on the problems and number of particles. For example, the spatial accuracy of SPH is improved by the combination of using more particles and increasing the neighbour number accordingly, which was analysed by \citet{Zhu2015}. In addition, it is well known that tackling strong shocks requires more $N_{\text{ngb}}$. In this paper, } \textcolor{black}{$N_{ngb}$ is set to 50, 228, and 80 in Section \ref{sec:hyd}, Section \ref{sec:pol}, and Section \ref{sec:kh}, respectively.}
\textcolor{black}{In the one-dimensional tests shown in Section \ref{sec:rie}, 
\textcolor{black}{the smoothing length is evaluated using equation (\ref{eq:condic}), then the density is updated in all schemes}. $N_{ngb}$ is set to 5.2 in Section \ref{sec:sod} and Section \ref{sec:vac}, and 8.0 in Section \ref{sec:strong}.} The kernel for one dimension is adopted by the 1D Wendland $C^4$ kernel defined as
\begin{equation}\label{eq:num_m3}
    W(r,h) = \frac{C_{norm}}{h}
    \begin{cases}
        (1-\frac{1}{2}\textcolor{black}{z})^5(2\textcolor{black}{z}^2 + \frac{5}{2}\textcolor{black}{z} + 1) & \text{if $0 \leq \textcolor{black}{z} < 2$,} \\
        0                                           & \text{if $2 \leq \textcolor{black}{z}$,}
    \end{cases}
\end{equation}
\textcolor{black}{where distance normalised by smoothing length $\textcolor{black}{z} = \frac{r}{h}$.}
For two or three dimensions, the 2D/3D Wendland $C^4$ kernel is employed and defined as follows:
\begin{equation}\label{eq:num_m4}
    W(q) = \frac{C_{norm}}{h^\nu}
    \begin{cases}
        (1-\frac{1}{2}\textcolor{black}{z})^6(\frac{35}{12}\textcolor{black}{z}^2 + 3\textcolor{black}{z} + 1) & \text{if $0 \leq \textcolor{black}{z} < 2$,} \\
        0                                             & \text{if $2 \leq \textcolor{black}{z}$,}
    \end{cases}
\end{equation}
where $\nu$ is the dimension and
\begin{equation}\label{eq:num_m5}
    C_{norm} = 
    \begin{cases}
        \frac{3}{4} & \text{if $\nu = 1$,} \\
        \frac{9}{4\pi} & \text{if $\nu = 2$,} \\
        \frac{495}{256\pi} & \text{if $\nu = 3$.}
    \end{cases}
\end{equation}
\citet{Dehnen2012} proved that the Wendland kernels, introduced by \citet{Wendland1995}, are stable to the paring instability at all neighbour numbers $N_{ngb}$ while having the compact support. \textcolor{black}{An exact Riemann solver, which uses iteration to find the exact solution, is used for GSPH / GDISPH.}

A table for the formula of the schemes we use is shown in Table \ref{table:scheme}. \textcolor{black}{GSPH Case 3-2 is GSPH Case 3-1 with $\bm{v}^*_{ij}$ replaced with $(\bm{v}_i+\bm{v}_j)/2$. The reason we introduce GSPH Case 3-2 is to show what physical effect $\bm{v}^*_{ij}$ gives by comparing the results between GSPH Case 3-2 and GSPH Case 3-1.} $\alpha_{AV} = 1$ is adopted as a default value for the schemes using the Monaghan artificial viscosity and $\alpha_{u} = 1$ for SSPH with ArtCond. \textcolor{black}{Note that $\alpha_{AV} = 0$ is used in Figure \ref{fig:Result_ShockTube_velpanel} , while various parameters including $\alpha_{AV} = 1$ are used in Section \ref{sec:pol}.}
For DISPH, we use the smoothed density by equation (\ref{eq:over3}) as the density in the artificial viscosity instead of using equation (\ref{eq:over16}) because it is more stable at the place where there are strong pressure gradients \citep{Saitoh_Makino2013}.
For all GDISPH, equation (\ref{eq:over16}) is used as the density in the inputs of the Riemann solver.
When plotting density, we use the smoothed density by equation (\ref{eq:over3}) for all schemes.
\textcolor{black}{Since the equation of DISPH is used to incorporate the Balsara switch in GDISPH, we think that a fair comparison between GDISPH and DISPH using the Balsara switch is not possible (e.g. There is a concern of an effect that is characteristic only of DISPH, not GDISPH,  can emerge in GDISPH.), so the results for Section \ref{sec:rie}, Section \ref{sec:hyd}, and Section \ref{sec:pol} are shown without the Balsara switch.
However, shear flow regions, where the artificial viscosity, GSPH, and our GDISPH can misidentify as shock regions, do not emerge physically in those tests and we have confirmed that the results with and without the Balsara switch were almost the same results.
Since there are the shear flow regions in the Kelvin-Helmholtz tests, the Balsara switch is incorporated in all schemes in Section \ref{sec:kh}, checking if our GDISPH with Balsara switch can work.}
\subsection{Riemann Problem Test}
\label{sec:rie}
Here, we show the results of the one-dimensional Riemann problem tests to check if the schemes have the ability to handle the shocks and the contact discontinuities correctly.

The setup is as follows. 
We set $\gamma = 1.4$ and give the initial internal energy to each particle to ensure the given initial $P$.
Equal-mass particles are used to generate the initial conditions and place them regularly within the domain of $-1\leq x < 1$.

\subsubsection{Sod's Shock Tube Test}
\label{sec:sod}
\begin{figure*}[!t]
     \centering
	\includegraphics[width=1\linewidth]{ 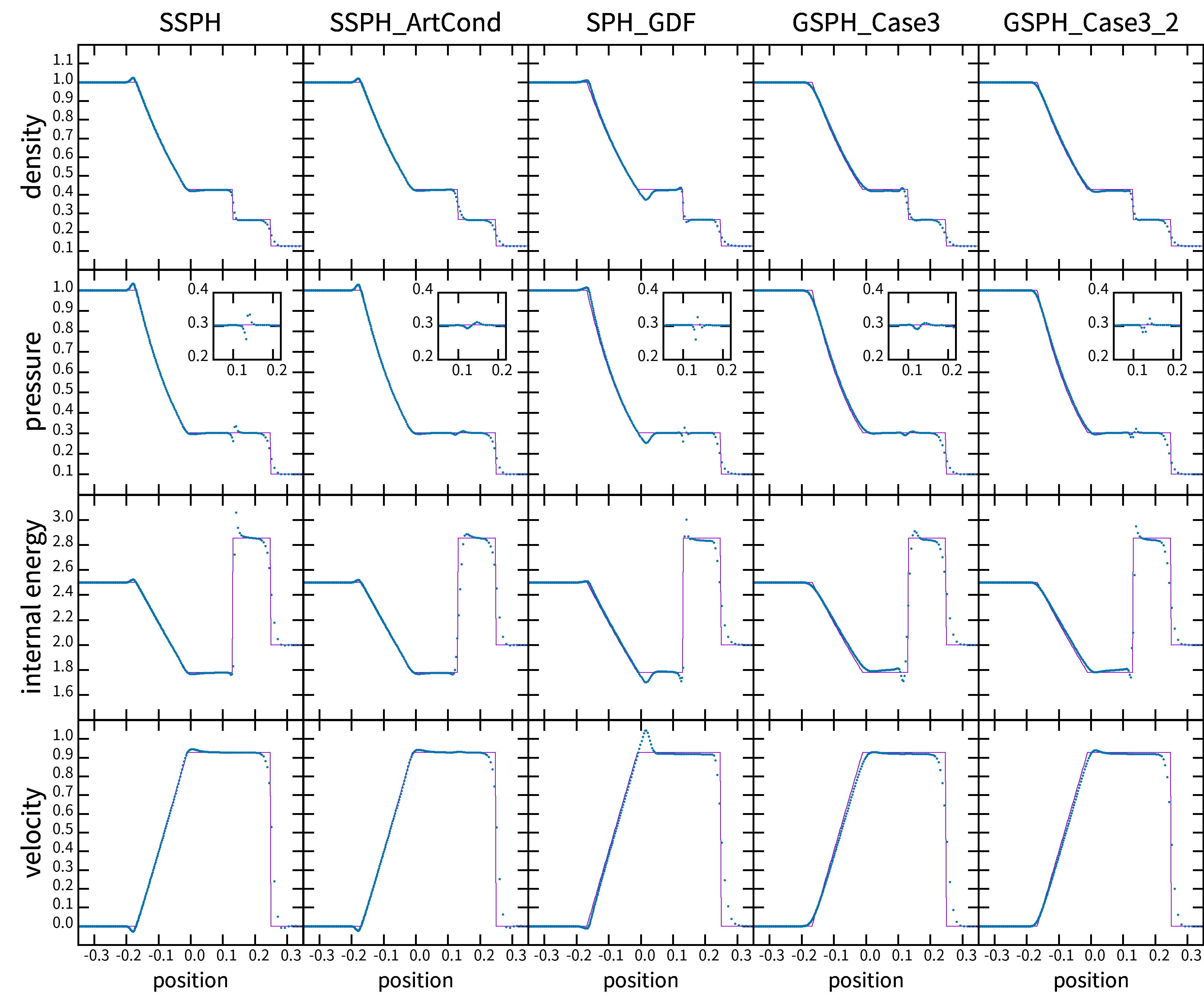}
    \caption{Results of the one-dimensional sod's shock tube tests at $t=0.14154$ with the SSPH, SSPH with ArtCond, SPH GDF, GSPH Case 3, and GSPH Case 3-2, from the first column to the fifth column, respectively.
    The density, pressure, internal energy, and velocity of each particle are plotted from the first row to the fourth row, respectively.
    The insets in the pressure panels are the close-up views around the contact discontinuity.
    The solid line indicates the analytical solution, while the dots indicate the physical quantities of each SPH particle.}
    \label{fig:Problem_ShockTube}
\end{figure*}

\begin{figure*}[!t]
    \centering
    \includegraphics[width=0.85\linewidth]{ 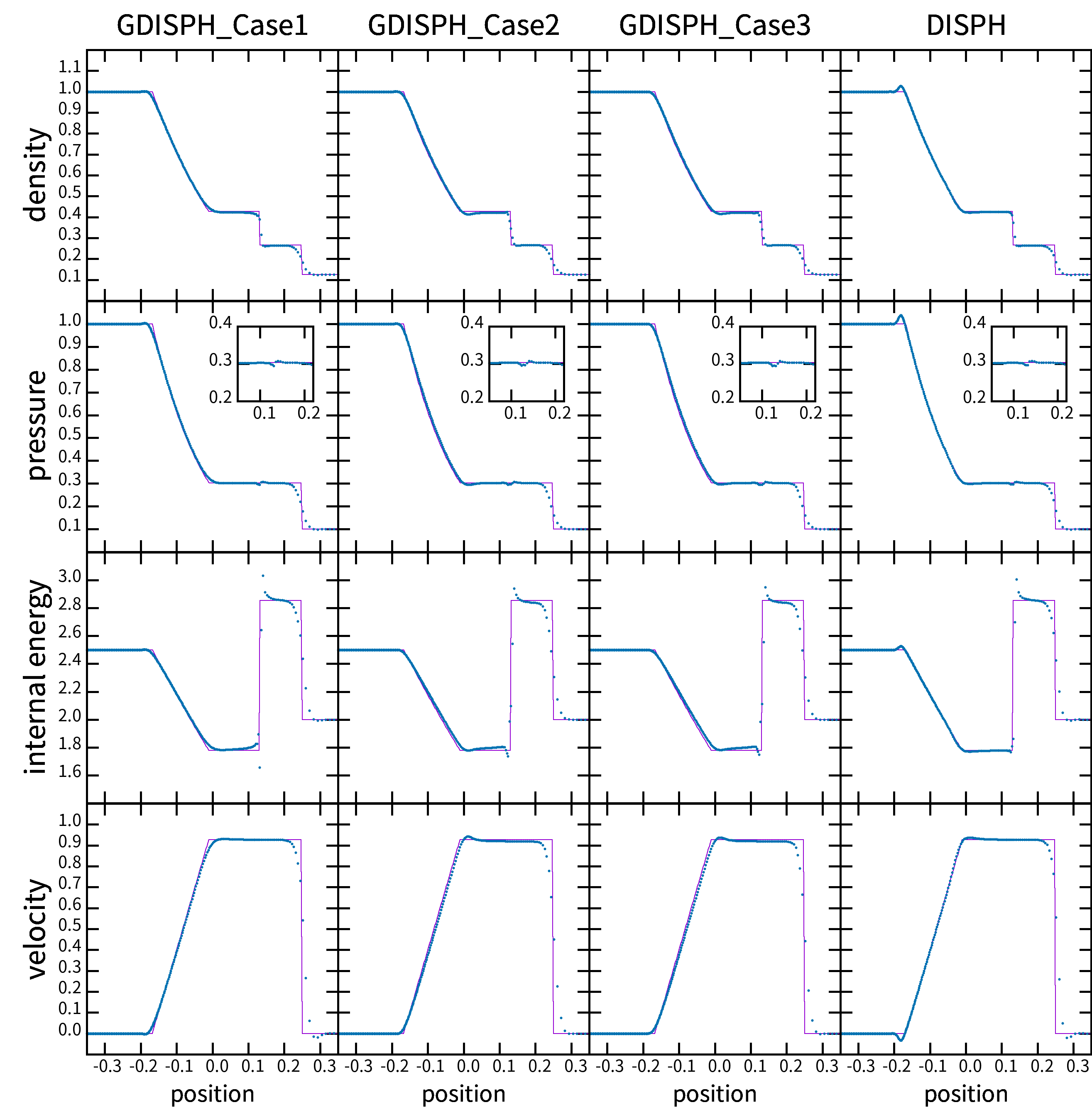}
    \caption{Same as Figure \ref{fig:Problem_ShockTube}, but with GDISPH Case 1, GDISPH Case 2, GDISPH Case 3 and DISPH.}
    \label{fig:Result_ShockTube}
\end{figure*}

\begin{figure}[!t]
    \includegraphics[width=\linewidth]{ 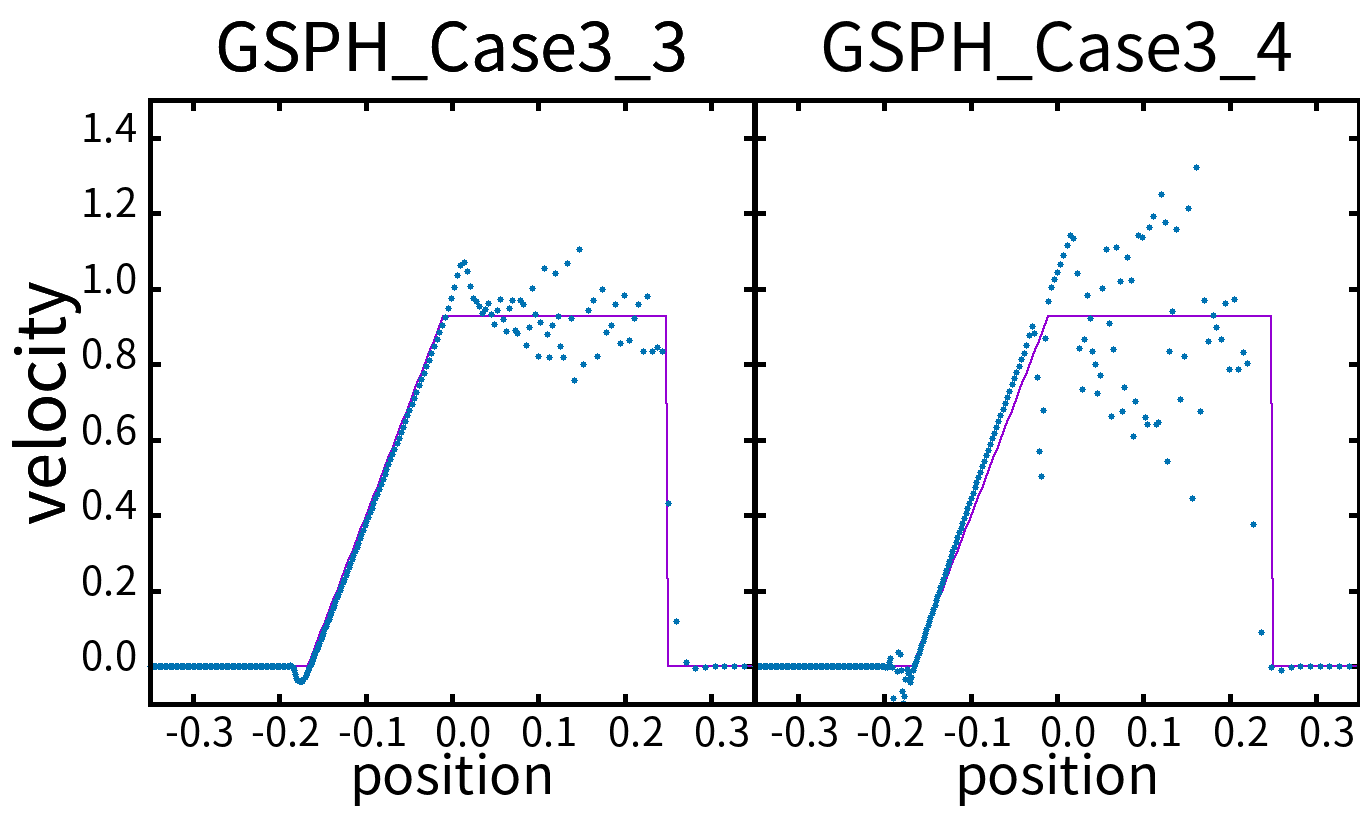}
    \includegraphics[width=\linewidth]{ 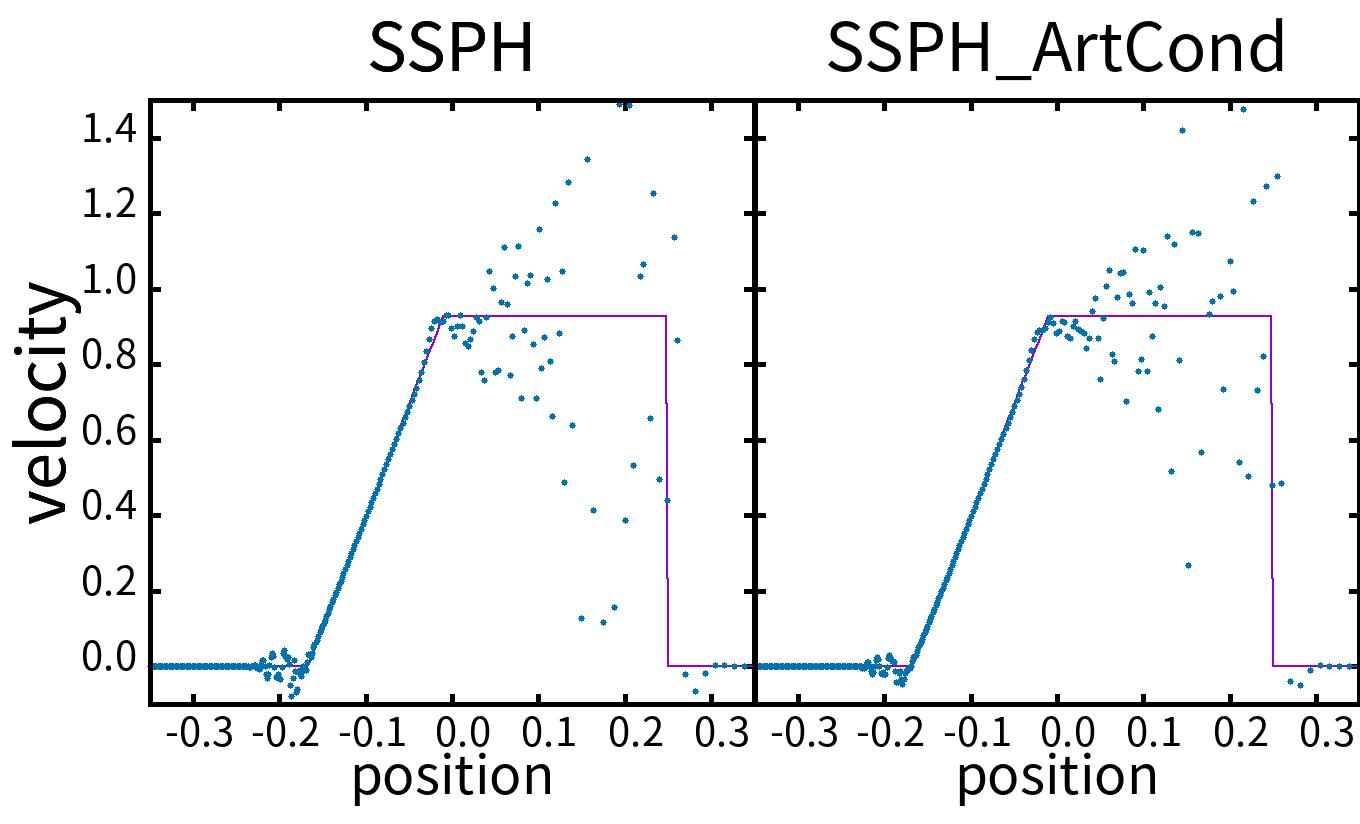}
    \caption{Same as Figure \ref{fig:Problem_ShockTube}, but with GSPH Case 3 with $P^*_{ij} = \frac{(P_i + P_j)}{2}$ (GSPH Case 3-3) in the upper left, GSPH Case 3 with $P^*_{ij} = \frac{(P_i + P_j)}{2}$ and $\bm{v}^*_{ij} = \frac{(\bm{v}_i + \bm{v}_j)}{2}$ (GSPH Case 3-4) in the upper right, SSPH in the lower left, and SSPH with ArtCond in the lower right.
    $\alpha_{AV} = 0$ is used.}
    \label{fig:Result_ShockTube_velpanel}
\end{figure}

The sod's shock tube test is the most basic test for numerical schemes for the compressible fluid
because expansion waves, contact discontinuities, and shock waves occur in the test.
The initial condition is given as follows:
\begin{equation}\label{eq:sod1}
    \begin{cases}
        \rho = 1.000, P = 1.000, v = 0.000 & \text{if $x \leq 0$,} \\
        \rho = 0.1250, P = 0.1000, v = 0.000 & \text{if $0 < x$.}
    \end{cases}
\end{equation}
We use \textcolor{black}{$N_{ngb} = 5.2$} and place $711$ and $87$ particles in the left and the right domains, respectively, to generate this initial condition.
\textcolor{black}{Note that while $\alpha_{AV}=1$ is used in Figure \ref{fig:Problem_ShockTube} and Figure \ref{fig:Result_ShockTube}, $\alpha_{AV}=0$ is used in Figure \ref{fig:Result_ShockTube_velpanel}.}

Figure \ref{fig:Problem_ShockTube} shows the results of the Sod's Shock Tube tests with SSPH, SSPH with ArtCond, SPH GDF, GSPH Case 3 and GSPH Case 3-2.
From the first row to the fourth row, 
the density, pressure, internal energy, and velocity of each particle are plotted by the blue dots. 
The purple lines show the analytical solutions.
The results of all the schemes \textcolor{black}{in the figure} are consistent with the analytical solution.
In SSPH, at the contact discontinuity where the density and internal energy are discontinuous, roughly 4 to 5 SPH particles sample the intermediate of the density discontinuity, while only 2 SPH particles sample the intermediate of the internal energy discontinuity. As a result, the inconsistency of smoothness arises, causing the pressure blip around the contact discontinuity. A similar result to SSPH is obtained for SPH GDF around the contact discontinuity. SSPH with ArtCond has more SPH particles sampling the intermediate of the internal energy discontinuity compared to SSPH, thanks to the additional dissipation term, \textcolor{black}{resulting in the successful suppression of the pressure blip compared to SSPH.}
A similar result to SSPH with ArtCond is obtained for GSPH Case 3 around the contact discontinuity, but GSPH Case 3-2. Compared to GSPH Case 3, GSPH Case 3-2 has fewer sampling particles around the intermediate of the internal energy discontinuity and a somewhat bigger pressure blip at the contact discontinuity.

The results of the Sod's Shock Tube tests with GDISPH Case 1, GDISPH Case 2, GDISPH Case 3, and DISPH are shown in Figure \ref{fig:Result_ShockTube}.
The results of all the schemes \textcolor{black}{in the figure} are consistent with the analytical solution, successfully suppressing the pressure blip without smoothing the internal energy around the internal energy discontinuity.

\textcolor{black}{As a whole}, our schemes, SSPH with ArtCond, and GSPH Case 3 still have a small variation in the pressure around the contact discontinuity as well as DISPH.
This is because of the change in the particle distribution at the contact discontinuity (see, \citet{Saitoh_Makino2013}).

For comparison, we execute additional tests \textcolor{black}{to examine what physical effects $P^*_{ij}$ gives in Godunov DISPH}.
In Figure \ref{fig:Result_ShockTube_velpanel}, we show the velocity results of the Sod's Shock Tube tests with GSPH Case 3 with $P^*_{ij} = \frac{(P_i + P_j)}{2}$ (GSPH Case 3-3), GSPH Case 3 with $P^*_{ij} = \frac{(P_i + P_j)}{2}$ and $\bm{v}^*_{ij} = \frac{(\bm{v}_i + \bm{v}_j)}{2}$ (GSPH Case 3-4), SSPH, and SSPH with ArtCond.
$\alpha_{AV} = 0$ is used.
It is clear that the behaviour of GSPH Case 3-3 and GSPH Case 3-4, in which the pressure solutions of the Riemann problem are not used, are similar to SSPH \textcolor{black}{with $\alpha_{AV}=0$} and SSPH with ArtCond with $\alpha_{AV}=0$. 
GSPH Case 3-3, in which only the velocity solution of the Riemann problem is used, relatively suppresses the post-shock oscillation compared to GSPH Case 3-4, in which none of the solutions are used. \textcolor{black}{Therefore we can deduce that $P^*_{ij}$ mainly, if not only, gives an effective viscosity.}

\subsubsection{Vacuum Test}
\label{sec:vac}
\begin{table}
	\centering
	\begin{tabular}{cc} 
		\hline
		Scheme & overestimation error at $x=0$\\
		\hline
            SPH GDF  & -56.4\%\\
            DISPH  & 19.4\%\\
            SSPH & 20.5\%\\
            SSPH with ArtCond & 49.6\%\\
            GDISPH Case 1 & 187\% \\
            GDISPH Case 3 & 206\%\\
            GSPH Case 3-2  & 207\% \\
            GDISPH Case 2  & 212\% \\
		GSPH Case 3  & 236\% \\
		\hline
	\end{tabular}
    \caption{Overestimation error of the internal energy from its analytic value at $x=0$ for all the schemes in the Vaccume tests (see Section \ref{sec:vac}). Sorted by smallest to largest.}
    \label{table:vac_error}
\end{table}

\begin{figure*}[!t]
    \centering
	\includegraphics[width=1\linewidth]{ 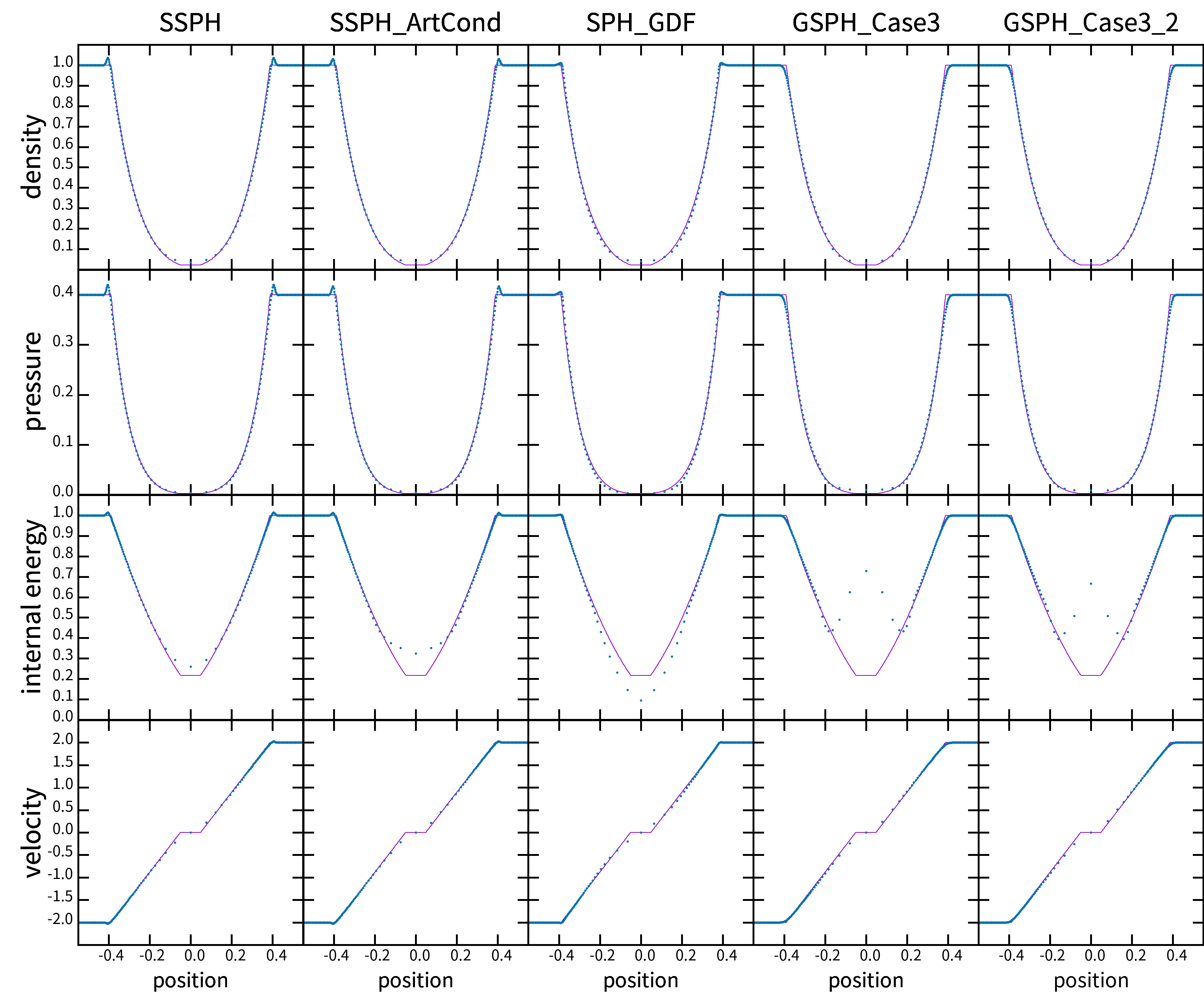}
    \caption{Same as Figure \ref{fig:Problem_ShockTube}, but for the one-dimensional vacuum tests at $t=0.14154$.}
    \label{fig:Problem_Vaccume}
\end{figure*}

\begin{figure*}[!t]
    \centering
	\includegraphics[width=0.85\linewidth]{ 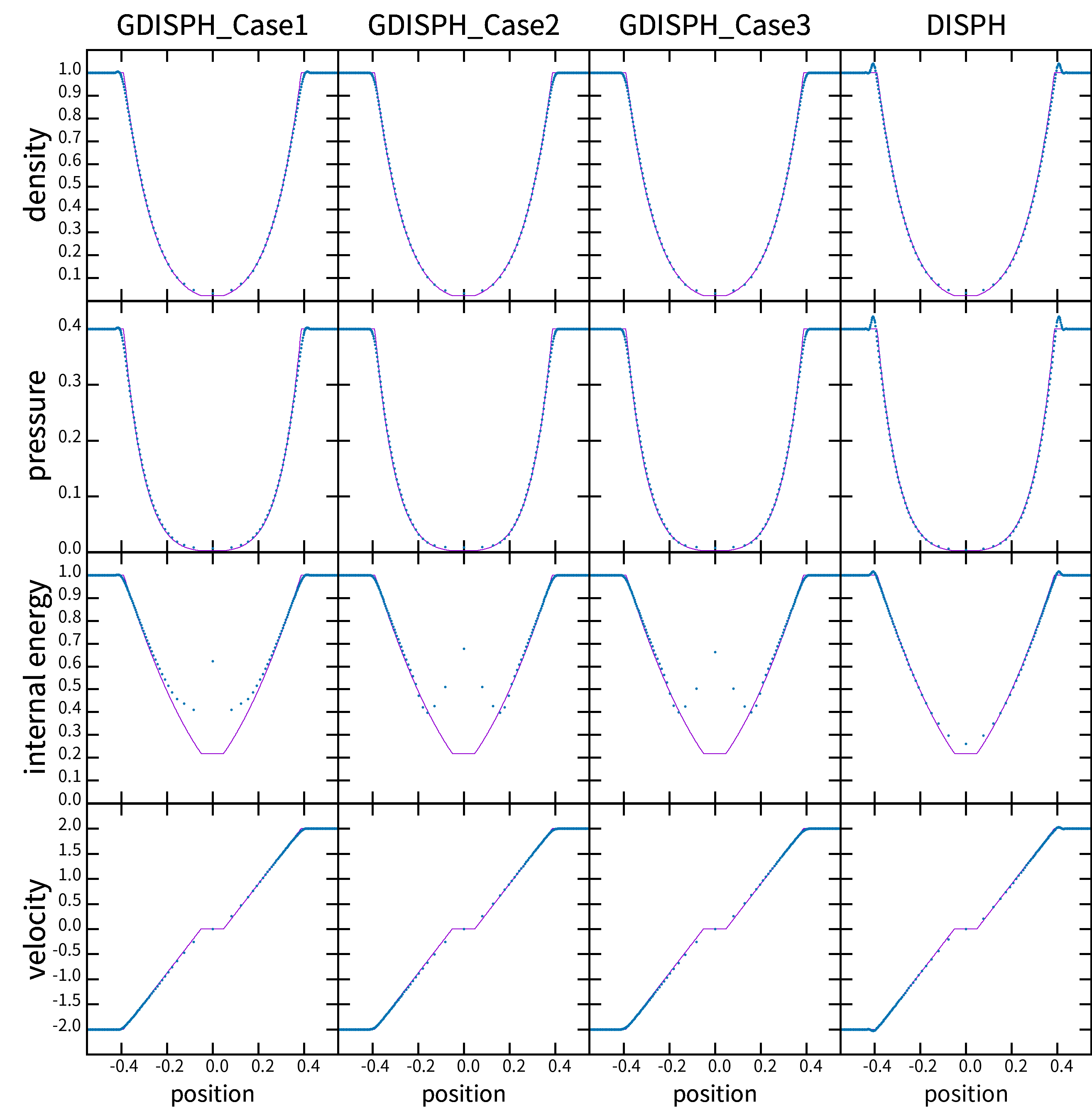}
    \caption{Same as Figure \ref{fig:Result_ShockTube}, but for the one-dimensional vacuum tests at $t=0.14154$.}
    \label{fig:Result_Vaccume}
\end{figure*}
Here, we see how the schemes handle the vacuum area.
The initial condition is given as follows:
\begin{equation}\label{eq:sod2}
    \begin{cases}
        \rho = 1.000, P = 0.4000, v = -2.000 & \text{if $x \leq 0$,} \\
        \rho = 1.000, P = 0.4000, v = 2.000 & \text{if $0 < x$.}
    \end{cases}
\end{equation}
The left side of the fluid moves toward the left, while the right side of the fluid moves toward the right.
Therefore, a vacuum area occurs around $x = 0$. \textcolor{black}{$N_{\text{ngb}}=5.2$} is used.
To represent this initial condition, place $400$ and $400$ particles in the left and right regions, respectively.

Figure \ref{fig:Problem_Vaccume} shows the results of the vacuum tests with
SSPH, SSPH with ArtCond, SPH GDF, GSPH Case 3, and GSPH Case 3-2, while Figure \ref{fig:Result_Vaccume} shows the results of  the tests with GDISPH Case 1, GDISPH Case 2,
GDISPH Case 3, and DISPH.
The overestimation error of the internal energy from its analytic value at $x=0$ is shown in Table \ref{table:vac_error}.
All of the schemes are able to reproduce the analytical solutions of all the physical quantities very well, except for the internal energy.
SSPH and DISPH have an overestimation error of $20\%$. SPH GDF underestimates the internal energy by about $56\%$, which is problematic because the internal energy is negative and causes a forced termination of the calculation.
SSPH with ArtCond overestimates the internal energy by about $50\%$ of the analytical solution compared to SSPH because artificial thermal conductivity transfers the energy from the outside to the inside.
A similar phenomenon can be observed between GSPH case 3 and GSPH case 3-2.
Comparing SSPH (DISPH) and GSPH (GDISPH), the scheme using the Riemann solver has a more significant error than the scheme using artificial viscosity \textcolor{black}{with properly adjusted $\alpha_{\text{AV}}$}. \textcolor{black}{All of the GDISPH and GSPH schemes have an overestimation error of above $187\%$, which is significantly worse than the other schemes (e.g. GDISPH Case 1 has 9 times bigger error than DISPH).  Among the schemes with the Riemann solver, our GDISPH Case 1 has the best performance in terms of the overestimation error, while GSPH Case 3 has $50\%$ more error than GDISPH Case 1.} Since the Godunov method also performs poorly in the vacuum regime, this poor performance can be attributed to using the Riemann solver (see \citet{Toro2009}).

\subsubsection{Strong Shock Test}
\label{sec:strong}
\begin{figure*}[!t]
\centering  
	\includegraphics[width=1\linewidth]{ 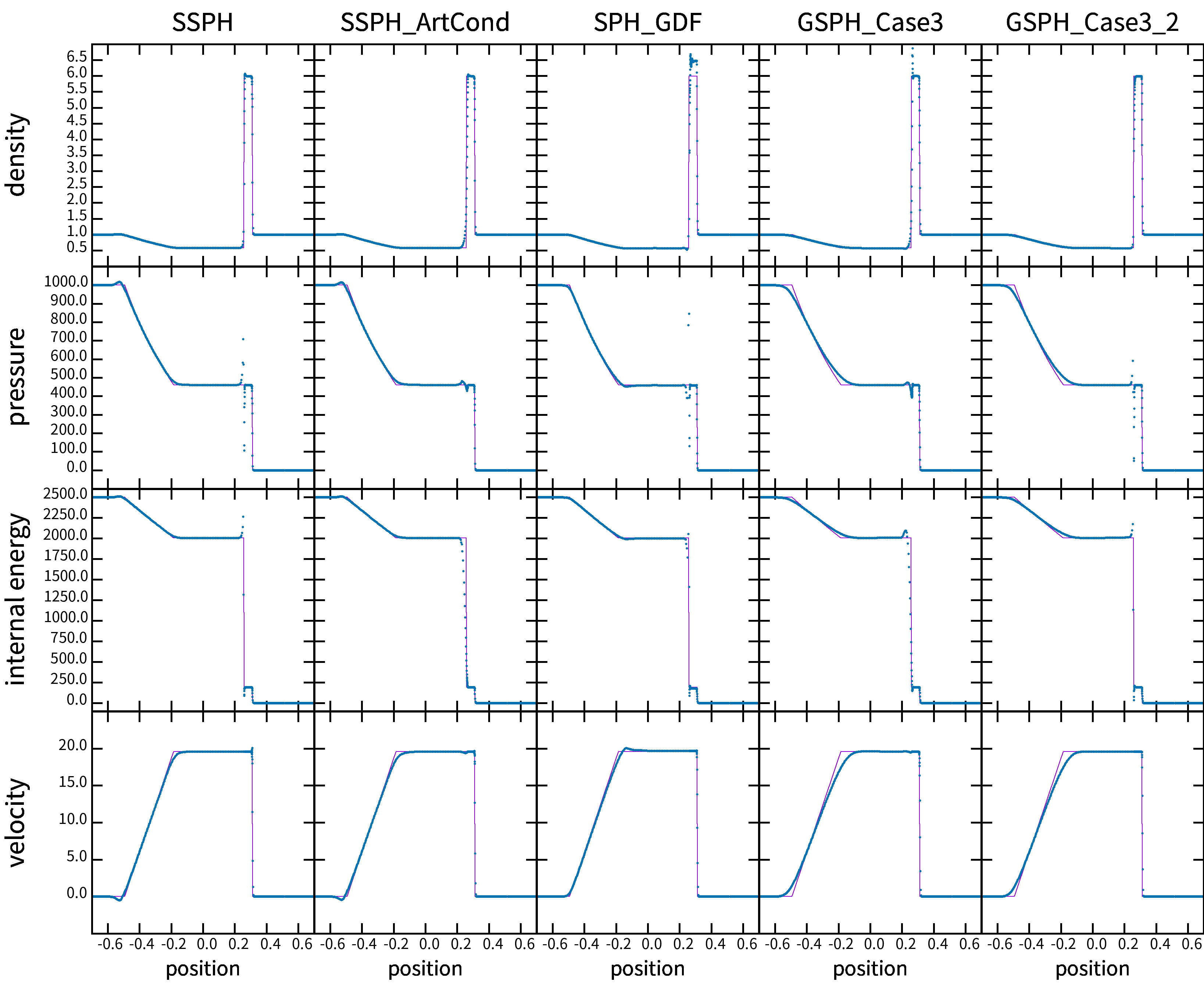}
    \caption{Same as Figure \ref{fig:Problem_ShockTube}, but for the one-dimensional strong shock tests at $t = 0.014154$.}
    \label{fig:Problem_Strong}
\end{figure*}

\begin{figure*}[!t]
\centering
	\includegraphics[width=0.85\linewidth]{ 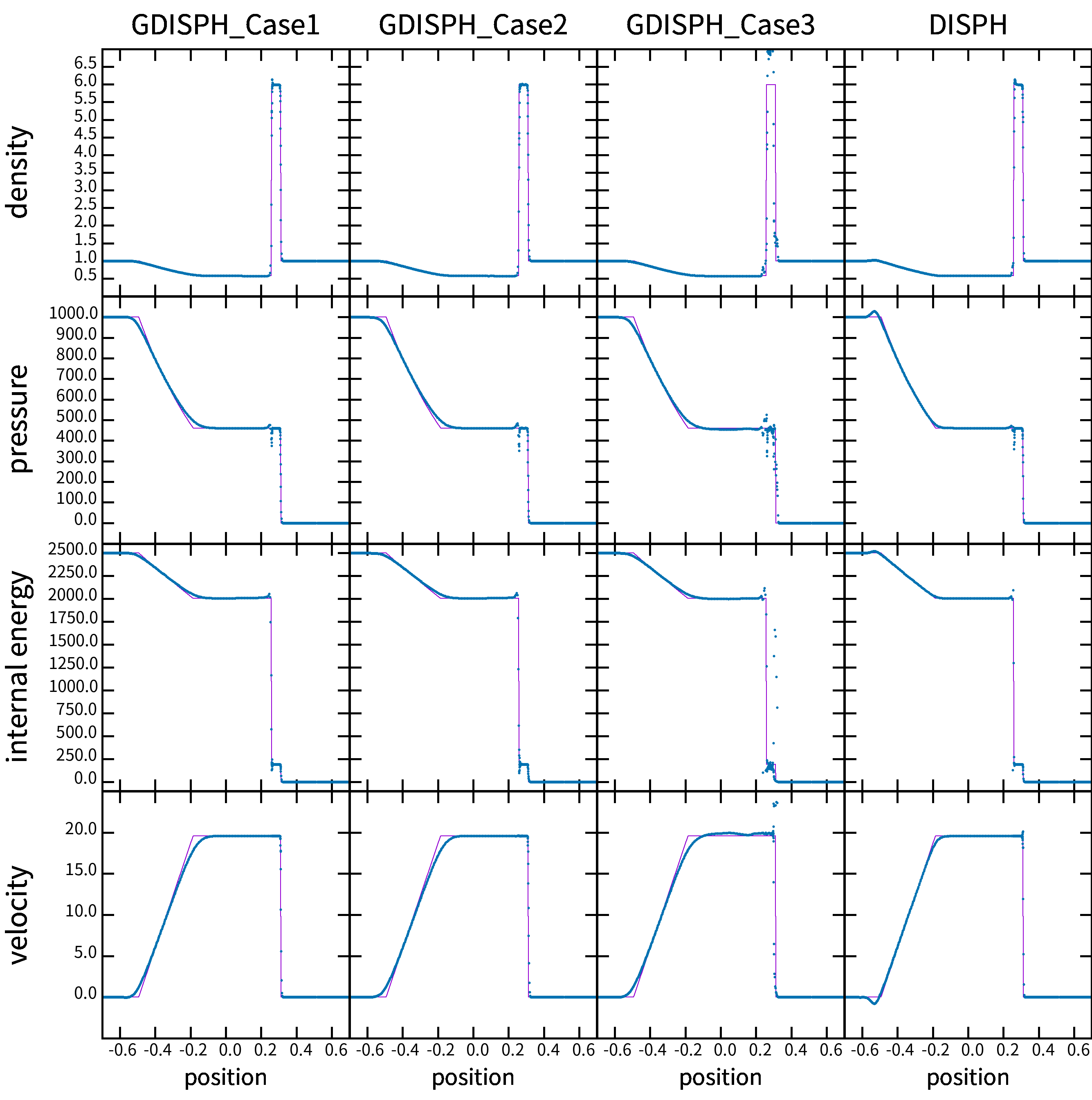}
    \caption{Same as Figure \ref{fig:Result_ShockTube}, but for the one-dimensional strong shock tests at $t = 0.014154$.}
    \label{fig:Result_Strong}
\end{figure*}

Here, we see how the schemes handle the strong shock.
The initial condition is given as follows:
\begin{equation}\label{eq:sod3}
    \begin{cases}
        \rho = 1.000, P = 1000, v = 0.000 & \text{if $x \leq 0$,} \\
        \rho = 1.000, P = 0.1000, v = 0.000 & \text{if $0 < x$.}
    \end{cases}
\end{equation}
There is a huge pressure difference in the initial condition, and 
the analytical ratio of densities before and after the shock wave is $5.99257$,
which is a close value to the strong shock limit of the density ratio: $(\gamma+1)/(\gamma-1)$.
Therefore this test would be hard for DISPH and GDISPH because 
they assume the differentiability of the pressure.
We use \textcolor{black}{$N_{\text{ngb}}=8.0$} and place $400$ and $400$ particles in the left and the right domain, respectively.

The results of the Strong Shock Tube tests with SSPH, SSPH with ArtCond, SPH GDF, GSPH Case 3, and GSPH Case 3-2 are presented in Figure \ref{fig:Problem_Strong}.
Overall, the results of all the schemes \textcolor{black}{in the figure excluding SPH GDF} are consistent with the analytical solution.
SPH GDF overestimates the density behind the shock by about 8\% of its analytical value.
GSPH Case 3 causes the jump of the density at the contact discontinuity.
Other than that, the characteristics of the results are similar to that of the Sod's Shock Tube Test (see Section \ref{sec:sod}): SSPH with ArtCond (GSPH Case 3) has smoother internal energy at the contact discontinuity than SSPH (GSPH Case 3-2), leading to the successful suppression of the pressure blip compared to SSPH (GSPH Case 3-2).

In Figure \ref{fig:Result_Strong}, we show the results of the Strong Shock tests with GDISPH Case 1, GDISPH Case 2, GDISPH Case 3, and DISPH.
GDISPH Case 3 cannot capture the shock correctly, causing the overestimation of the density by roughly 16\%, which is worse than SPH GDF, and the small oscillation in the velocity behind the shock.
The others \textcolor{black}{in the figure}, including our schemes: GDISPH Case 1 and GDISPH Case 2, successfully suppress the pressure blip, while reproducing the analytical solutions behind the shock very well.
\textcolor{black}{Comparing the successful scheme in terms of suppression of the pressure blip, SSPH with ArtCond and GSPH Case 3 have more accurate results around the contact discontinuity than all DISPH-type schemes.}

Contrary to our initial concern, DISPH and our schemes: GDISPH Case 1 and GDISPH Case 2 are able to deal with the initial large jump in pressure and the strong shock.

\subsection{Pressure Equilibrium Test}
\label{sec:hyd}
As is shown in Section \ref{sec:sod} and Section \ref{sec:strong}, some schemes are able to suppress the pressure blip, which is an unphysical repulsive force, in the Riemann problem tests at the contact discontinuities. 
The unphysical repulsive force effectively serves as the surface tension \citep{Saitoh_Makino2013}. To double-check if the effect of the unphysical repulsive force is actually suppressed for the schemes, we show the results of the two-dimensional pressure equilibrium test, which is first performed by \citet{Saitoh_Makino2013}.

We set the domain of $0\leq x,y < 1$ with periodic boundary conditions, use equal-mass particles, and place them regularly in a lattice manner. $\gamma$ is set to $5/3$.
The initial condition is given as follows:
\begin{equation}\label{eq:het1}
    \rho = 
    \begin{cases}
        4.00 & \text{if $0.25 \leq x \leq 0.75$ and $0.25 \leq y \leq 0.75$}, \\
        1.00 & \text{otherwise},
    \end{cases}
\end{equation}
\begin{equation}\label{eq:het2}
    P = 2.50,
\end{equation}
and
\begin{equation}\label{eq:het3}
    \bm{v} = 0.00.
\end{equation}
The contact discontinuity is the border between the \textcolor{black}{high-density} and \textcolor{black}{low-density} regions.
The number of particles in the dense region is $3969$ and that in the ambient is $3008$.
This system is \textcolor{black}{initially in pressure equilibrium}, 
so the analytical solution is the same as the initial condition at any time.
If the effective surface tension is working at the contact discontinuity, the shape of the \textcolor{black}{high-density} region turns into a circle in order to minimise the surface area of the contact discontinuity.
We use $N_{ngb} = 50$ as the neighbour number.
The slowest sound speed in this domain is the sound speed in the \textcolor{black}{high-density} region: $1.02$.
Therefore the time taken to cross the computational domain at the slowest sound speed in this domain $t^{s}_{cross}$ is about $1.0$.
The tests are performed up to $t=8.0$, which is about $8t^{s}_{cross}$.
\begin{figure*}[!t]
\centering
	\includegraphics[width=\linewidth]{ 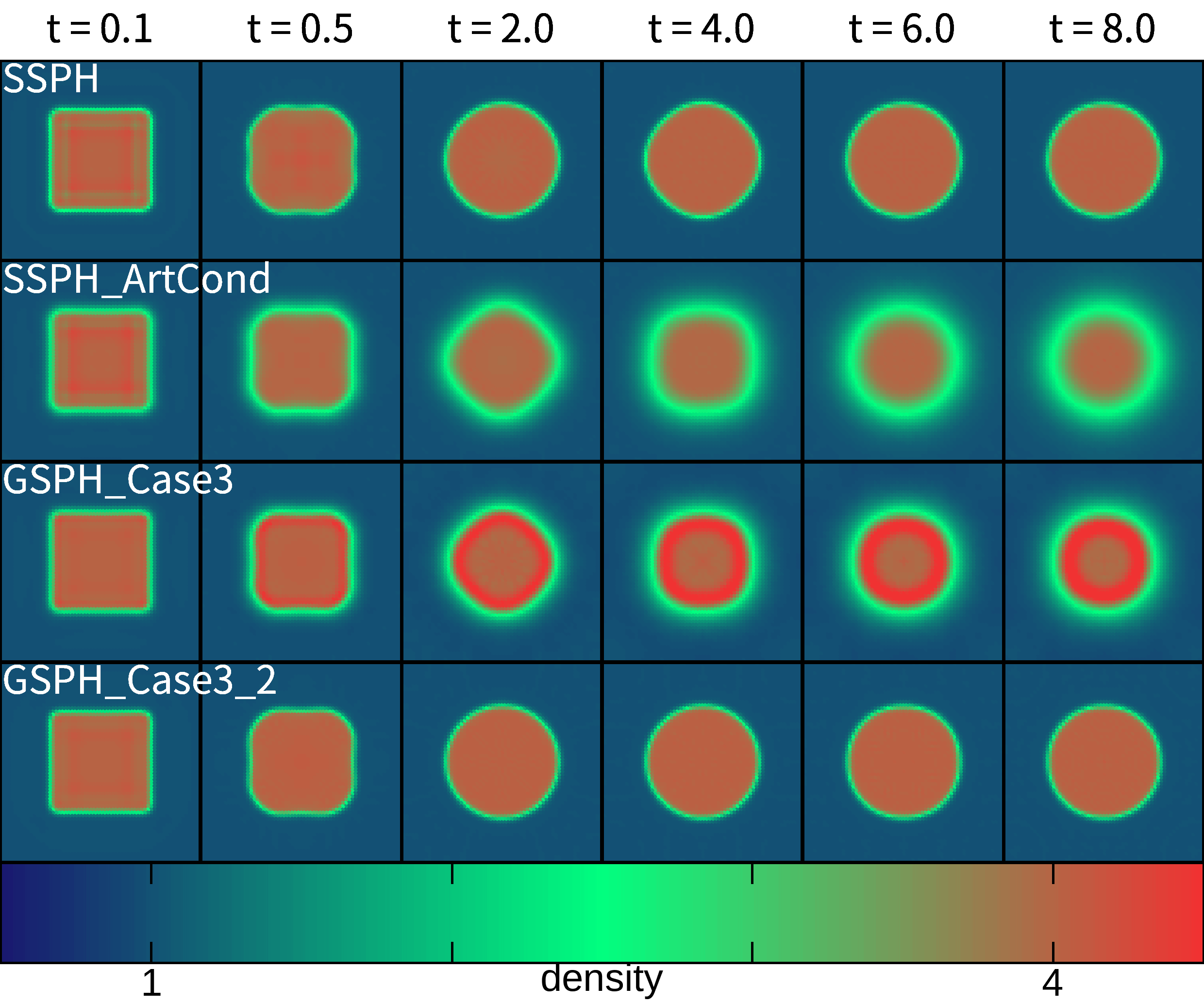}
    \caption{Density distributions of the pressure equilibrium tests 
    at $t = 0.1, 0.5, 2.0, 4.0, 6.0$ and $8.0$, respectively.
    The results of SSPH, SSPH with ArtCond, GSPH Case 3, and GSPH Case 3-2 are shown from the first row to the fourth row, respectively.
    }
    \label{fig:Problem_HE}
\end{figure*}
\begin{figure*}[!t]
\centering
	\includegraphics[width=\linewidth]{ 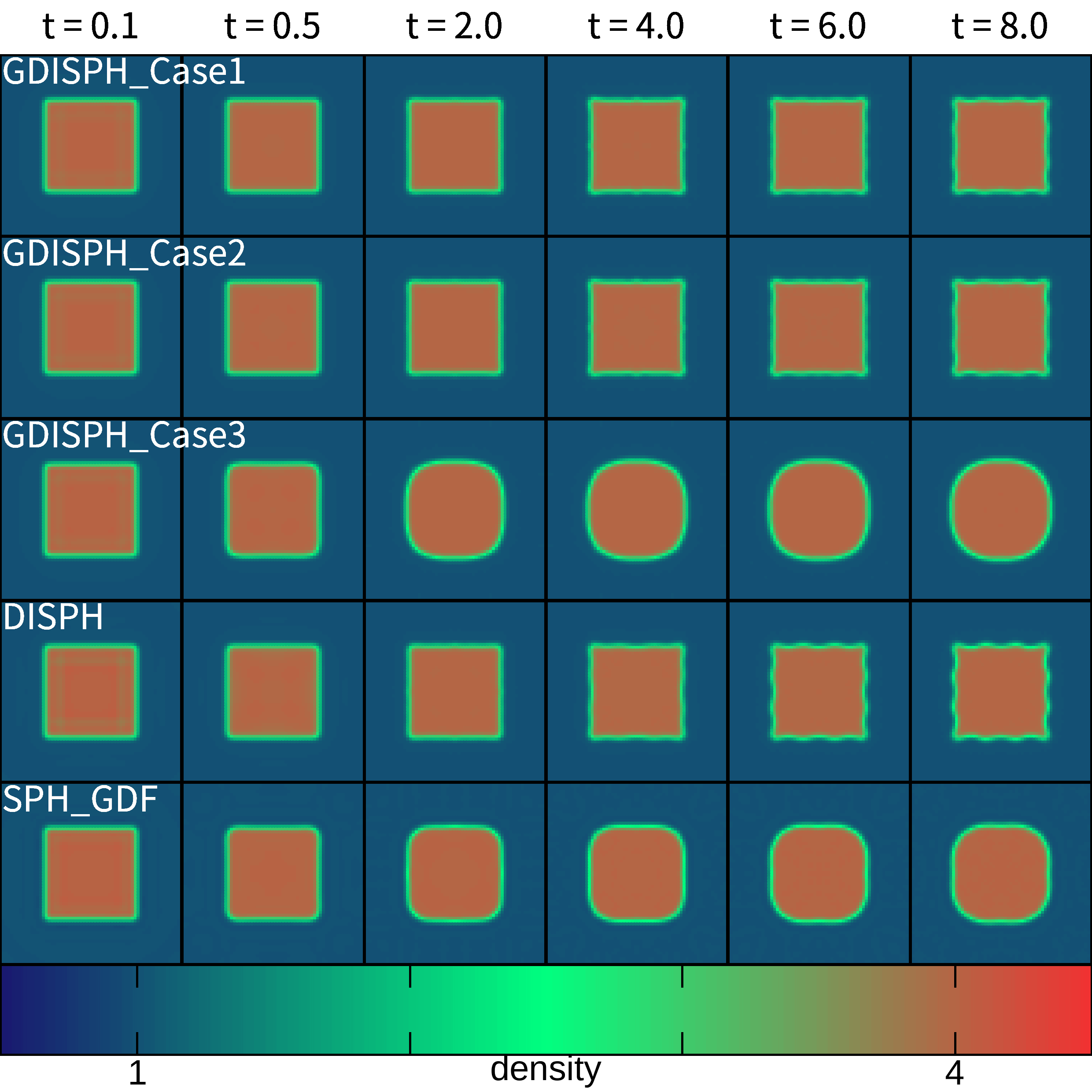}
    \caption{Same as Figure \ref{fig:Problem_HE}, but using GDISPH Case 1, GDISPH Case 2, GDISPH Case 3, DISPH, and SPH GDF.}
    \label{fig:Result_HE}
\end{figure*}

Figure \ref{fig:Problem_HE} presents the density distributions of the pressure equilibrium tests with SSPH, SSPH with ArtCond, GSPH Case 3, and GSPH Case 3-2 from the first row to the fourth row.
The snapshots at $t=0.1,0.5,2.0,4.0,6.0, $ and $8.0$ are shown from the first column to the sixth column.
All of the schemes turn the initial rectangle shape into almost a circle at $t=2t^s_{cross}$,
which suggests the existing influence of the effective surface tension.
In addition, SSPH with ArtCond blurs its border because of the artificial thermal conductivity adding the energy diffusion at the contact discontinuity.
At $t=8.0$, GSPH Case 3 has higher density values of about $4.2$ for the edges of the circle in the high-density region than the initial \textcolor{black}{high-density} region's value while blurring its contact discontinuity.
The results of GSPH Case 3-2 are quite similar to SSPH, rounding its shape but not blurring its boundaries like GSPH Case 3.

In Figure \ref{fig:Result_HE}, the density distributions of the pressure equilibrium tests with GDISPH Case 1, GDISPH Case 2, GDISPH Case 3, DISPH, and SPH GDF are shown.
It is clear that GDISPH Case 3 and SPH GDF round off the corner of their initial rectangle shape at $t=2.0$, but keep more initial square shape than a circle shape to $t=8t^s_{cross}$, which are better results than the schemes in Figure \ref{fig:Problem_HE}. On the other hand, DISPH and our schemes: GDISPH Case 1 and GDISPH Case 2 keep their rectangle shape up to $t=8t^s_{cross}$, which is much longer than the other schemes.
This suggests that the effective surface tension has little effect. However, careful observation shows that stationary waves develop along the discontinuity starting at $t=4t^{s}_{cross}$.

\subsection{Sedov-Taylor Test}
\label{sec:pol}
Even though our schemes: GDISPH Case 1 and GDISPH Case 2 are shown to be capable of handling the strong shock in one dimension in Section \ref{sec:strong}, we need to check if that is also true in three dimensions.
Here, the results of the Sedov-Taylor test is shown to see if our schemes can handle the strong shock in three dimensions.

We regularly place $128^3$ particles with a mass of $1/128^3$ in a three-dimensional simulation box with periodic boundary conditions of size $-1 \leq x,y,z < 1$. An initial density is $1.000$.
The thermal energy of unity is distributed within the radius of $0.05$ from the centre $(x, y, z) = (0.5, 0.5, 0.5)$ following the shape of the cubic spline kernel. As a result, $1098$ particles were injected with the thermal energy. 
Then, to set the thermal energy of the ambient matter, all particles are given $10^{-6}$ times the thermal energy of the central particle.
The initial velocity of \textcolor{black}{all} particles is set to $0.000$. We use $N_{ngb} = 228$ as the neighbour number. \textcolor{black}{$\gamma$ is set to $5/3$.}

\begin{figure*}[!t]
\centering
	\includegraphics[width=1\linewidth]{ 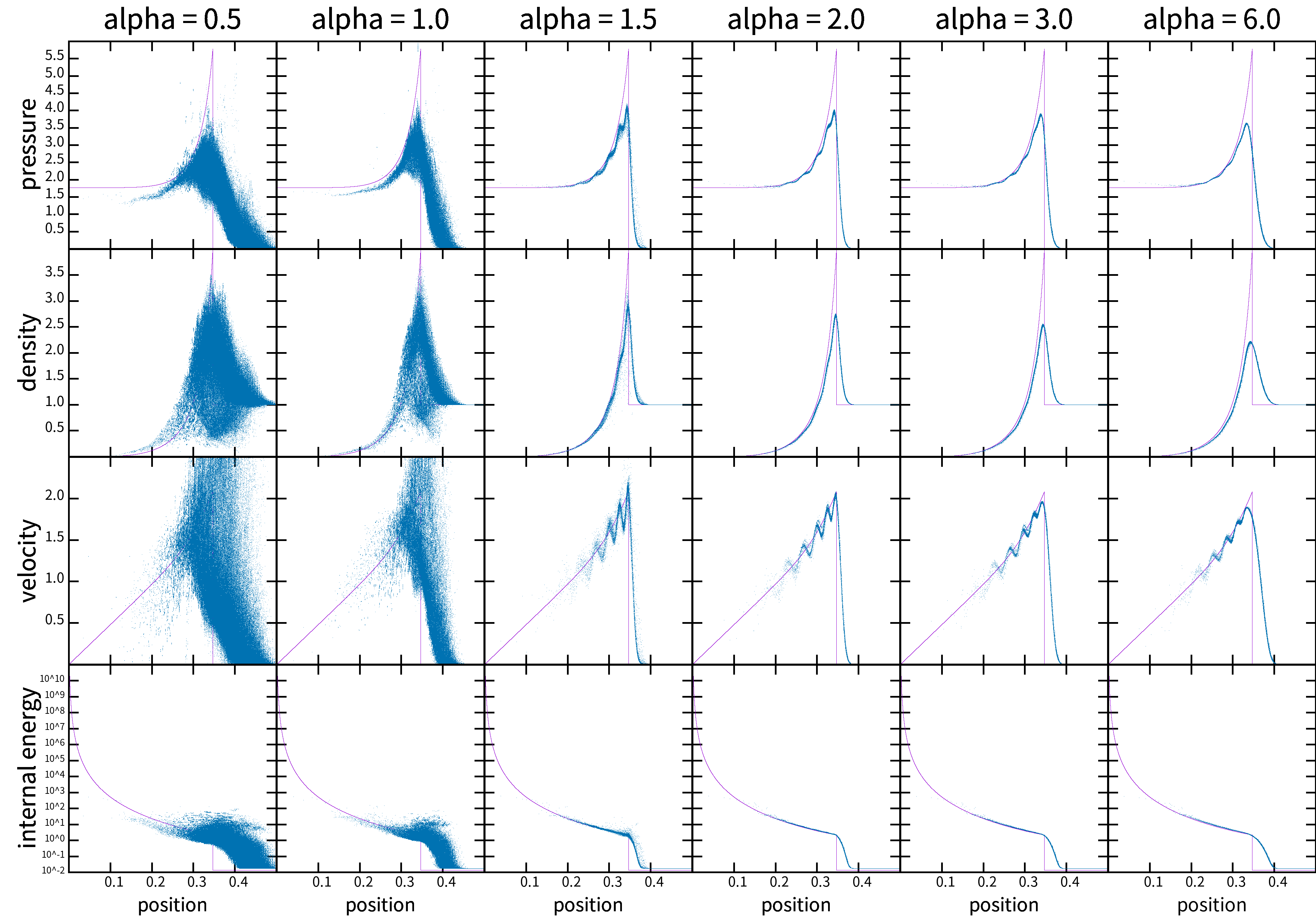}
    \caption{Profiles of the three-dimensional Sedov-Taylor tests with DISPH at $t = 0.05$.
    Pressure (first row), density (second row), velocity (third row), and internal energy (fourth row) are shown, respectively.
    $\alpha_{AV} = 0.5, 1.0, 1.5, 2.0, 3.0,$ and $6.0$ from the left column to the right column are shown, respectively.
    The $x$-axis represents the distance from $(x, y, z) = (0.5, 0.5, 0.5)$.
    The solid line indicates the analytical solution, while the dots indicate the physical quantities of each SPH particle at the position.
    }
    \label{fig:Problem_DISPH_Ple}
\end{figure*}

\begin{figure*}[!t]
\centering
	\includegraphics[width=0.80\linewidth]{ 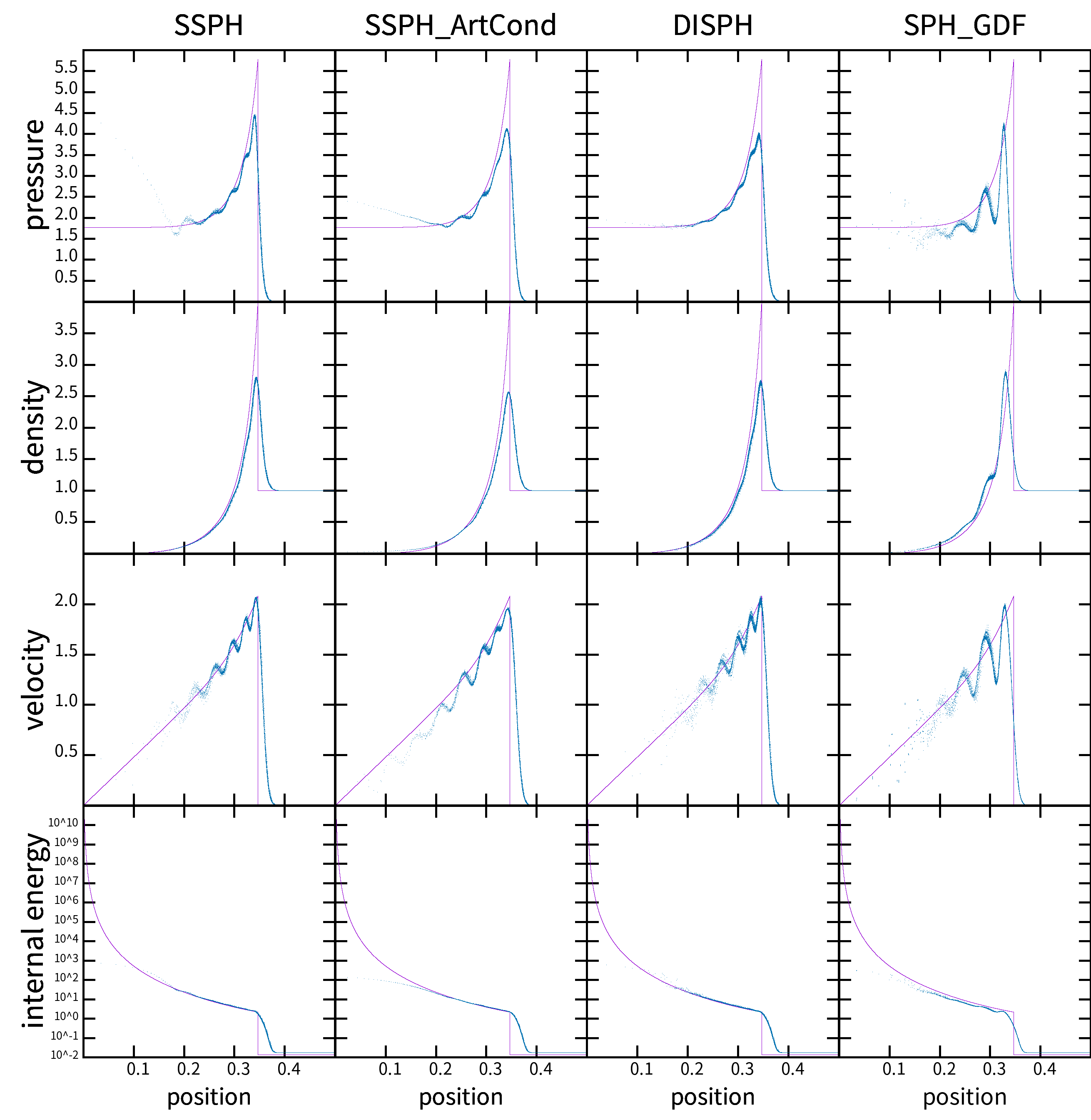}
    \caption{
        Profiles of the three-dimensional Sedov-Taylor tests with SSPH, SSPH with ArtCond, DISPH, and SPH GDF at $t = 0.05$.
        Pressure (first row), density (second row), velocity (third row), and internal energy (fourth row) are shown, respectively.
        $\alpha_{AV} = 2.0$ is used for all the schemes.
        The $x$-axis represents the distance from $(x, y, z) = (0.5, 0.5, 0.5)$.
        The solid line indicates the analytical solution, while the dots indicate the physical quantities of each SPH particle at the position.}
    \label{fig:Result_Ple0}
\end{figure*}

\begin{figure*}[!t]
\centering
	\includegraphics[width=0.95\linewidth]{ 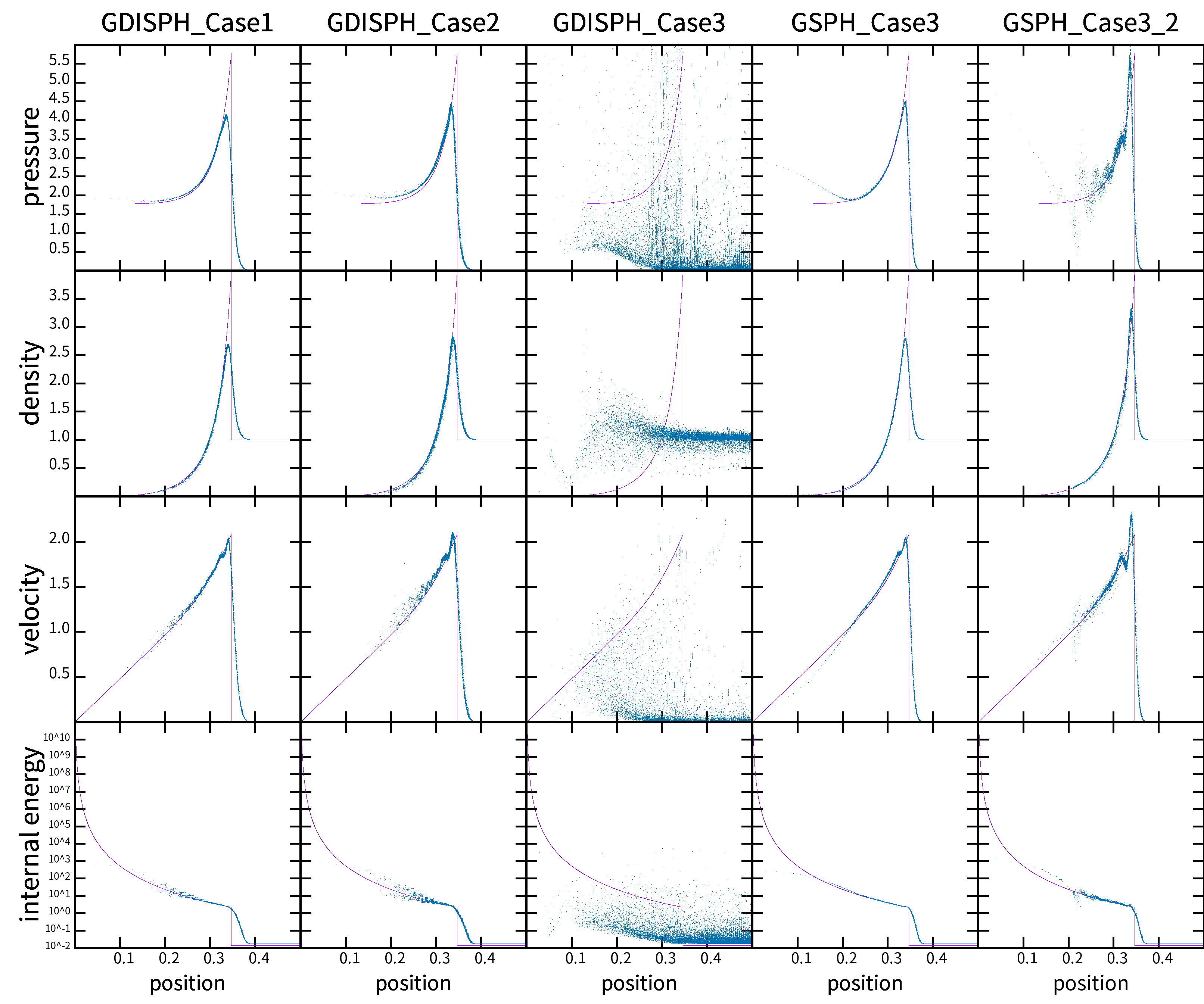}
    \caption{
    Same as Figure \ref{fig:Result_Ple0}, but with GDISPH Case 1, GDISPH Case 2, GDISPH Case 3, GSPH Case 3, and GSPH Case}
    \label{fig:Result_Ple1}
\end{figure*}

\begin{table}
	\centering
	\begin{tabular}{ccc} 
		\hline
		Scheme & density & pressure\\
		\hline
            \textcolor{black}{Analytical Solution} & \textcolor{black}{4.000} & \textcolor{black}{5.778}\\
            SSPH with $\alpha=1.5$ & 2.960 & 4.705\\
            SSPH with $\alpha=2$ & 2.811 & 4.458\\
            SSPH with $\alpha=3$ & 2.613 & 4.191\\
            SSPH with $\alpha=6$ & 2.256 & 3.783\\
            \textcolor{black}{SSPH with ArtCond with} $\textcolor{black}{\alpha=1.5}$ & \textcolor{black}{2.673} & \textcolor{black}{4.268}\\
            \textcolor{black}{SSPH with ArtCond with} $\textcolor{black}{\alpha=2}$ & \textcolor{black}{2.573} & \textcolor{black}{4.133}\\
            \textcolor{black}{SSPH with ArtCond with} $\textcolor{black}{\alpha=3}$ & \textcolor{black}{2.425} & \textcolor{black}{3.967}\\
            \textcolor{black}{SSPH with ArtCond with} $\textcolor{black}{\alpha=6}$ & \textcolor{black}{2.138} & \textcolor{black}{3.650}\\
            DISPH with $\alpha=1.5$ & 3.261 & 4.235\\
            DISPH with $\alpha=2$ & 2.759 & 4.034\\
            DISPH with $\alpha=3$ & 2.561 & 3.918\\
            DISPH with $\alpha=6$ & 2.228 & 3.645\\
            SPH GDF with $\alpha=1.5$ & 3.245 & 5.280\\
            SPH GDF with $\alpha=2$ & 2.924 & 4.273\\
            SPH GDF with $\alpha=3$ & 2.655 & 4.003\\
            SPH GDF with $\alpha=6$ & 2.267 & 3.663\\
            GSPH Case 3 & 2.811 & 4.519\\
            \textcolor{black}{GSPH Case 3-2} & \textcolor{black}{3.374} & \textcolor{black}{6.239}\\
            GDISPH Case1 & 2.705 & 4.161\\
            GDISPH Case2 & 2.8415 & 4.453\\
		\hline
	\end{tabular}
    \caption{The maximum value of the density and pressure around the tip of the shock wave for each scheme.}
    \label{table:pse_table}
\end{table}

\textcolor{black}{
In Figure \ref{fig:Problem_DISPH_Ple}, we present the profiles of physical quantities for the three-dimensional Sedov-Taylor tests with DISPH at $t = 0.05$. The results for pressure, density, velocity, and internal energy are shown along the panels from the top row to the bottom row, respectively, and for $\alpha_{AV} = 0.5, 1.0, 1.5, 2.0, 3.0,$ and $6.0$ are shown from the left column to the right column. 
The $x$-axis represents the distance from the centre.
The solid line indicates the analytical solution, while the dots indicate the physical quantities of each SPH particle at the position.
While DISPH with $\alpha_{AV} = 1.0$ is able to capture the strong shock in Section \ref{sec:strong}, it cannot capture the three-dimensional strong shock in this test, which suggests the need of fine-tuning depending on simulation problems.
Even though this test is quite severe for DISPH because of the large initial pressure gradient and strong shock wave,
the results with $\alpha_{AV} \geq 1.5$ are generally consistent with the analytical solution. However, the velocity oscillation both behind the shock and in the low-density region and the pressure oscillation in the \textcolor{black}{low-density} region occur at all $\alpha_{AV}$, and there is a slight error in the density profile behind the shock front.
Using bigger $\alpha_{AV}$ leads to blunting the whole solution, but the post-shock oscillations still exist behind the shock, which suggests the poor performance of the Monaghan's viscosity in this problem.
The results of other schemes with various $\alpha_{AV}$ can be seen in \ref{ap:sed_oth}, but the trend of those schemes is quite similar to that of DISPH. The maximum value of the density and pressure around the tip of the shock wave for each scheme is shown in Table \ref{table:pse_table}. While lower $\alpha_{AV}$ can be closer to the value of the analytical solution, the amplitude of the oscillation and noise becomes bigger. Therefore, in actual problem, we need to find a well-balanced parameter such that the maximum value around the tip of the shock wave is close to the analytical solution and yet there is little amplitude of the oscillation and noise.
}

\textcolor{black}{
In Figure \ref{fig:Result_Ple0}, we present the results of the Sedov-Taylor tests at $t=0.05$ with SSPH, SSPH with ArtCond, DISPH, and SPH GDF, from the first column to the fourth column.
\textcolor{black}{$\alpha_{AV}=2$ is chosen because in Figure \ref{fig:Problem_DISPH_Ple}, $\alpha_{AV}=2$ is the minimum value that successfully suppress the scattering behind the shock to some extent.}
All of the schemes in the figure have the velocity oscillation both behind the shock and in the low-density region, and the pressure oscillation in the low-density region. Especially, SPH GDF causes the large oscillation and the peak position around the shock front deviated from the analytical solution. SSPH, SSPH with ArtCond, and DISPH has a slight error in the density profile behind the shock front. SSPH and SSPH with ArtCond have a huge pressure error around the low-density region compared to DISPH. In SSPH with ArtCond, the velocity and internal energy around the centre are underestimated from the analytical solution.
}

\textcolor{black}{
Figure \ref{fig:Result_Ple1} shows the results with GDISPH Case 1, GDISPH Case 2, GDISPH Case 3, GSPH Case 3, and GSPH Case 3-2. The results of GDISPH Case 1 and GDISPH Case 2 are generally consistent with the analytical solution, without any arbitrary parameters for the shock, while GDISPH Case 3 cannot reproduce the analytical solution, causing some noisy behaviour. GSPH Case 3 effectively mitigates the oscillation of both velocity and pressure, both behind the shock front and in the low-density region, and a slight error in the density behind the shock does not occur, unlike SSPH or DISPH. However, GSPH Case 3 still has a considerable pressure difference in the \textcolor{black}{low-density} region and a slight error in the velocity in the \textcolor{black}{low-density} region.
While GSPH Case 3-2 has the closest maximum value of density and pressure around the tip of the shock front,
its velocity around the tip has huge error compared to the others and has huge noise in the low-density regions.
Our schemes: GDISPH Case 1 and GDISPH Case 2 also suppress the oscillation of pressure, both behind the shock and in the low-density regions, as well as the oscillation of velocity behind the shock.
Judging the strength of effective viscosity by the maximum value of the density around the tip of the shock wave, the strength of GDISPH and DISPH with $\alpha_{AV}=2$ is almost the same. Since GDISPH can suppress the oscillation but DISPH, we can conclude that the schemes with the Riemann solver can add effective viscosity in shock regions better than DISPH with Monaghan's viscosity.
We note that GDISPH Case 1 and GDISPH Case 2 still have a bit of poor performance in the pressure around the low-density region.
The poor performances are similar to DISPH \textcolor{black}{with $\alpha_{AV}$ above 1.5}, but are better than SSPH and GSPH. Also, both GDISPH Case 1 and GDISPH Case 2 have noise in the internal energy and the velocity in the low-density region. 
}

\textcolor{black}{
On the whole, GDISPH Case 1 and GDISPH Case 2 have better performance than SSPH, SSPH with ArtCond, SPH GDF, and DISPH behind the shock front like GSPH Case 3 in this test.
GDISPH Case 1 and GDISPH Case 2 still have poor performances to reproduce the analytical solution in the \textcolor{black}{low-density} region, but it is the same with the other schemes.
The reason all of the schemes have poor performances in the \textcolor{black}{low-density} region \textcolor{black}{is} because of the lack of resolution there.
Since SPH, in general, has a low-particle number density in the \textcolor{black}{low-density} regions, it is hard to solve problems around the regions precisely.
}

\subsection{Kelvin-Helmholtz Test}
\label{sec:kh}
\begin{figure*}[!t]
\centering
	\includegraphics[width=1\linewidth]{ 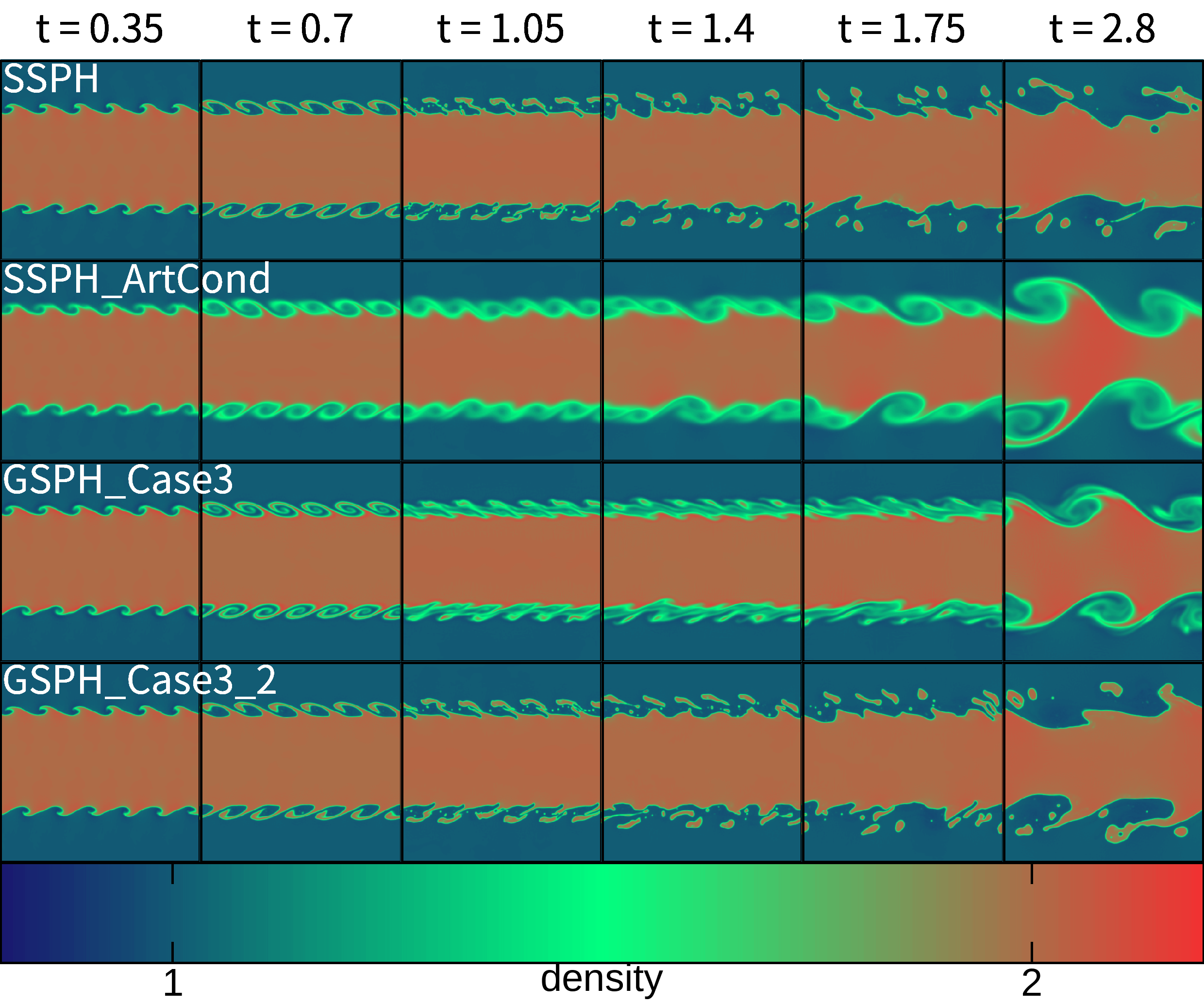}
    \caption{Density distributions of the two-dimensional Kelvin-Helmholtz tests with SSPH, SSPH with ArtCond, GSPH Case 3, and GSPH Case 3-2 from the first row to the forth row at $t = 0.35, 0.7, 
    1.05, 1.4, 1.75, $ and $2.8$ from the left column to the right column, respectively.
    }
    \label{fig:Result_kh11}
\end{figure*}
\begin{figure*}[!t]
\centering
	\includegraphics[width=1\linewidth]{ 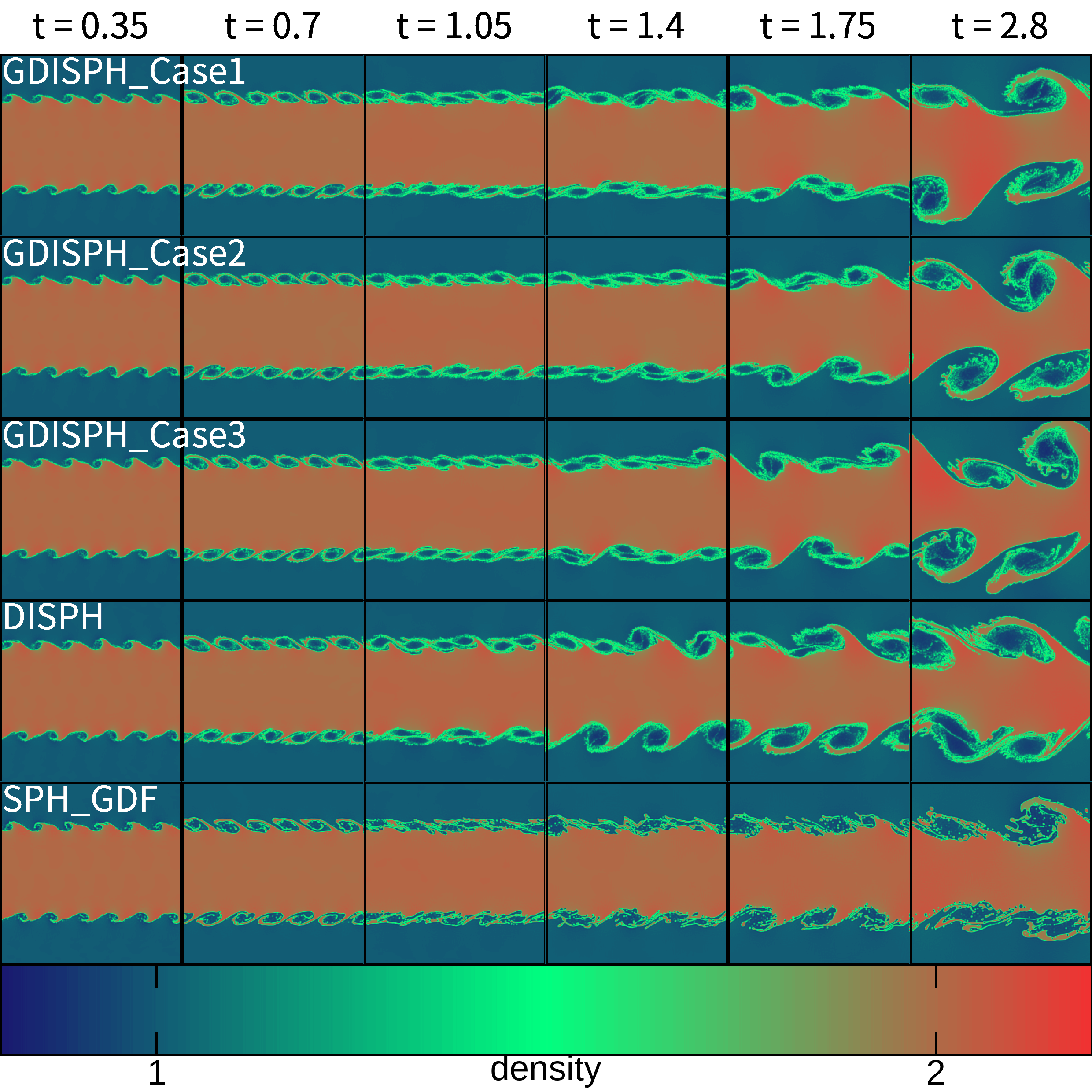}
    \caption{Same as Figure \ref{fig:Result_kh11}, but using GDISPH Case 1, GDISPH Case 2, GDISPH Case 3, DISPH, and SPH GDF.
    }
    \label{fig:Result_kh12}
\end{figure*}
\citet{Read2010} demonstrated that the long-term evolution of SPH in the Kelvin-Helmholtz instability eventually degrades, resulting in the 'gloopy' behaviour, pointing out that the behaviour is rather similar to the behaviour of fluids with explicit surface tensions. \textcolor{black}{Unlike the previous tests, shear flow regions, where the artificial viscosity, GSPH, and our GDISPH can misidentify as shock regions, emerge in this test.
Therefore, we incorporate the Balsara switch, which is a special treatment for the shear regions, into all schemes.
} We set the simulation box of $0\leq x,y < 1$ with periodic boundary conditions, use equal-mass particles, and place them regularly in a lattice manner. $\gamma$ is set to $5/3$.
The initial condition is given as follows:
\begin{equation}\label{eq:kh1}
    \rho = 
    \begin{cases}
        2.00 & \text{if $0.25 \leq y \leq 0.75$}, \\
        1.00 & \text{otherwise},
    \end{cases}
\end{equation}
\begin{equation}\label{eq:kh2}
    P = 2.50,
\end{equation}
and
\begin{equation}\label{eq:kh3}
    v_x = 
    \begin{cases}
        0.500  & \text{if $0.25 \leq y \leq 0.75$},\\
        -0.500 & \text{otherwise}.
    \end{cases}
\end{equation}
The velocity perturbation in the y-direction is as follows:
\begin{equation}\label{eq:kh4}
    v_y = w_0 sin(2\pi x / \lambda) \left \{ exp\left[ - \frac{(y-0.25)^2}{2\sigma^2}\right] + exp\left[ - \frac{(y-0.75)^2}{2\sigma^2}\right] \right\},
\end{equation}
where $w_0 = 0.025$, $\lambda = 1/6$, and $\sigma = 0.05/\sqrt{2}$. 
Therefore, the perturbations of the six wavelengths are added. 
In this test, the timescale of the growth of the Kelvin-Helmholtz instability is as follows:
\begin{equation}\label{eq:kh4-1}
    \tau_{kh} = \frac{\lambda (\rho_h + \rho_l)}{\sqrt{\rho_h \rho_l}|v_{x,h} - v_{x,l}|},
\end{equation}
where $\rho_h = 2$, $\rho_l = 1$, $v_{x,h} = 0.5$, and $v_{x,h} = -0.5$ in our test.
For our test setup, $\tau_{kh} = 0.35$.
The number of particles in the \textcolor{black}{high-density} region is $632\times316$ and that in the ambient is $447\times223$.
Note that in both regions, the particle spacing in the $x$-axis and $y$-axis direction are set to be the same, 
and the particle spacing in the \textcolor{black}{low-density} regions is $\sqrt{2}$ times bigger than that in the \textcolor{black}{high-density} region. \textcolor{black}{We use $N_{ngb} = 80$ as the neighbour number.}  \textcolor{black}{Without the Balsara swich, we confirmed that the KH instability did not grow significantly and remained almost unchanged from the initial conditions. This suggests the necessity of the switch in shear dominant-flows. This is consistent with the results given by \citet{Price2008} and \citet{Saitoh_Makino2013}.}

\textcolor{black}{
Figure \ref{fig:Result_kh11} shows the results of the Kelvin-Helmholtz test with SSPH, SSPH with ArtCond, GSPH Case 3, and GSPH Case 3-2 from the first row to the fourth row.
The density distributions are shown at $t = 0.35, 0.7, 1.05, 1.4, 1.75$, and $2.8$ from the first column to the sixth column, while Figure \ref{fig:Result_kh12} shows the results with GDISPH Case 1, GDISPH Case 2, GDISPH Case 3, DISPH, and SPH GDF.
Every scheme is able to grow the perturbation until $t=0.35$ and has the six-wavelength structures corresponding to the perturbations added to the initial conditions.
From $t=1.05$, SSPH and GSPH Case 3-2 have similar results with each other, breaking up the six-wavelength structures and creating some 'gloopy' structures. SSPH with ArtCond and GSPH Case 3 also have similar results with each other, succeeding in developing the perturbation but having extreme bluntness along the border between the low-density region and the high-density region. However, SSPH with ArtCond has a bigger development than GSPH Case 3.
All DISPH-type schemes show two-wavelength structures in the end, which is comparable to the DISPH results of \citet{Saitoh_Makino2013}, but the results are noisy and obviously different from each other. We discuss the noise and the difference in Section \ref{sec:conc}. The 'gloopy' structures do not emerge in DISPH and our GDISPH, suggesting that all DISPH-type schemes have little effect of effective surface tension in this test.
SPH GDF's result is intermediate between SSPH and DISPH-type schemes, having the 'gloopy' and 'noisy' structures, and it obviously fails to develop the structure around the bottom of the region.
While GDISPH Case 3 and SPH GDF have similar results in Figure \ref{fig:Result_HE}, which suggest having almost the same strength of effective surface tension, GDISPH Case 3 has better results in terms of the 'gloopy' structures. This is because, in shear flow regions, the effect of DISPH mainly emerges in GDISPH Case 3 when using our incorporation of the Balsara switch.
We note that while our incorporation of the Balsara switch into GDISPH (GSPH) is somewhat ad-hoc, it can work without any unfamiliar problems.
}

\section{Summary and Discussion}
\label{sec:conc}
In this paper, we have introduced the new hydrodynamic scheme named Godunov DISPH, which integrates the Riemann Solver into DISPH. 
Several realisations of the hydrodynamic equations under the SPH approximation have been proposed, and the performance of the resulting GDISPH Case 1, GDISPH Case 2, and GDISPH Case 3 have been evaluated.
As a result, our tests have confirmed that GDISPH Case 1 and GDISPH Case 2 can accurately handle the contact discontinuities as well as DISPH without any additional dissipation terms.
In addition, thanks to the use of the Riemann Solver, GDISPH Case 1 and GDISPH Case 2 could capture even the strong shocks without any manually tuned parameters like GSPH Case 3.
We also have devised a way to implement the Balsara switch into GDISPH.
The basic concept of this method can be applied to introducing the Balsara switch into SPHs such as Godunov SPH, where the artificial viscosity term is not separable.
While the method of deriving GDISPH Case 2 and GDISPH Case 3 is similar to that of \citet{Inutsuka2002}'s GSPH, 
GDISPH Case 1 is derived using an original method. As a result, GDISPH Case 1 succeeded in retaining the coefficient $g^{\textcolor{black}{\text{grad}}}$ that appears by considering the spatial derivative of smoothing length, allowing it to deal with temporal and spatial changes in smoothing length without any contradiction.
\textcolor{black}{GDISPH Case 1, out of all cases of GDISPH, is preferred because of its performance through the tests, reasonable derivation compared to the other cases, and success of retaining the coefficient $g^{\textcolor{black}{\text{grad}}}$.}

\textcolor{black}{
We have compared the performance of our schemes: Godunov DISPH and the schemes that were pointed out as the methods that can handle contact discontinuities better than SSPH: SSPH with ArtCond, SPH GDF, GSPH Case 3, and DISPH.
Throughout the Riemann problem test, DISPH, GDISPH Case 1, GDISPH Case 2, SSPH with ArtCond, and GSPH Case 3 can 
suppress the surface tension better than SSPH, reproducing the analytical solution.
In the pressure equilibrium test, DISPH, GDISPH Case 1, and GDISPH Case 2 can reproduce the contact discontinuities better than all of the other schemes, sustaining the pressure equilibrium longer than \textcolor{black}{any other scheme}. SSPH with ArtCond, GSPH Case 3, and SPH GDF fail to sustain the equilibrium.
In the KH test, all DISPH-type schemes have the prominent development of the instability, while SSPH with ArtCond and GSPH Case 3 have successful development but extreme blurriness.
Therefore, we concluded that DISPH, GDISPH Case 1, and GDISPH Case 2 have the best performance throughout all the test in terms of handling contact discontinuities.
}

Compared to SSPH, SSPH with ArtCond cause some characteristic behaviour: more SPH particles sampling the intermediate of the internal energy, \textcolor{black}{resulting in the better performance at the contact discontinuity in Section \ref{sec:sod} and \ref{sec:strong}}, more overestimation error of the internal energy (see Section \ref{sec:vac}), blurriness of its contact discontinuity (see Section \ref{sec:hyd}), \textcolor{black}{and \textcolor{black}{better development of the instability} but diffusive behaviour in the KH tests (see Section \ref{sec:kh})}. The reason for this difference is the use of the artificial thermal conductivity. Our tests confirmed that the same relation between SSPH and SSPH with ArtCond holds for the relation between GSPH Case 3-2 and GSPH Case 3, in which the former only uses the pressure solution of the Riemann problem and the latter uses the pressure and velocity solutions of that. 
In Figure \ref{fig:Result_ShockTube_velpanel}, even though GSPH Case 3-3 relatively suppresses the oscillation behind the shock compared to the other three schemes, it completely fails to capture the shock.
In addition, our GDISPH, in which only the pressure solution is used, does not cause the problem of the blurriness. 
According to these results, the use of the pressure solution may mainly give the effective viscosity, while that of the velocity solution may mainly give the effective thermal conductivity.

In the pressure equilibrium tests, DISPH, GDISPH Case 1, and GDISPH Case 2 give rise to the waves along the contact discontinuity in the later part of the calculation, which is the consistent result with \cite{Saitoh_Makino2013}.
While this can be seen as the downside of DISPH itself, we rather think this is coming from the zeroth-order error that SPH has in general \citep{Read2010}.
Since the difference between the high-density and low-density regions is expressed by the difference in particle number density, the physical values for particles near the boundary are slightly different for each particle. Whilst DISPH-type schemes, which can suppress the unphysical surface tensions, have superior treatment of the contact discontinuities, they may be affected by perturbative effects due to the particle distributions because of the absence of the unphysical surface tension that could possibly hold down the perturbation. The other schemes have the effect of the unphysical surface tension or the thermal conduction, which would be the reason for the absence of the wave. The numerical noise that DISPH-type schemes have in the KH tests may also be attributed to the perturbative effects due to the particle distributions.  The shorter the wavelength of the perturbation, the faster the growth of that by the KH instability. Since the wavelength is easily considered to be about the interparticle distance, the perturbative effects can be grown much faster than the seeded six-wavelength perturbation, causing the noise already at $t=\tau_{kh}$.
The reason for the difference in the final results between all DISPH-type schemes can be understood as follows: each scheme has a different momentum/energy equation, which makes different values of the zeroth-order error \textcolor{black}{from each other}, leading to different perturbative effects due to the particle distributions. 
The different perturbative effects cause different numerical noises, influencing the seeded six-wavelength perturbation differently, yielding distinct results. \textcolor{black}{SSPH and SSPH with ArtCond (GSPH Case 3) suppresses the numerical noise probably because of the unphysical surface tension and additional diffusion.
The more detailed discussion about the influence of the particle disorder and the artificial thermal conductivity to KH can be seen in \citet{Gilabert2022}.}

In the Sedov-Taylor tests, judging the bluntness of the overall solution by the maximum value of the density around the tip of the shock wave, \textcolor{black}{the solutions of DISPH} become blunted steadily from $\alpha_{AV}=2$ to $3$,$6$, but the amplitude of the oscillations behind the shock seems not to decrease accordingly. Even though we did not use parameters $\alpha_{AV}$ larger than $6.0$,
this fact suggests that further increasing the parameter does not completely suppress the oscillation behind the shock while overly blunting the overall solution, especially in the density and pressure. 
Suppose we define the "optimal parameters" as reproducing the solution exactly and eliminating all the unphysical oscillations. In that case, Monaghan's artificial viscosity is considered to have no such optimal parameters in our tests. This may indicate the performance limitation of Monaghan's artificial viscosity itself for the strong shock waves or pressure gradients.
On the other hand, the schemes with the Riemann solver suppress the oscillation at least behind the shock and reproduce the analytic solution well. Therefore we see the schemes with the Riemann solver add suitable, if not optimal, effective viscosity.
In Table \ref{table:pse_table}, comparing GSPH Case 3 (GDISPH Case 1 and GDISPH Case 2) and SSPH (DISPH) of the maximum density and pressure value around the shock wave, the strength of the effective viscosity of GSPH Case 3 (GDISPH Case 1 and GDISPH Case 2) is considered to be somewhere between $\alpha_{AV} = 1.5$ and $\alpha_{AV} = 3$ in a context of the Monaghan's artificial viscosity, and GDISPH Case 2 has less effective viscosity than GDISPH Case 1 in the test.

Although we have demonstrated the properties of our scheme in this paper, there is still much potential for improvement in GDISPH. For example, GDISPH can be improved to have higher-order spatial accuracy using interpolation functions such as the MUSCL or ENO-like methods, or approximate Riemann solvers or HLL-like methods can be applied to establish faster and more efficient methods. In addition, the practical use of GDISPH in realistic situations (e.g. formations of galaxy, stars, and so on) needs to be studied and compared with other successful schemes.
Our schemes still have the problem: the overestimation error and the numerical noise in the velocity and internal energy in low-density regions (see \ref{sec:vac} and \ref{sec:pol}), which one might not have to care much since SPH, in general, has less accuracy in the low-density region, and the results there are not trustworthy, so tackling the problem \textcolor{black}{should not be worthwhile.} \textcolor{black}{However, some methods like MFM (e.g. $gizmo$ \citep{Hopkins2015}) and moving mesh (e.g. $Arepo$ \citep{Springel2010A}), which are free from the spatial zeroth-order error, could handle the problem better than SPH.} Alternatively, we hope a completely new idea will break through such difficulties \textcolor{black}{within a framework of SPH}. 

\section*{Declaration of competing interest}
As the authors, we declare that there are no conflict of interests.

\section*{Data Availability}
The data underlying this article will be shared on reasonable request to the corresponding author.

\section*{Acknowledgements}
We thank \textcolor{black}{an anonymous referee for helpful comments and} Koki Otaki for useful discussions and insightful comments.
Numerical computations were performed with computational resources provided by the Multidisciplinary Cooperative Research Program in the Center for Computational Sciences, the University of Tsukuba. Masao Mori was supported by JSPS KAKENHI Grant Numbers JP20K04022.
\appendix

\section{One-dimensional Riemann Problem}
\label{ap:riemann}
\begin{figure}
    \begin{center}
        \includegraphics[width=\linewidth]{ 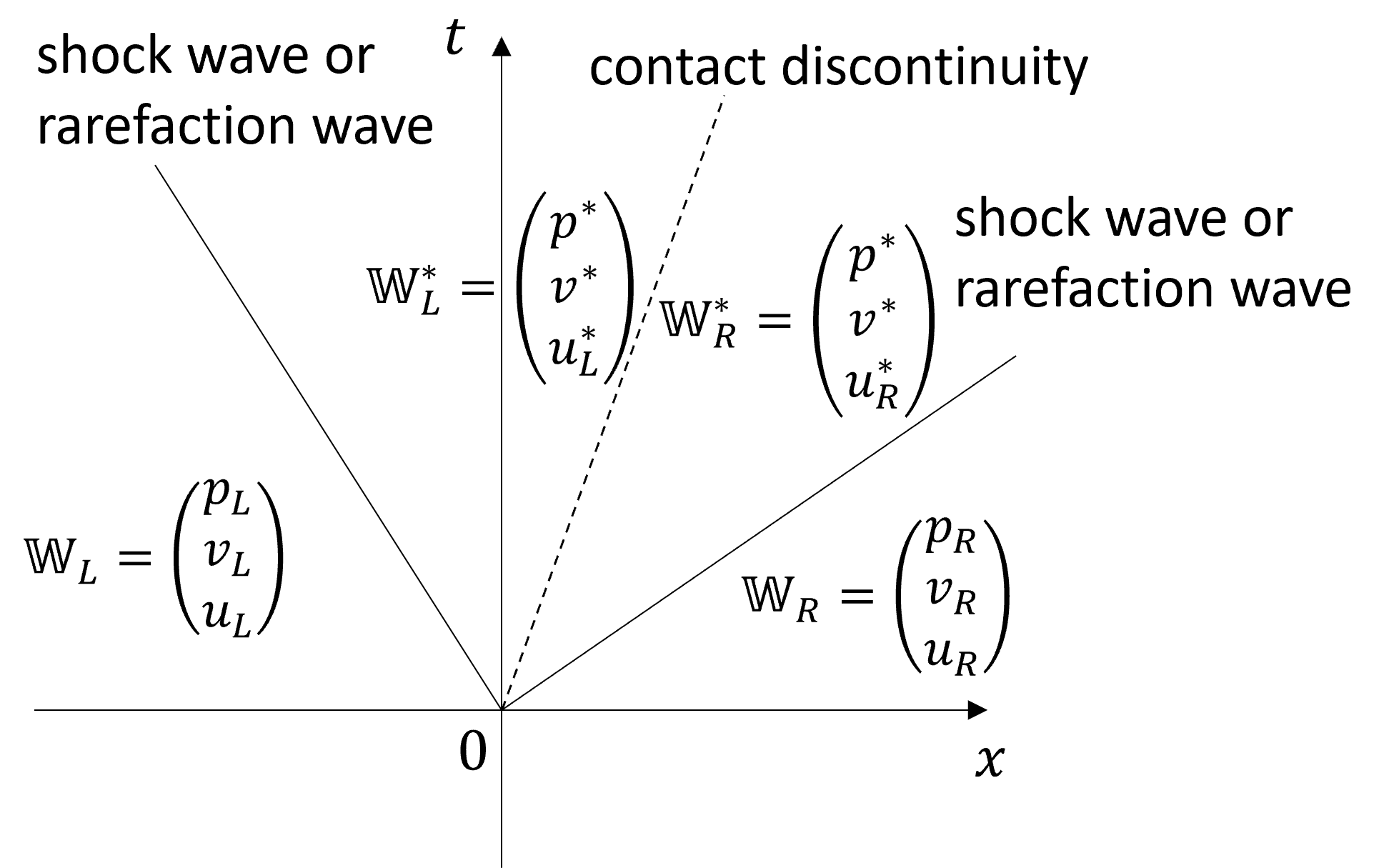}
        \caption{Solution types for the Riemann problem. The middle wave is always a contact discontinuity, sandwiched between the waves that can be either a shock wave or a rarefaction wave.
        }
    \label{fig:riemann}
    \end{center}
\end{figure}
Initial condition:
\begin{equation}\label{eq:initial}
    \textcolor{black}{\bm{\mathbb{W}}} =\begin{cases}
                \textcolor{black}{\bm{\mathbb{W}}_L} & \text{if $x < 0$,} \\
                \textcolor{black}{\bm{\mathbb{W}}_R} & \text{if $x \geq 0$,}
            \end{cases}
\end{equation}
where
\begin{equation}
    \textcolor{black}{\bm{\mathbb{W}}_L} = \begin{pmatrix}
                    p_L \\ 
                    v_L \\ 
                    u_L 
                \end{pmatrix},
    \textcolor{black}{\bm{\mathbb{W}}_R} = \begin{pmatrix}
                    p_R \\ 
                    v_R \\ 
                    u_R 
                \end{pmatrix},
\end{equation}
and $p$, $v$, and $u$ are the pressure, velocity, and specific internal energy, respectively.

The one-dimensional Riemann problem is the problem of solving the time development of one-dimensional fluids with the initial condition (\ref{eq:initial}).
The problem can be solved strictly.
The solution types for the problem at any given time are shown in Figure \ref{fig:riemann}.
The star region, which is surrounded by shock waves or rarefaction waves, always occurs.
In the star region, the velocity and the pressure are always constant.
The fluid on the left side from the contact discontinuity is the same fluid as the fluid that was at $t = 0$ and $x<0$,
while the fluid on the right side from that is the same fluid as the fluid that was at $t = 0$ and $x>0$.
Therefore, we can see the contact discontinuities as physical boundaries between the two fluids.
The detailed explanation for the solution of the one-dimensional Riemann problem can be found in \citet{VanLeer1979}.

\onecolumn
\allowdisplaybreaks
\section{Derivation of the momentum equation}
\label{ap:eom}
Here, we show the derivation of the momentum equation (\ref{eq:in7}).
We ignore the spatial derivative of the smoothing length.
\begin{equation}\label{eq:ap1}
    \begin{split}
        \left<\frac{d\bm{v}}{dt}\right> (\bm{r}_i) &= \int \frac{d\bm{v}(\bm{r})}{dt}W(|\bm{r}-\bm{r}_i|,h(\bm{r}))dV, \\
        &= -\int \frac{1}{\rho(\bm{r})} \nabla P(\bm{r}) W(|\bm{r}-\bm{r}_i|,h(\bm{r}))dV.
    \end{split}
\end{equation}
Considering $\frac{P(\bm{r})}{q(\bm{r})}$ is constant in the ideal gas, the RHS of equation (\ref{eq:ap1}) can be calculated as follows:
\begin{align}\label{eq:apin5}
    \notag&- \int \frac{1}{\rho(\bm{r})}\nabla P(\bm{r})W(|\bm{r}-\bm{r}_i|,h(\bm{r}))dV \\
    \notag&= - \int \frac{u(\bm{r})}{q(\bm{r})}\nabla P(\bm{r})W(|\bm{r}-\bm{r}_i|,h(\bm{r}))dV \\
    \notag&= -\int \left(\nabla \left(\frac{u(\bm{r})}{q(\bm{r})}P(\bm{r})\right) - \nabla \left(\frac{u(\bm{r})}{q(\bm{r})}\right)P(\bm{r})\right)W(|\bm{r}-\bm{r}_i|,h(\bm{r}))dV \\
    \notag&= -\int \nabla \left(\frac{u(\bm{r})}{q(\bm{r})}P(\bm{r})\right) W(|\bm{r} - \bm{r}_i|,h(\bm{r}))dV + \int \nabla \left(\frac{u(\bm{r})}{q(\bm{r})}\right) P(\bm{r})W(|\bm{r} - \bm{r}_i|,h(\bm{r}))dV \\
    \notag&= -\int \nabla \left(\frac{u(\bm{r})}{q(\bm{r})}P(\bm{r})\right) W(|\bm{r} - \bm{r}_i|,h(\bm{r}))dV  +\int \left(-\frac{u(\bm{r})}{q^2(\bm{r})} \frac{\partial q(\bm{r})}{\partial \bm{r}} + \frac{1}{q(\bm{r})} \frac{\partial u(\bm{r})}{\partial \bm{r}}\right) P(\bm{r})W(|\bm{r} - \bm{r}_i|,h(\bm{r}))dV \\
    \notag&= \int \left(\frac{u(\bm{r})}{q(\bm{r})}P(\bm{r})\right) \nabla W(|\bm{r_i}-\bm{r}|,h(\bm{r}))dV  +\int -\frac{P(\bm{r})u(\bm{r})}{q^2(\bm{r})} \frac{\partial q(\bm{r})}{\partial \bm{r}}W(|\bm{r} - \bm{r_i}|,h(\bm{r}))dV -\int  \left(\frac{u(\bm{r})}{q(\bm{r})}P(\bm{r})\right) \nabla W(|\bm{r} - \bm{r}_i|,h(\bm{r}))dV  \\
    \notag&= \int -\frac{P(\bm{r})u(\bm{r})}{q^2(\bm{r})} \frac{\partial q(\bm{r})}{\partial \bm{r}}W(|\bm{r} - \bm{r}_i|,h(\bm{r}))dV  \\
    &= \displaystyle \sum_{j=1}^N m_ju_j \int \frac{P(\bm{r})u(\bm{r})}{q^2(\bm{r})} \nabla_j W(|\bm{r}-\bm{r}_j|,h(\bm{r})) W(|\bm{r} - \bm{r}_i|,h(\bm{r}))dV.
\end{align}
As a result, the momentum equation is obtained by
\begin{equation}
    \begin{split}
        m_i \frac{d\bm{v}_i}{dt} &=  m_i \left<\frac{d\bm{v}}{dt}\right> (\bm{r}_i),\\
        &= m_i\displaystyle \sum_{j=1}^N m_ju_j \int \frac{P(\bm{r})u(\bm{r})}{q^2(\bm{r})} \nabla_j W(|\bm{r}-\bm{r}_j|,h(\bm{r})) W(|\bm{r} - \bm{r}_i|,h(\bm{r}))dV.
    \end{split}
\end{equation}
\section{Euler-Lagrange equation}
\label{ap:el}
Here, we show the derivation of equation (\ref{eq:in17}).
We again ignore the spatial dependence of the smoothing length.
\begin{equation}\label{eq:apin13}
    \begin{split}
        \frac{\partial L}{\partial \bm{r}_i} = &- \displaystyle \sum_k m_k \int \frac{\partial u}{\partial \bm{r}_i} W(|\bm{r}-\bm{r}_k|,h(\bm{r}))dV  - m_i\int u\frac{\partial}{\partial \bm{r}_i}W(|\bm{r}-\bm{r}_i|,h(\bm{r}))dV.
    \end{split}
\end{equation}
Since $\frac{P(\bm{r})}{q(\bm{r})}$ is constant in the ideal gas, the first term of the RHS of equation (\ref{eq:apin13}) becomes the following:
\begin{align}\label{eq:apin14}
    \notag&- \displaystyle \sum^N_{\textcolor{black}{k=1}} m_k \int \frac{\partial u}{\partial \bm{r}_i} W(|\bm{r}-\bm{r}_k|,h(\bm{r}))dV \\
    \notag&= - \displaystyle \sum^N_{\textcolor{black}{k=1}} m_k \int -P(\bm{r}) \frac{\partial}{\partial \bm{r}_i} \left ( \frac{u(\bm{r})}{q(\bm{r})} \right ) W(|\bm{r}-\bm{r}_k|,h(\bm{r}))dV\\
    \notag&= - \displaystyle \sum^N_{\textcolor{black}{k=1}} m_k \int -P(\bm{r}) \left(-\frac{u(\bm{r})}{q^2(\bm{r})} \frac{\partial q(\bm{r})}{\partial \bm{r}_i} \right. \left.+ \frac{1}{q(\bm{r})} \frac{\partial u(\bm{r})}{\partial \bm{r}_i}\right) W(|\bm{r}-\bm{r}_k|,h(\bm{r}))dV\\
    \notag&= - \displaystyle \sum^N_{\textcolor{black}{k=1}} m_k \int -P(\bm{r}) \left(-\frac{u(\bm{r})}{q^2(\bm{r})}  m_i u_i \nabla_i W(|\bm{r}-\bm{r}_i|,h(\bm{r}))\right. \left. + \frac{1}{q(\bm{r})} \frac{\partial u(\bm{r})}{\partial \bm{r}_i}\right) W(|\bm{r}-\bm{r}_k|,h(\bm{r}))dV\\
    \notag&= - \displaystyle \sum^N_{\textcolor{black}{k=1}} m_k \int  \left(\frac{P(\bm{r})u(\bm{r})}{q^2(\bm{r})} m_i u_i \nabla_i W(|\bm{r}-\bm{r}_i|,h(\bm{r})) W(|\bm{r}-\bm{r}_k|,h(\bm{r}))dV \right. - \left.\frac{P(\bm{r})}{q(\bm{r})} \frac{\partial u(\bm{r})}{\partial \bm{r}_i}W(|\bm{r}-\bm{r}_k|,h(\bm{r}))dV\right) \\
    \notag&= - \int \displaystyle \sum^N_{\textcolor{black}{k=1}} m_k \frac{P(\bm{r})u(\bm{r})}{q^2(\bm{r})}   m_i u_i \nabla_i W(|\bm{r}-\bm{r}_i|,h(\bm{r})) W(|\bm{r}-\bm{r}_k|,h(\bm{r}))dV +  \int \displaystyle \sum^N_{\textcolor{black}{k=1}} m_k \frac{P(\bm{r})}{q(\bm{r})} \frac{\partial u(\bm{r})}{\partial \bm{r}_i}W(|\bm{r}-\bm{r}_k|,h(\bm{r}))dV \\
    \notag&= - \int \displaystyle \sum^N_{\textcolor{black}{k=1}} m_k \frac{P(\bm{r})u(\bm{r})}{q^2(\bm{r})}   m_i u_i \nabla_i W(|\bm{r}-\bm{r}_i|,h(\bm{r})) W(|\bm{r}-\bm{r}_k|,h(\bm{r}))dV -  \int m_i \frac{P(\bm{r})}{q(\bm{r})} u(\bm{r}) \nabla_i W(|\bm{r}-\bm{r}_i|,h(\bm{r}))dV \\
    \notag&= - \int \displaystyle \sum^N_{\textcolor{black}{k=1}} m_k \frac{P(\bm{r})u(\bm{r})}{q^2(\bm{r})}   m_i u_i \nabla_i W(|\bm{r}-\bm{r}_i|,h(\bm{r})) W(|\bm{r}-\bm{r}_k|,h(\bm{r}))dV -  \int m_i \frac{P(\bm{r})}{q(\bm{r})} u(\bm{r}) \nabla_i W(|\bm{r}-\bm{r}_i|,h(\bm{r}))dV \\
    \notag&= - \int \displaystyle \sum^N_{\textcolor{black}{k=1}} m_k \frac{P(\bm{r})u(\bm{r})}{q^2(\bm{r})}   m_i u_i \nabla_i W(|\bm{r}-\bm{r}_i|,h(\bm{r})) W(|\bm{r}-\bm{r}_k|,h(\bm{r}))dV -  \int m_i \displaystyle \sum^N_{\textcolor{black}{k=1}} m_k u_k\frac{P(\bm{r})}{q^2(\bm{r})} u(\bm{r}) W(|\bm{r}-\bm{r}_k|,h(\bm{r})) \nabla_i W(|\bm{r}-\bm{r}_i|,h(\bm{r}))dV \\
    &= - \int \displaystyle \sum^N_{\textcolor{black}{k=1}} m_k \frac{P(\bm{r})u(\bm{r})}{q^2(\bm{r})}   m_i u_i \nabla_i W(|\bm{r}-\bm{r}_i|,h(\bm{r})) W(|\bm{r}-\bm{r}_k|,h(\bm{r}))dV -  \int m_i \displaystyle \sum^N_{\textcolor{black}{k=1}} m_k u_k\frac{P(\bm{r})}{q^2(\bm{r})} u(\bm{r}) W(|\bm{r}-\bm{r}_k|,h(\bm{r})) \nabla_i W(|\bm{r}-\bm{r}_i|,h(\bm{r}))dV,
\end{align}
and the second term of the RHS of equation (\ref{eq:apin13}) can be calculated as follows:
\begin{align}\label{eq:apin15}
    \notag &- m_i\int u\frac{\partial}{\partial \bm{r}_i}W(|\bm{r}-\bm{r}_i|,h)dV \\
    \notag &= m_i \int u \frac{\partial}{\partial \bm{r}} W(|\bm{r}-\bm{r}_i|,h)dV \\
    \notag &= - m_i \int \frac{\partial u(\bm{r})}{\partial \bm{r}} W(|\bm{r}-\bm{r}_i|,h)dV \\
    \notag &= - m_i \int -P(\bm{r})\frac{\partial}{\partial \bm{r}} \left( \frac{u(\bm{r})}{q(\bm{r})}\right)W(|\bm{r}-\bm{r}_i|,h)dV \\
    \notag &= - m_i \int  -P(\bm{r})\left(-\frac{u(\bm{r})}{q^2(\bm{r})} \frac{\partial q(\bm{r})}{\partial \bm{r}} + \frac{1}{q(\bm{r})} \frac{\partial u(\bm{r})}{\partial \bm{r}}\right)  W(|\bm{r}-\bm{r}_i|,h)dV\\
    \notag &= - m_i \int  -P(\bm{r})\left(-\frac{u(\bm{r})}{q^2(\bm{r})} \displaystyle \sum^N_{\textcolor{black}{j=1}} m_j u_j \nabla W(|\bm{r}-\bm{r}_j|,h)  + \frac{1}{q(\bm{r})} \frac{\partial u(\bm{r})}{\partial \bm{r}}\right) \cdot W(|\bm{r}-\bm{r}_i|,h)dV\\
    \notag &= - m_i \int  -P(\bm{r})\left(\frac{u(\bm{r})}{q^2(\bm{r})} \displaystyle \sum^N_{\textcolor{black}{j=1}} m_j u_j \nabla_j W(|\bm{r}-\bm{r}_j|,h)  + \frac{1}{q(\bm{r})} \frac{\partial u(\bm{r})}{\partial \bm{r}}\right)  \cdot W(|\bm{r}-\bm{r}_i|,h)dV\\
    \notag &=  \left( m_i \int P(\bm{r})\frac{u(\bm{r})}{q^2(\bm{r})} \displaystyle \sum^N_{\textcolor{black}{j=1}} m_j u_j \nabla_j W(|\bm{r}-\bm{r}_j|,h)W(|\bm{r}-\bm{r}_i|,h)dV \right. \left. + m_i \int \frac{P(\bm{r})}{q(\bm{r})} \frac{\partial u(\bm{r})}{\partial \bm{r}}W(|\bm{r}-\bm{r}_i|,h)dV\right)\\
    \notag &=  \left( m_i \int P(\bm{r})\frac{u(\bm{r})}{q^2(\bm{r})} \displaystyle \sum^N_{\textcolor{black}{j=1}} m_j u_j \nabla_j W(|\bm{r}-\bm{r}_j|,h)W(|\bm{r}-\bm{r}_i|,h)dV \right.+ \left. m_i \int \frac{P(\bm{r})}{q(\bm{r})} u(\bm{r}) \nabla_i W(|\bm{r}-\bm{r}_i|,h)dV\right)\\
    &=  \left( m_i \int P(\bm{r})\frac{u(\bm{r})}{q^2(\bm{r})} \displaystyle \sum^N_{\textcolor{black}{j=1}} m_j u_j \nabla_j W(|\bm{r}-\bm{r}_j|,h)W(|\bm{r}-\bm{r}_i|,h)dV  \right.\left. + m_i \int  \displaystyle \sum^N_{\textcolor{black}{j=1}} m_j u_j \frac{P(\bm{r})u(\bm{r})}{q^2(\bm{r})} W(|\bm{r}-\bm{r}_j|,h) \nabla_i W(|\bm{r}-\bm{r}_i|,h)dV\right),
\end{align}
Changing the subscript $k$ of equation (\ref{eq:apin14}) to the subscript $j$ and substituting equation (\ref{eq:apin14}) and equation (\ref{eq:apin15}) into equation (\ref{eq:apin13}) give
\begin{equation}\label{eq:apin16}
    \begin{split}
        &\frac{\partial L}{\partial \bm{r}_i} = - \displaystyle \sum^N_{\textcolor{black}{j=1}} m_i u_i m_j \int P \frac{u}{q^2} \nabla_i W(|\bm{r}-\bm{r}_i|,h) W(|\bm{r}-\bm{r}_j|,h)dV + \displaystyle \sum^N_{\textcolor{black}{j=1}} m_i m_j u_j \int P\frac{u}{q^2} W(|\bm{r}-\bm{r}_i|,h) \nabla_j W(|\bm{r}-\bm{r}_j|,h)dV.
    \end{split}
\end{equation}
As a result, the Euler-Lagrange equation gives the following equation:
\begin{equation}\label{eq:apin17}
    \begin{split}
        &m_i\frac{d\bm{v}_i}{dt} = - \displaystyle \sum^N_{\textcolor{black}{j=1}} m_i u_i m_j \int P \frac{u}{q^2} \nabla_i W(|\bm{r}-\bm{r}_i|,h) W(|\bm{r}-\bm{r}_j|,h)dV + \displaystyle \sum^N_{\textcolor{black}{j=1}} m_i m_j u_j \int P\frac{u}{q^2} W(|\bm{r}-\bm{r}_i|,h) \nabla_j W(|\bm{r}-\bm{r}_j|,h)dV
    \end{split}
\end{equation}
\section{Derivation of the energy equation}
\label{ap:ee}
Here, we show the derivation of equation (\ref{eq:in21}), ignoring the spatial dependence of the smoothing length.

\begin{equation}\label{eq:apin18}
    \begin{split}
        \left< \frac{d\bm{u}}{dt} \right> (\bm{r}_i) &= \int \frac{du(\bm{r})}{dt} W(|\bm{r} - \bm{r}_i|,h(\bm{r}))dV, \\
        &= - \int \frac{P(\bm{r})}{\rho(\bm{r})} \nabla \cdot \bm{v}(\bm{r}) W(|\bm{r} - \bm{r}_i|,h(\bm{r}))dV.
    \end{split}
\end{equation}
We assume that the following approximations holds:
\begin{equation}\label{eq:apin18-1}
    \begin{split}
        \int \frac{1}{\rho(\bm{r})} &\left[\bm{v}(\bm{r})\cdot \nabla P(\bm{r}) \right]W(|\bm{r}-\bm{r}_i|,h(\bm{r}))dV = \int \frac{1}{\rho(\bm{r})} \left[\bm{v}_i \cdot \nabla P(\bm{r}) \right]W(|\bm{r}-\bm{r}_i|,h(\bm{r}))dV + \mathcal{O}(h^2),
    \end{split}
\end{equation}
\begin{equation}\label{eq:apin18-2}
    \begin{split}        
        &\int \frac{P(\bm{r})u(\bm{r})}{q(\bm{r})} \nabla \left(\left[ \bm{v}(\bm{r}) - \bm{v}_i \right] W(|\bm{r}-\bm{r}_i|,h(\bm{r})) \right) dV  = \int \frac{P(\bm{r})u(\bm{r})}{q(\bm{r})} \left[ \bm{v}(\bm{r}) - \bm{v}_i \right] \nabla W(|\bm{r}-\bm{r}_i|,h(\bm{r})) dV + \mathcal{O}(h^2).\\
    \end{split}
\end{equation}
Note that the approximation of equation (\ref{eq:apin18-1}) is also used in \citet{Inutsuka2002}.
Using that $\frac{P(\bm{r})}{q(\bm{r})}$ is constant in the ideal gas,
the RHS of equation (\ref{eq:apin18}) can be calculated as follows:
\begin{align}\label{eq:apin19}
    \notag&\int \frac{du(\bm{r})}{dt}W(|\bm{r}-\bm{r}_i|,h(\bm{r}))dV \\
    \notag&=\int  - \frac{P(\bm{r})}{\rho(\bm{r})} \nabla \cdot \bm{v}(\bm{r})W(|\bm{r}-\bm{r}_i|,h(\bm{r}))dV\\
    \notag&= - \int \frac{1}{\rho(\bm{r})} \left[\nabla \cdot P(\bm{r})\bm{v(\bm{r})}\right]W(|\bm{r}-\bm{r}_i|,h(\bm{r}))dV + \int \frac{1}{\rho(\bm{r})} \left[\bm{v}(\bm{r})\cdot \nabla P(\bm{r}) \right]W(|\bm{r}-\bm{r}_i|,h(\bm{r}))dV\\
    \notag&\approx - \int \frac{1}{\rho(\bm{r})} \left[\nabla \cdot P(\bm{r})\bm{v(\bm{r})}\right]W(|\bm{r}-\bm{r}_i|,h(\bm{r}))dV + \int \frac{1}{\rho(\bm{r})} \left[\bm{v}_i\cdot \nabla P(\bm{r}) \right]W(|\bm{r}-\bm{r}_i|,h(\bm{r}))dV\\
    \notag&= \int P(\bm{r})\left[ \bm{v}(\bm{r}) - \bm{v}_i \right] \cdot  \nabla \left[ \frac{1}{\rho(\bm{r})}W(|\bm{r}-\bm{r}_i|,h(\bm{r})) \right] dV\\
    \notag&= \int P(\bm{r})\left[ \bm{v}(\bm{r}) - \bm{v}_i \right] \cdot  \nabla \left[ \frac{u(\bm{r})}{q(\bm{r})}W(|\bm{r}-\bm{r}_i|,h(\bm{r})) \right] dV\\
    \notag&= \int P(\bm{r})\left[ \bm{v}(\bm{r}) - \bm{v}_i \right] \cdot  \left[ \nabla \left(\frac{u(\bm{r})}{q(\bm{r})} \right) W(|\bm{r}-\bm{r}_i|,h(\bm{r})) \right.\left.+  \left(\frac{u(\bm{r})}{q(\bm{r})} \right) \nabla W(|\bm{r}-\bm{r}_i|,h(\bm{r}))\right] dV\\
    \notag&= \int P(\bm{r})\left[ \bm{v}(\bm{r}) - \bm{v}_i \right] \cdot  \left[ \left(-\frac{u(\bm{r})}{q^2(\bm{r})} \frac{\partial q(\bm{r})}{\partial \bm{r}} \right.\right.\left.\left. + \frac{1}{q(\bm{r})} \frac{\partial u(\bm{r})}{\partial \bm{r}}\right)  W(|\bm{r}-\bm{r}_i|,h(\bm{r})) \right.\left.+  \frac{u(\bm{r})}{q(\bm{r})} \nabla W(|\bm{r}-\bm{r}_i|,h(\bm{r}))\right] dV\\
    \notag&= \int \left[ \left(-\left[ \bm{v}(\bm{r}) - \bm{v}_i \right] \cdot \frac{P(\bm{r})u(\bm{r})}{q^2(\bm{r})} \displaystyle \sum^N_{\textcolor{black}{j=1}}m_j u_j \nabla W(|\bm{r}-\bm{r}_j|,h(\bm{r})) \right. \right.\left.\left. + \left[ \bm{v}(\bm{r}) - \bm{v}_i \right] \cdot \frac{P(\bm{r})}{q(\bm{r})} \frac{\partial u(\bm{r})}{\partial \bm{r}}\right)  W(|\bm{r}-\bm{r}_i|,h(\bm{r}))  dV\right .\\
    \notag& \left.\quad \quad \quad \quad \quad \quad  + \left[ \bm{v}(\bm{r}) - \bm{v}_i \right] \cdot \frac{P(\bm{r})u(\bm{r})}{q^2(\bm{r})} \cdot \right. \left.\displaystyle \sum^N_{\textcolor{black}{j=1}} m_j u_j W(|\bm{r}-\bm{r}_j|,h(\bm{r})) \nabla W(|\bm{r}-\bm{r}_i|,h(\bm{r}))dV\right] \\
    \notag&= \int \left[ \left(-\left[ \bm{v}(\bm{r}) - \bm{v}_i \right] \cdot \frac{P(\bm{r})u(\bm{r})}{q^2(\bm{r})}\cdot  \right.\right.\left.\left. \displaystyle \sum^N_{\textcolor{black}{j=1}}m_j u_j \nabla W(|\bm{r}-\bm{r}_j|,h(\bm{r})) W(|\bm{r}-\bm{r}_i|,h(\bm{r})) \right.\right.\left.\left. -  \frac{P(\bm{r})u(\bm{r})}{q(\bm{r})} \nabla \left(\left[ \bm{v}(\bm{r}) - \bm{v}_i \right] W(|\bm{r}-\bm{r}_i|,h(\bm{r})) \right) \right) dV\right . \\
    \notag& \quad \quad \quad \quad \quad \quad \left. + \left[ \bm{v}(\bm{r}) - \bm{v}_i \right] \cdot \left(\frac{P(\bm{r})u(\bm{r})}{q^2(\bm{r})} \right) \cdot \right. \left.\displaystyle \sum^N_{\textcolor{black}{j=1}} m_j u_j W(|\bm{r}-\bm{r}_j|,h(\bm{r})) \nabla W(|\bm{r}-\bm{r}_i|,h(\bm{r}))\right] dV\\
    \notag&\approx \int \left[ \left(-\left[ \bm{v}(\bm{r}) - \bm{v}_i \right] \cdot \frac{P(\bm{r})u(\bm{r})}{q^2(\bm{r})} \cdot \right.\right. \left.\left.\displaystyle \sum^N_{\textcolor{black}{j=1}}m_j u_j \nabla W(|\bm{r}-\bm{r}_j|,h(\bm{r})) W(|\bm{r}-\bm{r}_i|,h(\bm{r})) \right.\right.\left.\left. -\frac{P(\bm{r})u(\bm{r})}{q(\bm{r})} \left(\left[ \bm{v}(\bm{r}) - \bm{v}_i \right]\nabla  W(|\bm{r}-\bm{r}_i|,h(\bm{r})) \right) \right) dV\right .\\
    \notag& \left. \quad \quad \quad \quad \quad \quad +\left[ \bm{v}(\bm{r}) - \bm{v}_i \right] \cdot \left(\frac{P(\bm{r})u(\bm{r})}{q^2(\bm{r})} \right) \cdot \right.\left. \displaystyle \sum^N_{\textcolor{black}{j=1}} m_j u_j W(|\bm{r}-\bm{r}_j|,h(\bm{r})) \nabla W(|\bm{r}-\bm{r}_i|,h(\bm{r}))dV\right] \\
    &= \displaystyle \sum^N_{\textcolor{black}{j=1}}m_j u_j \int \frac{P(\bm{r})u(\bm{r})}{q^2(\bm{r})} \left[ \bm{v}(\bm{r}) - \bm{v}_i \right] \cdot \nabla_j W(|\bm{r}-\bm{r}_j|,h(\bm{r})) W(|\bm{r}-\bm{r}_i|,h(\bm{r}))dV.
\end{align}
Then, the energy equation is given by 
\begin{equation}\label{eq:in2033}
    \begin{split}
        m_i\frac{du_i}{dt} &= m_i\left<\frac{du}{dt}\right> (\bm{r}_i), \\
        &=m_i \displaystyle \sum^N_{\textcolor{black}{j=1}}m_j u_j \int \frac{P(\bm{r})u(\bm{r})}{q^2(\bm{r})} \left[ \bm{v}(\bm{r}) - \bm{v}_i \right] \cdot \nabla_j W(|\bm{r}-\bm{r}_j|,h(\bm{r})) W(|\bm{r}-\bm{r}_i|,h(\bm{r}))dV  
    \end{split}
\end{equation}

\begin{multicols}{2}

\section{Results of the Sedov-Taylor tests with various artificial viscosity parameter}
\label{ap:sed_oth}
\begin{figure*}[!b]
\centering
	\includegraphics[width=0.86\linewidth]{ 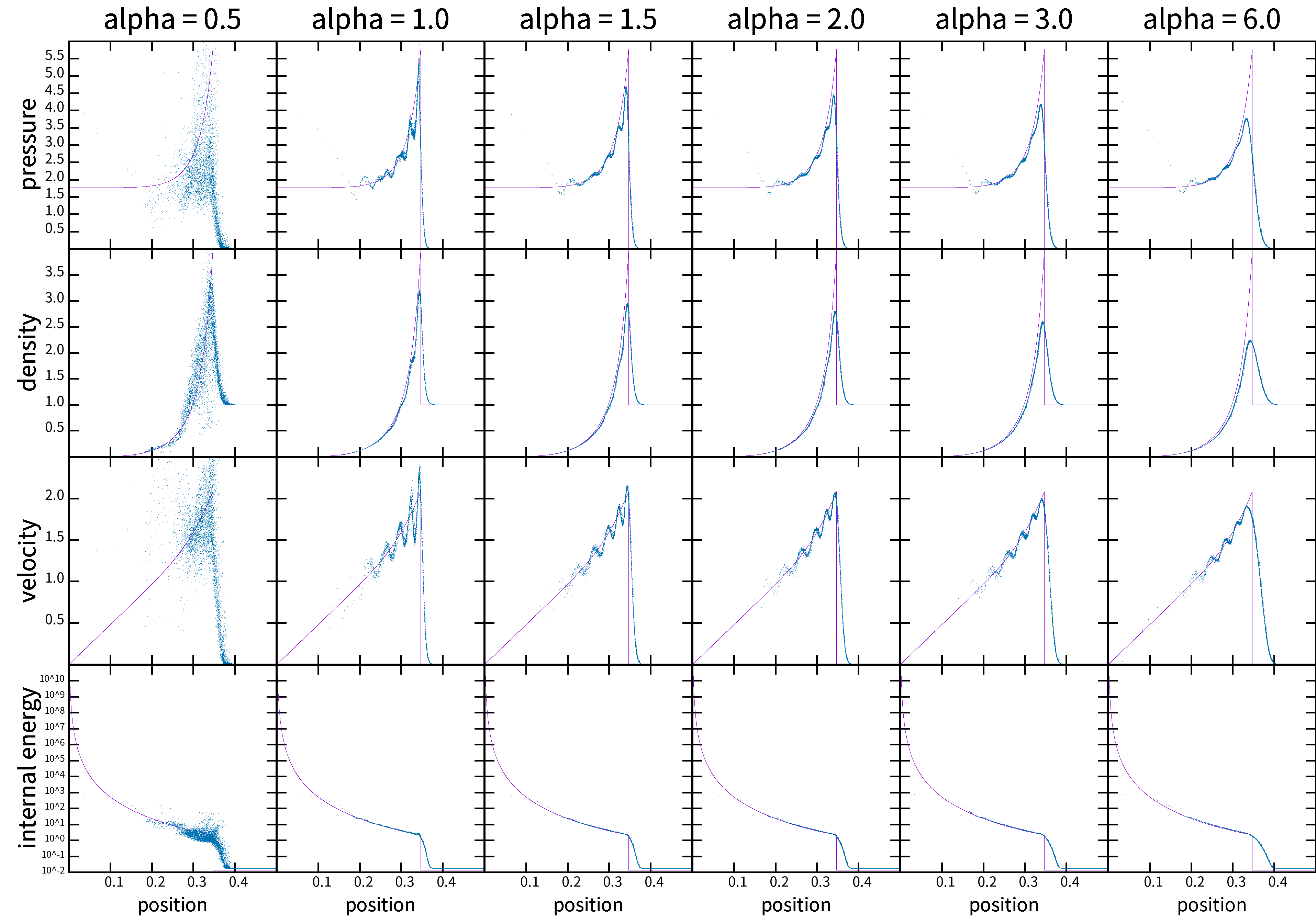}
    \caption{Same as Figure \ref{fig:Problem_DISPH_Ple}, but with SSPH.
    }
    \label{fig:Problem_SSPH_Ple}
\end{figure*}

\begin{figure*}[!t]
\centering
	\includegraphics[width=0.86\linewidth]{ 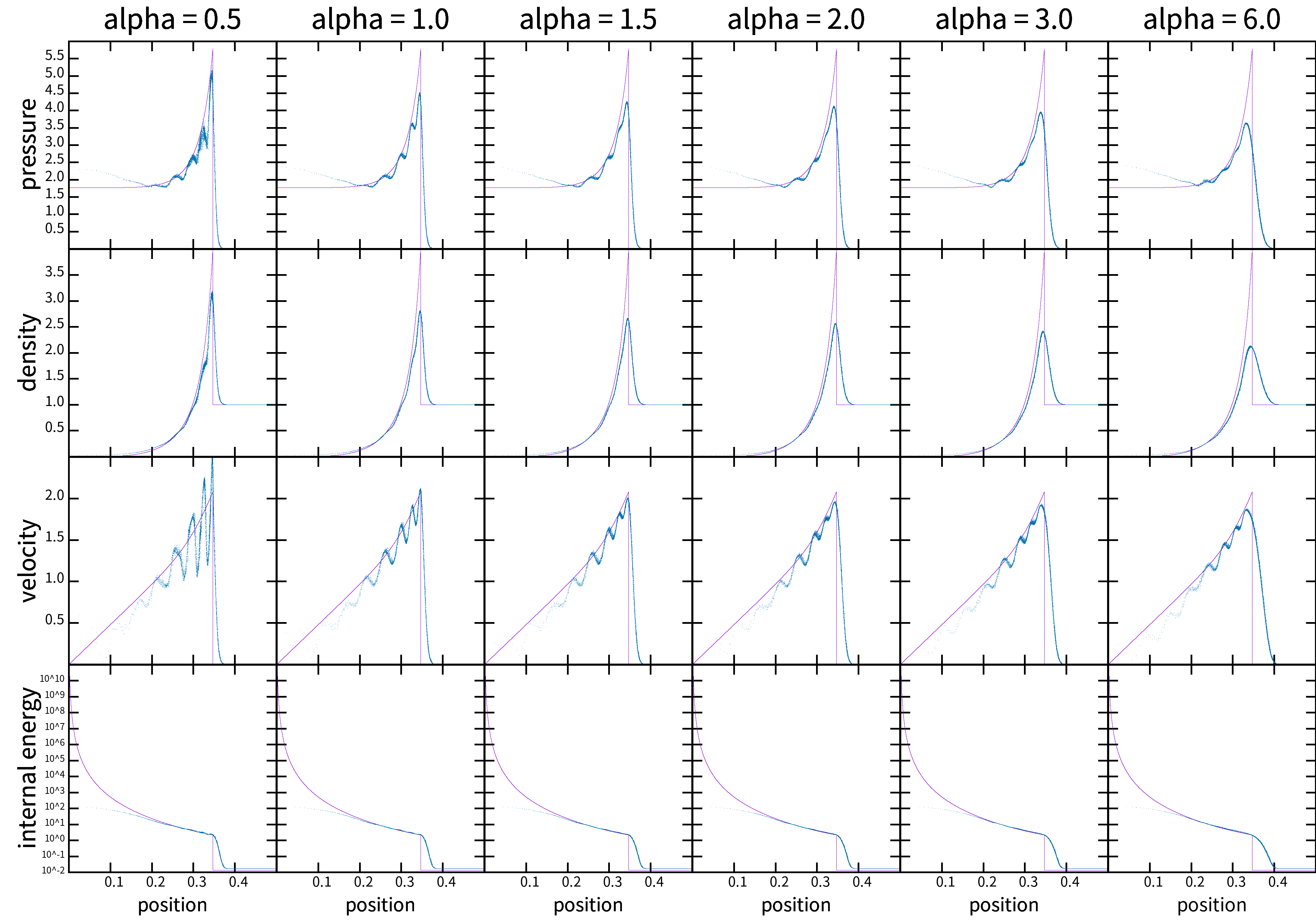}
    \caption{
        Same as Figure \ref{fig:Problem_DISPH_Ple}, but with SSPH with ArtCond.
    }
    \label{fig:Problem_SSPH_ArtCond_Ple}
\end{figure*}

\begin{figure*}[!t]
\centering
	\includegraphics[width=0.86\linewidth]{ 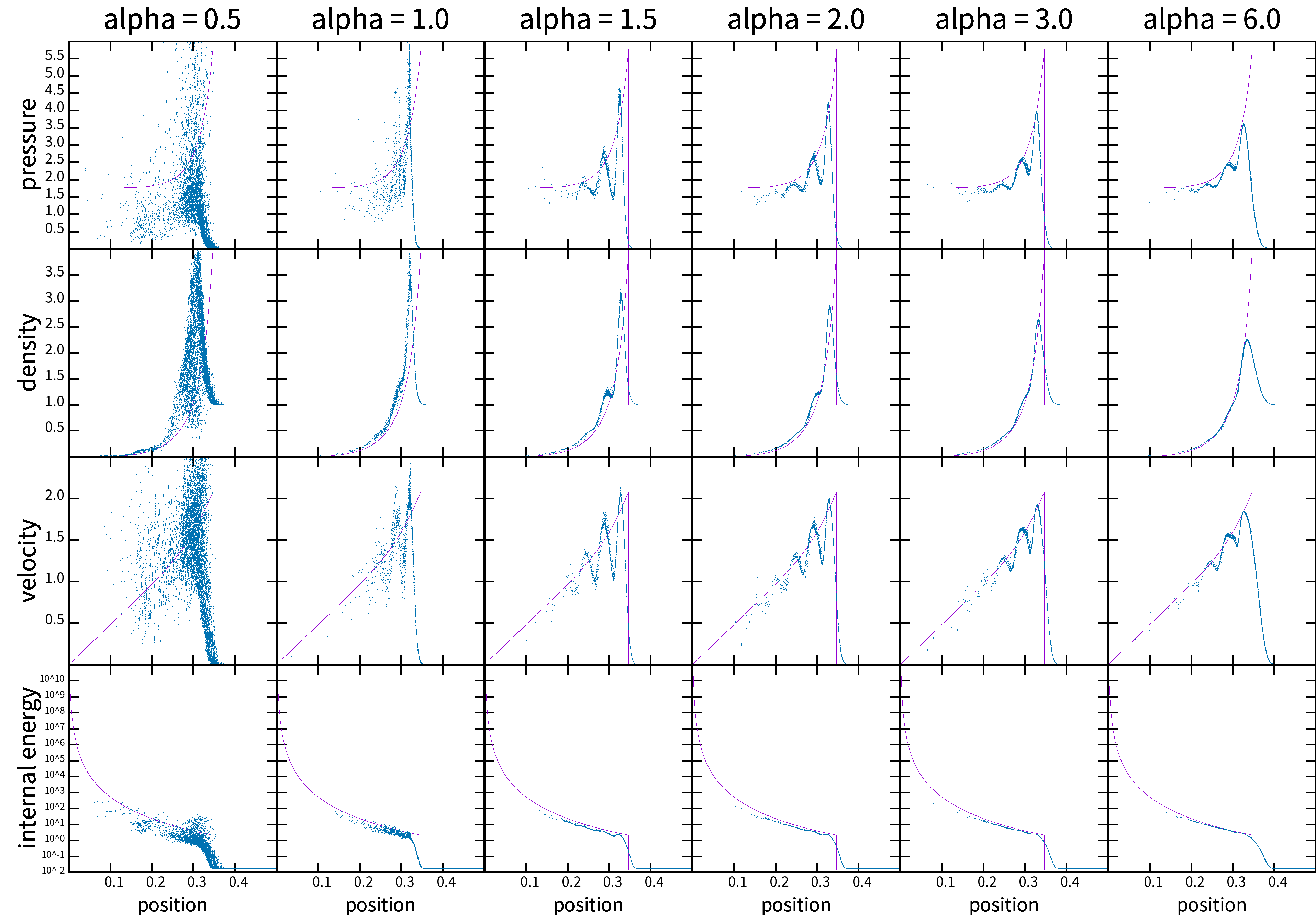}
    \caption{
        Same as Figure \ref{fig:Problem_DISPH_Ple}, but with SPH GDF.
    }
    \label{fig:Problem_SPH_GDF_Ple}
\end{figure*}

\textcolor{black}{
In Figure \ref{fig:Problem_SSPH_Ple}, we present the profiles of physical quantities for the three-dimensional Sedov-Taylor tests with SSPH at $t = 0.05$. The results for pressure, density, velocity, and internal energy are shown along the panels from the top row to the bottom row, respectively, and for $\alpha_{AV} = 0.5, 1.0, 1.5, 2.0, 3.0,$ and $6.0$ are shown from the left column to the right column. 
The $x$-axis represents the distance from the centre.
The solid line indicates the analytical solution, while the dots indicate the physical quantities of each SPH particle at the position.
The results with $\alpha_{AV} \geq 1.0$ are generally consistent with the analytical solution, but the velocity oscillation behind the shock and the pressure oscillation in the low-density region occur at all $\alpha_{AV}$, and there is a slight error in the density profile behind the shock front and a huge pressure error around the low-density area at all $\alpha_{AV}$.
}

\textcolor{black}{
Figure \ref{fig:Problem_SSPH_ArtCond_Ple} shows the results of SSPH with ArtCond.
The trend of the results is the same with SSPH but has less pressure error around the low-density region. However, the velocity and internal energy around the centre are underestimated from the analytical solution.
}

\textcolor{black}{
The results with SPH GDF are shown in Figure \ref{fig:Problem_SPH_GDF_Ple}.
Compared to the results of SSPH, SPH GDF has a considerable velocity and pressure oscillation both behind the shock front and in the \textcolor{black}{low-density} region, and the peak position around the shock front deviated from the analytical solution.
Even the results for $\alpha_{AV} = 6.0$ still show large post-shock oscillations, while the density results lose sharpness near the shock front due to too much artificial viscosity.
In addition, the internal energy behind the shock has more errors than SSPH.
}

\bibliographystyle{elsarticle-harv} 
\bibliography{cite}

\begin{thebibliography}{33}
\expandafter\ifx\csname natexlab\endcsname\relax\def\natexlab#1{#1}\fi
\providecommand{\url}[1]{\texttt{#1}}
\providecommand{\href}[2]{#2}
\providecommand{\path}[1]{#1}
\providecommand{\DOIprefix}{doi:}
\providecommand{\ArXivprefix}{arXiv:}
\providecommand{\URLprefix}{URL: }
\providecommand{\Pubmedprefix}{pmid:}
\providecommand{\doi}[1]{\href{http://dx.doi.org/#1}{\path{#1}}}
\providecommand{\Pubmed}[1]{\href{pmid:#1}{\path{#1}}}
\providecommand{\bibinfo}[2]{#2}
\ifx\xfnm\relax \def\xfnm[#1]{\unskip,\space#1}\fi
\bibitem[{Balsara(1995)}]{Balsara1995}
\bibinfo{author}{Balsara, D.S.}, \bibinfo{year}{1995}.
\newblock \bibinfo{title}{Von neumann stability analysis of smoothed particle hydrodynamics—suggestions for optimal algorithms}.
\newblock \bibinfo{journal}{Journal of Computational Physics} \bibinfo{volume}{121}, \bibinfo{pages}{357--372}.
\newblock \URLprefix \url{https://doi.org/10.1016/S0021-9991(95)90221-X}.
\bibitem[{Beck et~al.(2015)Beck, Murante, Arth, Remus, Teklu, Donnert, Planelles, Beck, Förster, Imgrund, Dolag and Borgani}]{Beck2016}
\bibinfo{author}{Beck, A.M.}, \bibinfo{author}{Murante, G.}, \bibinfo{author}{Arth, A.}, \bibinfo{author}{Remus, R.S.}, \bibinfo{author}{Teklu, A.F.}, \bibinfo{author}{Donnert, J.M.F.}, \bibinfo{author}{Planelles, S.}, \bibinfo{author}{Beck, M.C.}, \bibinfo{author}{Förster, P.}, \bibinfo{author}{Imgrund, M.}, \bibinfo{author}{Dolag, K.}, \bibinfo{author}{Borgani, S.}, \bibinfo{year}{2015}.
\newblock \bibinfo{title}{{An improved SPH scheme for cosmological simulations}}.
\newblock \bibinfo{journal}{Monthly Notices of the Royal Astronomical Society} \bibinfo{volume}{455}, \bibinfo{pages}{2110--2130}.
\newblock \URLprefix \url{https://doi.org/10.1093/mnras/stv2443}.
\bibitem[{Cha and Whitworth(2003)}]{Cha_Whitworth2003}
\bibinfo{author}{Cha, S.H.}, \bibinfo{author}{Whitworth, A.P.}, \bibinfo{year}{2003}.
\newblock \bibinfo{title}{Implementations and tests of godunov-type particle hydrodynamics}.
\newblock \bibinfo{journal}{Monthly Notices of the Royal Astronomical Society} \bibinfo{volume}{340}, \bibinfo{pages}{73--90}.
\newblock \URLprefix \url{https://doi.org/10.1046/j.1365-8711.2003.06266.x}.
\bibitem[{Cullen and Dehnen(2010)}]{Cullen_Dehnen2010}
\bibinfo{author}{Cullen, L.}, \bibinfo{author}{Dehnen, W.}, \bibinfo{year}{2010}.
\newblock \bibinfo{title}{Inviscid smoothed particle hydrodynamics}.
\newblock \bibinfo{journal}{Monthly Notices of the Royal Astronomical Society} \bibinfo{volume}{408}, \bibinfo{pages}{669--683}.
\newblock \URLprefix \url{https://doi.org/10.1111/j.1365-2966.2010.17158.x}.
\bibitem[{Dehnen and Aly(2012)}]{Dehnen2012}
\bibinfo{author}{Dehnen, W.}, \bibinfo{author}{Aly, H.}, \bibinfo{year}{2012}.
\newblock \bibinfo{title}{Improving convergence in smoothed particle hydrodynamics simulations without pairing instability}.
\newblock \bibinfo{journal}{Monthly Notices of the Royal Astronomical Society} \bibinfo{volume}{425}, \bibinfo{pages}{1068--1082}.
\newblock \URLprefix \url{https://doi.org/10.1111/j.1365-2966.2012.21439.x}.
\bibitem[{Gingold and Monaghan(1977)}]{Gingold_Monaghan1977}
\bibinfo{author}{Gingold, R.A.}, \bibinfo{author}{Monaghan, J.J.}, \bibinfo{year}{1977}.
\newblock \bibinfo{title}{Smoothed particle hydrodynamics: theory and application to non-spherical stars}.
\newblock \bibinfo{journal}{Monthly notices of the royal astronomical society} \bibinfo{volume}{181}, \bibinfo{pages}{375--389}.
\newblock \URLprefix \url{https://doi.org/10.1093/mnras/181.3.375}.
\bibitem[{Hopkins(2015)}]{Hopkins2015}
\bibinfo{author}{Hopkins, P.F.}, \bibinfo{year}{2015}.
\newblock \bibinfo{title}{{A new class of accurate, mesh-free hydrodynamic simulation methods}}.
\newblock \bibinfo{journal}{Monthly Notices of the Royal Astronomical Society} \bibinfo{volume}{450}, \bibinfo{pages}{53--110}.
\newblock \URLprefix \url{https://doi.org/10.1093/mnras/stv195}.
\bibitem[{Hosono et~al.(2016a)Hosono, Saitoh and Makino}]{Hosono2016}
\bibinfo{author}{Hosono, N.}, \bibinfo{author}{Saitoh, T.R.}, \bibinfo{author}{Makino, J.}, \bibinfo{year}{2016}a.
\newblock \bibinfo{title}{A {COMPARISON} {OF} {SPH} {ARTIFICIAL} {VISCOSITIES} {AND} {THEIR} {IMPACT} {ON} {THE} {KEPLERIAN} {DISK}}.
\newblock \bibinfo{journal}{The Astrophysical Journal Supplement Series} \bibinfo{volume}{224}, \bibinfo{pages}{32}.
\newblock \URLprefix \url{https://doi.org/10.3847/0067-0049/224/2/32}.
\bibitem[{Hosono et~al.(2016b)Hosono, Saitoh, Makino, Genda and Ida}]{Hosono20161b}
\bibinfo{author}{Hosono, N.}, \bibinfo{author}{Saitoh, T.R.}, \bibinfo{author}{Makino, J.}, \bibinfo{author}{Genda, H.}, \bibinfo{author}{Ida, S.}, \bibinfo{year}{2016}b.
\newblock \bibinfo{title}{The giant impact simulations with density independent smoothed particle hydrodynamics}.
\newblock \bibinfo{journal}{Icarus} \bibinfo{volume}{271}, \bibinfo{pages}{131--157}.
\newblock \URLprefix \url{https://doi.org/10.1016/j.icarus.2016.01.036}.
\bibitem[{Inutsuka(2002)}]{Inutsuka2002}
\bibinfo{author}{Inutsuka, S.i.}, \bibinfo{year}{2002}.
\newblock \bibinfo{title}{Reformulation of smoothed particle hydrodynamics with riemann solver}.
\newblock \bibinfo{journal}{Journal of Computational Physics} \bibinfo{volume}{179}.
\newblock \URLprefix \url{https://doi.org/10.1006/jcph.2002.7053}.
\bibitem[{Iwasaki and Inutsuka(2011)}]{Iwasaki2011}
\bibinfo{author}{Iwasaki, K.}, \bibinfo{author}{Inutsuka, S.i.}, \bibinfo{year}{2011}.
\newblock \bibinfo{title}{Smoothed particle magnetohydrodynamics with a riemann solver and the method of characteristics}.
\newblock \bibinfo{journal}{Monthly Notices of the Royal Astronomical Society} \bibinfo{volume}{418}, \bibinfo{pages}{1668--1688}.
\newblock \URLprefix \url{https://doi.org/10.1111/j.1365-2966.2011.19588.x}.
\bibitem[{Lucy(1977)}]{Lucy1977}
\bibinfo{author}{Lucy, L.B.}, \bibinfo{year}{1977}.
\newblock \bibinfo{title}{A numerical approach to the testing of the fission hypothesis}.
\newblock \bibinfo{journal}{The astronomical journal} \bibinfo{volume}{82}, \bibinfo{pages}{1013--1024}.
\newblock \URLprefix \url{https://doi.org/10.1086/112164}.
\bibitem[{Marin-Gilabert et~al.(2022)Marin-Gilabert, Valentini, Steinwandel and Dolag}]{Gilabert2022}
\bibinfo{author}{Marin-Gilabert, T.}, \bibinfo{author}{Valentini, M.}, \bibinfo{author}{Steinwandel, U.P.}, \bibinfo{author}{Dolag, K.}, \bibinfo{year}{2022}.
\newblock \bibinfo{title}{{The role of physical and numerical viscosity in hydrodynamical instabilities}}.
\newblock \bibinfo{journal}{Monthly Notices of the Royal Astronomical Society} \bibinfo{volume}{517}, \bibinfo{pages}{5971--5991}.
\newblock \URLprefix \url{https://doi.org/10.1093/mnras/stac3042}.
\bibitem[{Monaghan(1992)}]{Monaghan1992}
\bibinfo{author}{Monaghan, J.J.}, \bibinfo{year}{1992}.
\newblock \bibinfo{title}{Smoothed particle hydrodynamics}.
\newblock \bibinfo{journal}{Annual review of astronomy and astrophysics} \bibinfo{volume}{30}, \bibinfo{pages}{543--574}.
\newblock \URLprefix \url{https://doi.org/10.1146/annurev.aa.30.090192.002551}.
\bibitem[{Monaghan(1997)}]{Monaghan1997}
\bibinfo{author}{Monaghan, J.J.}, \bibinfo{year}{1997}.
\newblock \bibinfo{title}{Sph and riemann solvers}.
\newblock \bibinfo{journal}{Journal of Computational Physics} \bibinfo{volume}{136}, \bibinfo{pages}{298--307}.
\newblock \URLprefix \url{https://doi.org/10.1006/jcph.1997.5732}.
\bibitem[{Monaghan and Gingold(1983)}]{Monaghan1983}
\bibinfo{author}{Monaghan, J.J.}, \bibinfo{author}{Gingold, R.A.}, \bibinfo{year}{1983}.
\newblock \bibinfo{title}{Shock simulation by the particle method sph}.
\newblock \bibinfo{journal}{Journal of computational physics} \bibinfo{volume}{52}, \bibinfo{pages}{374--389}.
\newblock \URLprefix \url{https://doi.org/10.1016/0021-9991(83)90036-0}.
\bibitem[{Murante et~al.(2011)Murante, Borgani, Brunino and Cha}]{Murante2011}
\bibinfo{author}{Murante, G.}, \bibinfo{author}{Borgani, S.}, \bibinfo{author}{Brunino, R.}, \bibinfo{author}{Cha, S.H.}, \bibinfo{year}{2011}.
\newblock \bibinfo{title}{Hydrodynamic simulations with the godunov smoothed particle hydrodynamics}.
\newblock \bibinfo{journal}{Monthly Notices of the Royal Astronomical Society} \bibinfo{volume}{417}, \bibinfo{pages}{136--153}.
\newblock \URLprefix \url{https://doi.org/10.1111/j.1365-2966.2011.19021.x}.
\bibitem[{Price(2008)}]{Price2008}
\bibinfo{author}{Price, D.J.}, \bibinfo{year}{2008}.
\newblock \bibinfo{title}{Modelling discontinuities and kelvin--helmholtz instabilities in sph}.
\newblock \bibinfo{journal}{Journal of Computational Physics} \bibinfo{volume}{227}, \bibinfo{pages}{10040--10057}.
\newblock \URLprefix \url{https://doi.org/10.1016/j.jcp.2008.08.011}.
\bibitem[{Price et~al.(2018)Price, Wurster, Tricco, Nixon, Toupin, Pettitt, Chan, Mentiplay, Laibe, Glover et~al.}]{Price2018}
\bibinfo{author}{Price, D.J.}, \bibinfo{author}{Wurster, J.}, \bibinfo{author}{Tricco, T.S.}, \bibinfo{author}{Nixon, C.}, \bibinfo{author}{Toupin, S.}, \bibinfo{author}{Pettitt, A.}, \bibinfo{author}{Chan, C.}, \bibinfo{author}{Mentiplay, D.}, \bibinfo{author}{Laibe, G.}, \bibinfo{author}{Glover, S.}, et~al., \bibinfo{year}{2018}.
\newblock \bibinfo{title}{Phantom: A smoothed particle hydrodynamics and magnetohydrodynamics code for astrophysics}.
\newblock \bibinfo{journal}{Publications of the Astronomical Society of Australia} \bibinfo{volume}{35}.
\newblock \URLprefix \url{https://doi.org/10.1017/pasa.2018.25}.
\bibitem[{Read et~al.(2010)Read, Hayfield and Agertz}]{Read2010}
\bibinfo{author}{Read, J.}, \bibinfo{author}{Hayfield, T.}, \bibinfo{author}{Agertz, O.}, \bibinfo{year}{2010}.
\newblock \bibinfo{title}{Resolving mixing in smoothed particle hydrodynamics}.
\newblock \bibinfo{journal}{Monthly Notices of the Royal Astronomical Society} \bibinfo{volume}{405}, \bibinfo{pages}{1513--1530}.
\newblock \URLprefix \url{https://doi.org/10.1111/j.1365-2966.2010.16577.x}.
\bibitem[{Ritchie and Thomas(2001)}]{Ritchie2001}
\bibinfo{author}{Ritchie, B.W.}, \bibinfo{author}{Thomas, P.A.}, \bibinfo{year}{2001}.
\newblock \bibinfo{title}{Multiphase smoothed-particle hydrodynamics}.
\newblock \bibinfo{journal}{Monthly Notices of the Royal Astronomical Society} \bibinfo{volume}{323}, \bibinfo{pages}{743--756}.
\newblock \URLprefix \url{https://doi.org/10.1046/j.1365-8711.2001.04268.x}.
\bibitem[{Saitoh and Makino(2013)}]{Saitoh_Makino2013}
\bibinfo{author}{Saitoh, T.R.}, \bibinfo{author}{Makino, J.}, \bibinfo{year}{2013}.
\newblock \bibinfo{title}{A {DENSITY}-{INDEPENDENT} {FORMULATION} {OF} {SMOOTHED} {PARTICLE} {HYDRODYNAMICS}}.
\newblock \bibinfo{journal}{The Astrophysical Journal} \bibinfo{volume}{768}, \bibinfo{pages}{44}.
\newblock \URLprefix \url{https://doi.org/10.1088/0004-637x/768/1/44}.
\bibitem[{Saitoh and Makino(2016)}]{Saitoh_Makino2016}
\bibinfo{author}{Saitoh, T.R.}, \bibinfo{author}{Makino, J.}, \bibinfo{year}{2016}.
\newblock \bibinfo{title}{Santa barbara cluster comparison test with disph}.
\newblock \bibinfo{journal}{The Astrophysical Journal} \bibinfo{volume}{823}, \bibinfo{pages}{144}.
\newblock \URLprefix \url{https://dx.doi.org/10.3847/0004-637X/823/2/144}.
\bibitem[{Schaller et~al.(2016)Schaller, Gonnet, Chalk and Draper}]{Schaller2016}
\bibinfo{author}{Schaller, M.}, \bibinfo{author}{Gonnet, P.}, \bibinfo{author}{Chalk, A.B.G.}, \bibinfo{author}{Draper, P.W.}, \bibinfo{year}{2016}.
\newblock \bibinfo{title}{Swift: Using task-based parallelism, fully asynchronous communication, and graph partition-based domain decomposition for strong scaling on more than 100,000 cores}, in: \bibinfo{booktitle}{Proceedings of the Platform for Advanced Scientific Computing Conference}, \bibinfo{publisher}{Association for Computing Machinery}, \bibinfo{address}{New York, NY, USA}.
\newblock \URLprefix \url{https://doi.org/10.1145/2929908.2929916}, \DOIprefix\doi{10.1145/2929908.2929916}.
\bibitem[{Springel(2005)}]{Springel2005}
\bibinfo{author}{Springel, V.}, \bibinfo{year}{2005}.
\newblock \bibinfo{title}{The cosmological simulation code gadget-2}.
\newblock \bibinfo{journal}{Monthly notices of the royal astronomical society} \bibinfo{volume}{364}, \bibinfo{pages}{1105--1134}.
\newblock \URLprefix \url{https://doi.org/10.1111/j.1365-2966.2005.09655.x}.
\bibitem[{Springel(2010a)}]{Springel2010A}
\bibinfo{author}{Springel, V.}, \bibinfo{year}{2010}a.
\newblock \bibinfo{title}{{E pur si muove: Galilean-invariant cosmological hydrodynamical simulations on a moving mesh}}.
\newblock \bibinfo{journal}{Monthly Notices of the Royal Astronomical Society} \bibinfo{volume}{401}, \bibinfo{pages}{791--851}.
\newblock \URLprefix \url{https://doi.org/10.1111/j.1365-2966.2009.15715.x}.
\bibitem[{Springel(2010b)}]{Springel2010}
\bibinfo{author}{Springel, V.}, \bibinfo{year}{2010}b.
\newblock \bibinfo{title}{Smoothed particle hydrodynamics in astrophysics}.
\newblock \bibinfo{journal}{Annual Review of Astronomy and Astrophysics} \bibinfo{volume}{48}, \bibinfo{pages}{391--430}.
\newblock \URLprefix \url{https://doi.org/10.1146/annurev-astro-081309-130914}.
\bibitem[{Springel and Hernquist(2002)}]{Springel2002}
\bibinfo{author}{Springel, V.}, \bibinfo{author}{Hernquist, L.}, \bibinfo{year}{2002}.
\newblock \bibinfo{title}{{Cosmological smoothed particle hydrodynamics simulations: the entropy equation}}.
\newblock \bibinfo{journal}{Monthly Notices of the Royal Astronomical Society} \bibinfo{volume}{333}, \bibinfo{pages}{649--664}.
\newblock \URLprefix \url{https://doi.org/10.1046/j.1365-8711.2002.05445.x}.
\bibitem[{Toro(2009)}]{Toro2009}
\bibinfo{author}{Toro, E.F.}, \bibinfo{year}{2009}.
\newblock \bibinfo{title}{The HLL and HLLC Riemann Solvers}. \bibinfo{publisher}{Springer Berlin Heidelberg}, \bibinfo{address}{Berlin, Heidelberg}.
\newblock pp. \bibinfo{pages}{315--344}.
\newblock \URLprefix \url{https://doi.org/10.1007/b79761\_10}.
\bibitem[{{van Leer}(1979)}]{VanLeer1979}
\bibinfo{author}{{van Leer}, B.}, \bibinfo{year}{1979}.
\newblock \bibinfo{title}{Towards the ultimate conservative difference scheme. v. a second-order sequel to godunov's method}.
\newblock \bibinfo{journal}{Journal of Computational Physics} \bibinfo{volume}{32}, \bibinfo{pages}{101--136}.
\newblock \URLprefix \url{https://doi.org/10.1016/0021-9991(79)90145-1}.
\bibitem[{Wadsley et~al.(2017)Wadsley, Keller and Quinn}]{Wadsley2017}
\bibinfo{author}{Wadsley, J.W.}, \bibinfo{author}{Keller, B.W.}, \bibinfo{author}{Quinn, T.R.}, \bibinfo{year}{2017}.
\newblock \bibinfo{title}{Gasoline2: a modern smoothed particle hydrodynamics code}.
\newblock \bibinfo{journal}{Monthly Notices of the Royal Astronomical Society} \bibinfo{volume}{471}, \bibinfo{pages}{2357--2369}.
\newblock \URLprefix \url{https://doi.org/10.1093/mnras/stx1643}.
\bibitem[{Wendland(1995)}]{Wendland1995}
\bibinfo{author}{Wendland, H.}, \bibinfo{year}{1995}.
\newblock \bibinfo{title}{Piecewise polynomial, positive definite and compactly supported radial functions of minimal degree}.
\newblock \bibinfo{journal}{Advances in computational Mathematics} \bibinfo{volume}{4}, \bibinfo{pages}{389--396}.
\newblock \URLprefix \url{https://doi.org/10.1007/BF02123482}.
\bibitem[{Zhu et~al.(2015)Zhu, Hernquist and Li}]{Zhu2015}
\bibinfo{author}{Zhu, Q.}, \bibinfo{author}{Hernquist, L.}, \bibinfo{author}{Li, Y.}, \bibinfo{year}{2015}.
\newblock \bibinfo{title}{{NUMERICAL} {CONVERGENCE} {IN} {SMOOTHED} {PARTICLE} {HYDRODYNAMICS}}.
\newblock \bibinfo{journal}{The Astrophysical Journal} \bibinfo{volume}{800}, \bibinfo{pages}{6}.
\newblock \URLprefix \url{https://doi.org/10.1088/0004-637x/800/1/6}.

\end{thebibliography}





\end{multicols}
\end{document}